\input epsf
\documentstyle[epsfig,eqsecnum,aps,prd,tighten,preprint,floats]{revtex}

\draft 

\begin{document}

\title{Detecting a stochastic background of gravitational radiation:\\
Signal processing strategies and sensitivities}
\author{Bruce Allen%
\thanks{Electronic address: ballen@dirac.phys.uwm.edu}
and Joseph D.~Romano%
\thanks{Current address: 
LIGO Project, California Institute of Technology, MS 18-34,
Pasadena, CA 91125 (jromano@ligo.caltech.edu); and 
Department of Physics and Astronomy, Northwestern University, 
Evanston, IL 60208 (j-romano@nwu.edu)}}
\address{Department of Physics, University of Wisconsin-Milwaukee, 
Milwaukee WI 53201}

\maketitle
\begin{abstract}
We analyze the signal processing required for the optimal detection 
of a stochastic background of gravitational radiation using laser 
interferometric detectors.
Starting with basic assumptions about the statistical properties of 
a stochastic gravity-wave background, we derive expressions 
for the optimal filter function and signal-to-noise ratio for the
cross-correlation of the outputs of two gravity-wave detectors.
Sensitivity levels required for detection are then calculated.
Issues related to: (i) calculating the signal-to-noise ratio for 
arbitrarily large stochastic backgrounds, (ii) performing the data 
analysis in the presence of nonstationary detector noise, (iii) 
combining data from multiple detector pairs to increase the sensitivity
of a stochastic background search, (iv) correlating the outputs of 
4 or more detectors, and (v) allowing for the possibility of correlated 
noise in the outputs of two detectors are discussed.
We briefly describe a computer simulation that was used to 
``experimentally'' verify the theoretical calculations derived in the 
paper, and which mimics the generation and detection of a simulated 
stochastic gravity-wave signal in the presence of simulated detector noise.
Numerous graphs and tables of numerical data for the five major 
interferometers (LIGO-WA, LIGO-LA, VIRGO, GEO-600, and TAMA-300) are 
also given.
This information consists of graphs of the noise power spectra, 
overlap reduction functions, and optimal filter functions; also included
are tables of the signal-to-noise ratios and sensitivity levels 
for cross-correlation measurements between different detector pairs.  
The treatment given in this paper should be accessible to both
theorists involved in data analysis and experimentalists involved
in detector design and data acquisition.

\end{abstract}

\narrowtext\noindent 

\section{Introduction}
\label{sec:introduction}

The design and construction of a number of new and more sensitive
detectors of gravitational radiation is currently underway.  
These include the two 
LIGO detectors being built in Hanford, WA and
Livingston, LA by a joint Caltech/MIT collaboration \cite{science92}, 
the VIRGO detector being built near Pisa, Italy by an Italian/French 
collaboration\cite{virgo}, the GEO-600 detector being built in Hanover,
Germany by an Anglo/German collaboration\cite{geo600}, and the TAMA-300 
detector being built near Tokyo, Japan \cite{tama300}.  
There are also several resonant bar detectors currently in operation, 
and several more refined bar and interferometric detectors presently 
in the planning and proposal stages.

The operation of these detectors will have a major impact on the field
of gravitational physics.
For the first time, there will be a significant amount of experimental 
data to be analyzed, and the ``ivory tower" relativists will be forced 
to interact with a broad range of experimenters and data analysts to 
extract the interesting physics from the data stream.  
``Known'' sources such as coalescing neutron star (or black hole) 
binaries, pulsars, supernovae, and other periodic and transient (or burst) 
sources should all be observable with gravity-wave detectors.
One might also be able to detect a faint stochastic background of 
gravitational radiation, produced very shortly after the big bang.
These detections may happen soon after the detectors go ``on-line'' 
or they may require a decade of further work to increase the sensitivity 
of the instruments.
But it is fairly safe to say that eventually, when their sensitivity passes 
some threshold value, the gravity-wave detectors {\em will\,} find sources.
Even more exciting is the prospect that the detectors will discover 
{\em new\,} sources of gravitational radiation---sources which are 
different from those mentioned above, and which we had not expected to
find.
It promises to be an exciting time.

The subject of this paper is a {\em stochastic\,} (i.e., random) background 
of gravitational radiation, first studied in detail by Michelson\cite{mich}, 
Christensen\cite{chris}, and Flanagan\cite{flan}.
Roughly speaking, it is the type of gravitational radiation produced by 
an extremely large number of weak, independent, and unresolved gravity-wave 
sources.
The radiation is stochastic in the sense that it can be characterized
only statistically.
As mentioned above, a stochastic background of gravitational radiation might 
be the result of processes that took place very shortly after the big bang.
But since we know very little about the state of the universe at that time, 
it is impossible to say with any certainty.  
A stochastic background of gravitational radiation might also arise from more 
recent processes (e.g., radiation from many unresolved binary star systems), 
and this more recent contribution might overwhelm the parts of the 
background that contain information about the state of the early universe.
In any case, the properties of the radiation will be very dependent upon 
the source.
For example, one would expect a stochastic background of cosmological 
origin to be highly isotropic, whereas that produced by white dwarf 
binaries in our own galaxy would be highly anisotropic.
We will just have to wait and see what the detectors reveal before we can 
decide between these two possibilities.

This paper will focus on issues related to the {\em detection\,} of a
stochastic background of gravitational radiation.
(We will not talk much about possible sources.)
We give a complete and comprehensive treatment of the problem of 
detecting a stochastic background, which should be accessible to both 
theorists involved in the data analysis and experimentalists involved 
in detector design and data acquisition.

The outline of the paper is as follows:

In Sec.~\ref{sec:properties}, we begin by describing the properties of 
a stochastic background of gravitational radiation---its spectrum, 
statistical assumptions, and current observational constraints.

In Sec.~\ref{sec:detection}, we describe how one can correlate the
outputs of two gravity-wave detectors to detect (or put an upper limit 
on) a stochastic gravity-wave signal.
Section~\ref{subsec:overlap} includes a detailed derivation of the overlap 
reduction function that covers the case where the two arms of a detector 
are not perpendicular (e.g., GEO-600) and corrects a typographical error
that appears in the literature.
Section~\ref{subsec:optimal_filtering} includes a rigorous derivation of the 
optimal signal processing strategy.
Most of the material in Secs.~\ref{sec:properties} and \ref{sec:detection} 
has already appeared in the literature.
Interested readers should see Ref.~\cite{leshouches} for more details, 
if desired.

In Sec.~\ref{sec:detection_etc}, we ask the questions:
(i) ``How do we decide, from the experimental data, if we've detected a 
stochastic gravity-wave signal?''
(ii) ``Assuming that a stochastic gravity-wave signal is present, 
how do we estimate its strength?''
(iii) ``Assuming that a stochastic gravity-wave signal is present, 
what is the minimum value of $\Omega_0$ required to detect it 95\% of 
the time?''
This leads to a discussion of signal detection, parameter estimation,
and sensitivity levels for stochastic background searches, adopting a
frequentist point of view.
The calculation of $\Omega_0^{95\%,5\%}$ in 
Sec.~\ref{subsec:sensitivity_levels} 
corrects an error that has appeared in the literature.

In Sec.~\ref{sec:complications}, we return to the problem of detection by 
addressing a series of subtle issues initially ignored in 
Sec.~\ref{sec:detection}.
These include: (i) calculating the the signal-to-noise ratio for
arbitrarily large stochastic backgrounds, 
(ii) performing the data analysis 
in the presence of nonstationary detector noise, (iii) combining data 
from multiple detector pairs to increase the sensitivity of a stochastic 
background search, (iv) correlating the outputs of 4 or more detectors,
and (v) allowing for the possibility of correlated noise in the outputs 
of two detectors.
The material presented in these sections extends the initial treatment 
of these issues given, for example, in Refs.~\cite{chris,flan}.

Section~\ref{sec:numerical_data} consists of a series of graphs and
tables of numerical data for the five major interferometers
(LIGO-WA, LIGO-LA, VIRGO, GEO-600, TAMA-300).
The noise power spectra, 
overlap reduction functions, optimal filter functions, signal-to-noise 
ratios, and sensitivity levels for cross-correlation measurements
between different detector pairs (not just LIGO) are given.
This information allows us to determine the optimal way of combining 
data from multiple detector pairs when searching for a stochastic background
of gravitational radiation.
Section~\ref{subsec:nps} also includes a graph of the ``enhanced'' LIGO 
detector noise curves, which track the projected performance of the LIGO 
detector design over the next decade.

In Sec.~\ref{sec:computer_simulation}, we describe a computer simulation 
that mimics the generation and detection of a simulated stochastic 
gravity-wave signal in the presence of simulated detector noise. 
The simulation was used to verify some of the theoretical calculations
derived in the previous sections.

Section~\ref{sec:conclusion} concludes the paper with a brief summary
and lists some topics for future work.

Note: Throughout the paper, we use $c$ to denote the speed of light and 
$G$ to denote Newton's gravitational constant 
($c=2.998\times 10^{10}\ {\rm cm/sec}$ and $G=6.673\times 10^{-8}\ 
{\rm cm}^3/{\rm gm}\cdot{\rm sec}^2$).

\section{The Stochastic Background - Properties}
\label{sec:properties}

\subsection{Spectrum}
\label{subsec:spectrum}

A stochastic background of gravitational radiation is a random 
gravity-wave signal produced by a large number of weak, independent, and 
unresolved gravity-wave sources.
In many ways it is analogous to the Cosmic Microwave Background 
Radiation (CMBR) \cite{kolbturner}, which is a stochastic background of 
{\em electromagnetic\,} radiation.
As with the CMBR, it is useful to characterize the spectral properties
of the gravitational background by specifying how the energy is distributed 
in frequency.
Explicitly, one introduces the dimensionless quantity
\begin{equation}
\Omega_{\rm gw}(f):={1\over\rho_{\rm critical}}\ 
{d\rho_{\rm gw}\over d\ln f}\ ,
\label{e:Omega_gw}
\end{equation}
where $d\rho_{\rm gw}$ is the energy density of the gravitational
radiation contained in the frequency range $f$ to $f+df$, and
$\rho_{\rm critical}$ is the critical energy density required (today) 
to close the universe:
\begin{equation} 
\rho_{\rm critical} = {3 c^2 H_0^2\over 8 \pi G} 
\approx 1.6\times 10^{-8}\ h_{100}^2\ {\rm ergs\over cm^3}\ .
\label{e:rho_crit}
\end{equation}
$H_0$ is the Hubble expansion rate (today):
\widetext%
\begin{equation} 
H_0 = h_{100} \cdot 100\ {\rm  km\over sec\cdot Mpc} 
= 3.2 \times 10^{-18}\ h_{100}\ {\rm 1 \over sec}  
= 1.1 \times 10^{-28}\ c \cdot h_{100}\ {\rm 1 \over cm}\ ,
\label{e:H_0}
\end{equation}
\narrowtext\noindent%
and $h_{100}$ is a dimensionless factor, included to account 
for the different values of $H_0$ that are quoted in the literature.%
\footnote{$h_{100}$ almost certainly lies within the range $1/2<h_{100}<1$.}
It is this dimensionless function of frequency, $\Omega_{\rm gw}(f)$,
that we will use to describe the spectrum of a stochastic background of 
gravitational radiation.
It follows directly from the above definitions that 
$\Omega_{\rm gw}(f)\ h_{100}^2$ is {\em independent\,} of the actual Hubble 
expansion rate.
For this reason, we will often focus attention on this quantity, rather
than on $\Omega_{\rm gw}(f)$ alone.

Two remarks are in order:

(i) There appears to be be some confusion about $\Omega_{\rm gw}(f)$ in
the literature.
Some authors assume that $\Omega_{\rm gw}(f)$ is constant---i.e., independent 
of frequency.
Although this is true for some cosmological models, it is not true for all
of them.
The important point is that {\em any\,} spectrum of gravitational radiation
can be described by an appropriate $\Omega_{\rm gw}(f)$.
With the correct dependence on frequency, $\Omega_{\rm gw}(f)$
can describe a flat spectrum, a blackbody spectrum, or any other 
distribution of energy with frequency.

(ii) $\Omega_{\rm gw}(f)$ is the ratio of the stochastic gravity-wave 
energy density contained in a bandwidth $\Delta f=f$ to the total energy 
density required to close the universe.
For the CMBR, one can define an analogous quantity:
\begin{equation}
\Omega_{\rm em}(f):={1\over\rho_{\rm critical}}\ 
{d\rho_{\rm em}\over d\ln f}\ .
\label{e:Omega_em}
\end{equation}
Since the 2.73 K blackbody spectrum has a peak value of
$\Omega_{\rm em}(f)\ h_{100}^2\approx 10^{-5}$ 
at $f=10^{12}$ Hz, the CMBR
contains (in the vicinity of $10^{12}$ Hz) approximately $10^{-5}$ of 
the total energy density required to close the universe.
A similar interpretation applies to $\Omega_{\rm gw}(f)$.

\subsection{Statistical assumptions}
\label{subsec:statistical_assumptions}

The spectrum $\Omega_{\rm gw}(f)$ completely specifies the stochastic
background of gravitational radiation provided we make enough additional
assumptions.
We will assume that the stochastic background is: (i) isotropic, (ii) 
unpolarized, (iii) stationary, and (iv) Gaussian.
Since these properties might not hold in general, it is 
worthwhile to consider each one of them in turn.

(i) Since it is now well established that the CMBR is highly isotropic 
\cite{kolbturner}, it is not unreasonable to assume that a stochastic 
background of gravitational radiation is also isotropic. 
But this assumption might not be true.  
For example, as mentioned in Sec.~\ref{sec:introduction}, 
if the dominant source of the stochastic gravity-wave background is a large 
number of unresolved white dwarf binary star systems within our own galaxy, 
then the stochastic background will have a distinctly {\em anisotropic\,} 
distribution, which forms a ``band in the sky" distributed roughly in 
the same way as the Milky Way galaxy. 
It is also possible for a stochastic gravity-wave background of 
cosmological origin to be anisotropic, although one would then
have to explain why the CMBR is isotropic but the gravity-wave background 
is not.
In either case, such anisotropies {\em can} be searched for in the data
stream.
(See Ref.~\cite{allenottewill} for details.)

(ii) The second assumption is that the stochastic gravity-wave background 
is unpolarized.
This means that the gravitational radiation incident on a detector has 
statistically equivalent ``plus'' and ``cross'' polarization components.
We see no strong reason why this should not be the case.

(iii) The assumption that the stochastic background is stationary (i.e., 
that all statistical quantities depend only upon the difference between 
times, and not on the choice of time origin) is almost certainly justified.  
This is because the age of the universe is roughly 20 orders of magnitude 
larger than the characteristic period of the waves that LIGO, VIRGO, etc.\ 
can detect, and 9 orders of magnitude larger than the 
longest realistic observation times.
It seems very unlikely that a stochastic background of gravitational
radiation would have statistical properties that vary over either of 
these time-scales.
But unlike the stochastic gravity-wave background, the noise intrinsic to 
the detectors {\em will\,} change over the course of the observation times.
This poses a problem for the data analysis, which we initially ignore 
in Sec.~\ref{sec:detection}.
We return to this problem in Sec.~\ref{subsec:nonstationary_detector_noise} 
where we discuss nonstationary detector noise. 

(iv) The final assumption is that the stochastic gravity-wave 
background is a Gaussian random process.
This means that the joint probability density function of the gravitational 
strains $h_i(t_i),h_j(t_j),\cdots\ $ in detectors $i,j,\cdots\ $
is a multivariate Gaussian (i.e., normal) distribution. 
In this case, the mean values $\langle h_i(t)\rangle$ and the second-order 
moments $\langle h_i(t_i)h_j(t_j)\rangle$ completely specify the statistical 
properties of the signal.
For many early-universe processes, or even for more recent sources of 
a gravity-wave background, this is a reasonable assumption.
It can be justified by the central limit theorem, which says that the sum 
of a large number of statistically independent random variables is a 
Gaussian random variable, independent of the probability distributions of 
the original variables.
This will be the case for the stochastic background if it is the sum of 
gravity-wave signals produced by a large number of independent gravity-wave 
sources.
This assumption will not be true, however, if the stochastic
background is the sum of the radiation produced, e.g.,  by 
only a few unresolved binary star systems radiating in a given frequency
interval at any instant of time.
(See, e.g., Ref.~\cite{Blair}.)

The above four properties form the basis for the statistical analysis 
that we will give in the following sections.
We will assume that they hold throughout, unless we explicitly state 
otherwise.

\subsection{Expectation value}
\label{subsec:expectation_value}

Using the definition of the spectrum $\Omega_{\rm gw}(f)$ and the 
statistical assumptions described in the previous subsection, 
we can derive a useful result for the
expectation value of the Fourier amplitudes of a stochastic background 
of gravitational radiation.
This result will be needed in Sec.~\ref{sec:detection} when we discuss 
signal detection and optimal filtering.

The starting point of the derivation is a plane wave expansion for 
the gravitational metric perturbations in a transverse, traceless gauge:
\widetext%
\begin{equation}
h_{ab}(t,\vec x)=\sum_A\int_{-\infty}^\infty df\ \int_{S^2} 
d\hat\Omega\ h_A(f,\hat \Omega)\ e^{i2\pi f(t-\hat \Omega\cdot\vec x/c)}\ 
e_{ab}^A(\hat \Omega)\ .
\label{e:h_ab}
\end{equation}
\narrowtext\noindent%
Here $\hat\Omega$ is a unit vector specifying a direction on the 
two-sphere, with wave vector $\vec k:=2\pi f\hat\Omega/c$.
Also, $e_{ab}^A(\hat\Omega)$ are the spin-two polarization 
tensors for the ``plus'' and ``cross'' polarization states $A=+,\times$.
Explicitly, 
\begin{eqnarray}
e_{ab}^+(\hat\Omega) & = &\hat m_a\hat m_b - \hat n_a\hat n_b\ ,
\label{e:e_ab^+}\\
e_{ab}^\times(\hat\Omega) & = &\hat m_a\hat n_b + \hat n_a\hat m_b\ ,
\label{e:e_ab^x}
\end{eqnarray}
where
\begin{eqnarray}
\hat \Omega &=&\cos\phi\sin\theta\ \hat x + \sin\phi\sin\theta\ 
\hat y + \cos\theta\ \hat z\ ,
\label{e:Omega}\\
\hat m& = & \sin\phi\ \hat x - \cos\phi\ \hat y\ , 
\label{e:m}\\
\hat n& = & \cos\phi\cos\theta\ \hat x + \sin\phi\cos\theta\ \hat y 
- \sin\theta\ \hat z\ ,
\label{e:n}
\end{eqnarray}
and $(\theta,\phi)$ are the standard polar and azimuthal angles on
the two-sphere.
The Fourier amplitudes $h_A(f,\hat \Omega)$ are arbitrary complex 
functions that satisfy
$h_A(-f,\hat \Omega) = h_A^*(f,\hat \Omega)$, 
where $*$ denotes complex conjugation.
This last relation follows as a consequence of the reality of 
$h_{ab}(t,\vec x)$.

The assumptions that the stochastic background is isotropic, unpolarized,
and stationary imply that the expectation value
(i.e., ensemble average) of the Fourier amplitudes $h_A(f,\hat\Omega)$
satisfies:
\widetext%
\begin{equation}
\langle h_A^*(f, \hat \Omega) h_{A'}(f',\hat \Omega')\rangle=
\delta^2(\hat \Omega ,\hat \Omega')\delta_{AA'}\delta(f-f')\ H(f)\ ,
\label{e:ev1}
\end{equation}
\narrowtext\noindent%
where $\delta^2(\hat\Omega,\hat\Omega'):=\delta(\phi-\phi')
\delta(\cos\theta-\cos\theta')$ 
is the covariant Dirac delta function on the two-sphere, and $H(f)$ is a 
real, non-negative function, satisfying $H(f)=H(-f)$.%
\footnote{If the stochastic background is anisotropic, we should replace 
$H(f)$ by a function that depends on $\hat\Omega$ in addition to $f$.
If the stochastic background is polarized, we should replace $H(f)$ 
by a function that depends on the polarization $A=+,\times$ as well.}
If we further assume that the stochastic background has zero mean, then
\begin{equation}
\langle h_A(f,\hat\Omega)\rangle=0\ .
\label{e:ev2} 
\end{equation}
Finally, since we are assuming that the stochastic background is 
Gaussian, the expectation values (\ref{e:ev1}) and 
(\ref{e:ev2}) {\em completely\,} specify its statistical properties.

$H(f)$ is related to the spectrum $\Omega_{\rm gw}(f)$ of the stochastic 
gravity-wave background.
This follows from the expression 
\begin{equation} 
\rho_{\rm gw} = {c^2 \over 32 \pi G}\  
\langle \dot h_{ab}(t,\vec x) \dot h^{ab}(t,\vec x) \rangle
\label{e:rho_gw}
\end{equation}
for the energy density in gravitational waves (see, e.g., p.955
of Ref.~\cite{MTW}).
By differentiating the plane wave expansion (\ref{e:h_ab}) with 
respect to $t$, forming the contraction in Eq.~(\ref{e:rho_gw}), and
calculating the expectation value using (\ref{e:ev2}), we find
\begin{equation}
\rho_{\rm gw}={4\pi^2 c^2\over G}\int_0^\infty df\ f^2 H(f)\quad
\left(=:\int_0^\infty df\ {d\rho_{\rm gw}\over df}\ \right)\ .
\label{e:rho_gw2}
\end{equation}
Using Eqs.~(\ref{e:Omega_gw}) and (\ref{e:rho_crit})
for $\Omega_{\rm gw}(f)$, then yields
\begin{equation}
H(f)={3H_0^2\over 32\pi^3}\ |f|^{-3}\ \Omega_{\rm gw}(|f|)\ .
\label{e:H(f)}
\end{equation}
Thus,
\widetext%
\begin{equation}
\langle h_A^*(f,\hat\Omega)h_{A'}(f',\hat\Omega')\rangle =
{3H_0^2\over 32\pi^3}\ \delta^2(\hat \Omega ,\hat \Omega')
\delta_{AA'}\delta(f-f')\ |f|^{-3}\ \Omega_{\rm gw}(|f|)\ ,
\label{e:hAhA'}
\end{equation}
\narrowtext\noindent%
which is the desired result.

\subsection{Observational constraints}
\label{subsec:observational_constraints}

At present, there are three observational constraints on the stochastic
gravity-wave spectrum $\Omega_{\rm gw}(f)$.
These constraints are quite weak in the frequency range of interest for 
ground-based interferometers 
($1\ {\rm Hz} < f < 10^3\ {\rm Hz}$) 
and for proposed space-based detectors 
($10^{-4}\ {\rm Hz} < f <10^{-1} \ {\rm Hz}$).
There are tighter constraints on the spectrum in two frequency ranges, 
and one ``wideband'' but very weak constraint.
In this paper, we simply state the constraints.
For a more complete discussion, see Ref.~\cite{leshouches} and the 
references mentioned therein.

(i) The strongest observational constraint on $\Omega_{\rm gw}(f)$ comes 
from the high degree of isotropy observed in the CMBR.
In particular, the one-year\cite{cobea,cobeb}, two-year\cite{cobe2}, and
four-year\cite{cobe4} data sets from the Cosmic Background Explorer 
(COBE) satellite place very strong restrictions on $\Omega_{\rm gw}(f)$ 
at very low frequencies:
\widetext%
\begin{equation} 
\Omega_{\rm gw}(f)\  h_{100}^2 < 7 \times 10^{-11} 
\left({H_0\over f} \right)^2 
\quad{\rm for}\quad  
H_0 < f < 30 H_0\ .
\end{equation}
\narrowtext\noindent%
Note that the above constraint does not apply to any gravitational 
wave, but only to those of cosmological origin that were already present 
at the time of last scattering of the CMBR.
Also, since $H_0=3.2\times 10^{-18}\ h_{100}\ {\rm Hz}$, 
this limit applies only over a narrow band of frequencies 
$(10^{-18}\ {\rm Hz}<f<10^{-16}\ {\rm Hz})$, 
which is far below any frequency band accessible to investigation by either 
earth-based or space-based interferometers.
Thus, although this constraint is severe, it is not directly relevant for 
any of the present-day gravity-wave experiments.

(ii) The second observational constraint comes from almost a decade of 
monitoring the radio pulses arriving from a number of stable millisecond 
pulsars \cite{Taylor}.  
These pulsars are remarkably stable clocks, and the regularity of their 
pulses places tight constraints on $\Omega_{\rm gw}(f)$ at frequencies on 
the order of the inverse of the observation time of the pulsars 
($\sim 10^{-8}$ Hz):
\begin{equation}
\Omega_{\rm gw}(f=10^{-8}\ {\rm Hz})\ h_{100}^2 < 10^{-8}\ .
\end{equation}
Like the constraint on the stochastic gravity-wave background from the 
isotropy of the CMBR, 
the millisecond pulsar timing constraint is irrelevant for
current gravity-wave experiments.
The frequency $f=10^{-8}$ Hz is 10 orders of magnitude smaller than the 
band of frequencies accessible to LIGO, VIRGO, etc., and 4 orders of 
magnitude smaller than that for proposed space-based detectors.

(iii) The third and final observational constraint on $\Omega_{\rm gw}(f)$ 
comes from the standard model of big-bang nucleosynthesis \cite{kolbturner}.
This model provides remarkably accurate fits to the observed abundances
of the light elements in the universe, tightly constraining a number of
key cosmological parameters.  
One of the parameters constrained in this way is the expansion 
rate of the universe at the time of nucleosynthesis.  
This places a constraint on the energy density of the universe 
at that time, which in turn constrains the energy density in a cosmological 
background of gravitational radiation:
\begin{equation}
\int_{f>10^{-8}\ {\rm Hz}} d\ln f\ \Omega_{\rm gw }(f)\ h_{100}^2 
< 10^{-5}\ .
\end{equation}
Although this bound constrains the spectrum of gravitational radiation 
$\Omega_{\rm gw}(f)$ over a broad range of frequencies, it is not 
very restrictive.

\section{The Stochastic Background - Detection}
\label{sec:detection}

In this section, we begin our detailed discussion of the detection of 
a stochastic background of gravitational radiation.
We explain how one can correlate the outputs of two gravity-wave
detectors to detect (or put an upper limit on) a stochastic 
background signal.
In Sec.~\ref{subsec:overlap}, we give a detailed derivation of the overlap 
reduction function that covers the case where the two arms of a detector 
are not perpendicular (e.g., GEO-600).
In Sec.~\ref{subsec:optimal_filtering}, 
we give a rigorous derivation of the optimal signal processing strategy.
The statistical assumptions that we will use for the stochastic 
gravity-wave background are those described in 
Sec.~\ref{subsec:statistical_assumptions}.
In addition, we will assume that the noise intrinsic to the detectors are:
(i) stationary, (ii) Gaussian, (iii) statistically independent of one 
another and of the stochastic gravity-wave background, and (iv) much 
larger in magnitude than the stochastic gravity-wave background.
The modifications that are necessary when one relaxes most of these 
assumptions will be discussed in Sec.~\ref{sec:complications}.

\subsection{Coincident and coaligned detectors}
\label{subsec:coincident_coaligned}

To begin, let us consider the simplest possible case.
Let us suppose that we have two {\em coincident} and {\em coaligned} 
gravity-wave detectors with outputs 
\begin{eqnarray}
s_1(t)&:=&h_1(t) + n_1(t)\ ,\\
s_2(t)&:=&h_2(t) + n_2(t)\ .
\end{eqnarray}
Here $h_1(t)$ and $h_2(t)$ denote the gravitational strains in the two 
detectors due to the stochastic background, and $n_1(t)$ and $n_2(t)$ denote
the noise intrinsic to the first and second detector, respectively.%
\footnote{We will assume throughout that the detector outputs are not 
whitened.}
Since we are assuming that the two detectors are coincident and 
coaligned (i.e., have identical locations and arm orientations), the 
gravitational strains are identical:
\begin{equation}
h(t):=h_1(t)=h_2(t)\ .
\end{equation}  
But the noise $n_1(t)$ and $n_2(t)$ are {\em not\,} equal to one 
another.
As mentioned above, we will assume that they are
stationary, Gaussian, statistically independent of one another and of
the gravitational strain, and much larger in magnitude than the 
gravitational strains.%
\footnote{The assumption that the noise intrinsic to the detectors are 
statistically independent of one another is unrealistic for the case of
coincident and coaligned detectors.
But it is a reasonable assumption for widely-separated detector sites.
(See Sec.~\ref{subsec:correlated_detector_noise} for more details.)} 

Given the detector outputs $s_1(t)$ and $s_2(t)$, we can form a product 
``signal" $S$ by multiplying them together and integrating over time:
\begin{equation} 
S:= \int_{-T/2}^{T/2}dt\  s_1(t) s_2(t)\ .
\end{equation}
This quantity is proportional to the (zero-lag) cross-correlation of 
$s_1(t)$ and $s_2(t)$ for an observation time $T$.
Since $s_1(t)$ and $s_2(t)$ are random variables, so too is $S$.
It has a mean value 
\begin{equation}
\mu:=\langle S\rangle\ ,
\label{e:mu_label}
\end{equation}
and variance 
\begin{equation}
\sigma^2:=\langle S^2\rangle-\langle S\rangle^2\ ,
\label{e:sigma2_label}
\end{equation}
which are related to 
the variances of $n_1(t)$, $n_2(t)$, and $h(t)$.%
\footnote{The mean values of $n_1(t)$, $n_2(t)$, and $h(t)$ are equal 
to zero, either by assumption or by definition.}
The goal is to calculate $\mu$ and $\sigma$, and then to construct the 
signal-to-noise ratio
\begin{equation}
{\rm SNR}:={\mu\over\sigma}\ .
\end{equation}
As we shall see in Sec.~\ref{sec:detection_etc}, the value of 
the signal-to-noise ratio enters the decision rule for the detection
of a stochastic gravity-wave signal.

Let us start with the mean value $\mu$.
By definition,
\widetext%
\begin{eqnarray}
\mu
&:=&\langle S\rangle=\int_{-T/2}^{T/2} dt\ \langle s_1(t) s_2(t)\rangle\\
&=&\int_{-T/2}^{T/2} dt\ \langle h^2(t)+h(t)n_2(t)+n_1(t)h(t)
+n_1(t)n_2(t)\rangle\\
&=&\int_{-T/2}^{T/2} dt\ \langle h^2(t)\rangle\\
&=&T\ \langle h^2(t)\rangle=:T\ \sigma_h^2\ ,
\end{eqnarray}
\narrowtext\noindent%
where $\sigma_h^2$ denotes the (time-independent) variance of $h(t)$.%
\footnote{The dimensions of $\sigma_h^2$ and $\sigma^2$ are different:
$\sigma_h^2$ has dimensions of ${\rm strain}^2$, while
$\sigma^2$ has dimensions of ${\rm strain}^4\cdot{\rm sec}^2$.
(See, e.g., Eq.~(\ref{e:var_1a}).)}
Note that we used the statistical independence of $n_1(t)$, $n_2(t)$, and
$h(t)$ to obtain the third line, and the stationarity of $h(t)$ to obtain
the last.

To express the variance $\sigma_h^2:=\langle h^2(t)\rangle$ in terms of 
the frequency spectrum $\Omega_{\rm gw}(f)$, we will make use of the plane 
wave expansion (\ref{e:h_ab}) and the expectation value (\ref{e:hAhA'}).
Since
\begin{equation}
h(t):=h_{ab}(t,\vec x_0)\ 
{1\over 2}\left(\hat X^a\hat X^b - \hat Y^a\hat Y^b\right)
\label{e:h(t)}
\end{equation}
(where $\vec x_0$ is the common position vector of the central station
of the two coincident and coaligned detectors, and $\hat X^a$ and 
$\hat Y^a$ are unit vectors pointing in the directions of the detector 
arms),%
\footnote{$\vec x_0$ and $\hat X^a$, $\hat Y^a$ are actually functions 
of {\em time\,} due to the Earth's motion with respect to the cosmological 
rest frame.
They can be treated as constants, however, since: (i) the velocity of the 
Earth with respect to the cosmological rest frame is small compared to
the speed of light, and (ii) the distance that the central stations and
arms move during the correlation time between the two detectors is small 
compared to the arm length.
(The correlation time equals zero for coincident and coaligned detectors;
it equals the light travel time between the two detectors when the 
detectors are spatially separated.)
See Ref.~\cite{allenottewill} for more details.}
it follows that
\widetext%
\begin{eqnarray}
\sigma_h^2
&=&\sum_A\sum_{A'}\int_{S^2}d\hat\Omega \int_{S^2}d\hat\Omega'
\int_{-\infty}^{infty}df\int_{-\infty}^{\infty}df'\ 
\langle h_A^*(f,\hat\Omega) h_{A'}(f',\hat\Omega')\rangle 
\nonumber\\
&&\quad\quad\times\ e^{-i2\pi f(t- \hat\Omega\cdot\vec x_0/c)}\ 
e^{i2\pi f'(t- \hat\Omega'\cdot\vec x_0/c)}\ 
F^A(\hat\Omega)F^{A'}(\hat\Omega')\ ,
\end{eqnarray}
\narrowtext\noindent%
where
\begin{equation} 
F^A(\hat\Omega):= e_{ab}^A(\hat\Omega)\
{1\over 2}\left(\hat X^a\hat X^b - \hat Y^a\hat Y^b\right)\ 
\label{e:F^A}
\end{equation}
is the response of either detector to a zero frequency, unit amplitude,
$A=+,\times$ polarized gravitational wave.
Using (\ref{e:hAhA'}) for the expectation value 
$\langle h_A^*(f,\hat\Omega) h_{A'}(f',\hat\Omega')\rangle$, 
the above expression for $\sigma_h^2$ simplifies to
\widetext%
\begin{eqnarray}
\sigma_h^2&=&{3 H_0^2\over 32\pi^3}\ 
\int_{-\infty}^\infty df\ |f|^{-3}\ \Omega_{\rm gw}(|f|)\ 
\sum_{A}\int_{S^2}d\hat\Omega\ F^A(\hat\Omega)F^A(\hat\Omega)\\ 
&=&{3 H_0^2\over 20\pi^2}\ 
\int_{-\infty}^\infty df\ |f|^{-3}\ \Omega_{\rm gw}(|f|)\ ,
\end{eqnarray}
\narrowtext\noindent%
where we used 
\begin{equation}
\sum_{A}\int_{S^2}d\hat\Omega\ F^A(\hat\Omega)F^A(\hat\Omega)
={8\pi\over 5}
\label{e:8pi5}
\end{equation}
to obtain the last line.
Thus, for coincident and coaligned detectors, the mean value of the 
cross-correlation signal $S$ is
\begin{equation}
\mu={3 H_0^2\over 20\pi^2}\ T\ \int_{-\infty}^\infty df\ 
|f|^{-3}\ \Omega_{\rm gw}(|f|)\ .
\label{e:mu}
\end{equation}
This is the first of our desired results.

\begin{figure}[htb!]
\begin{center}
{\epsfig{file=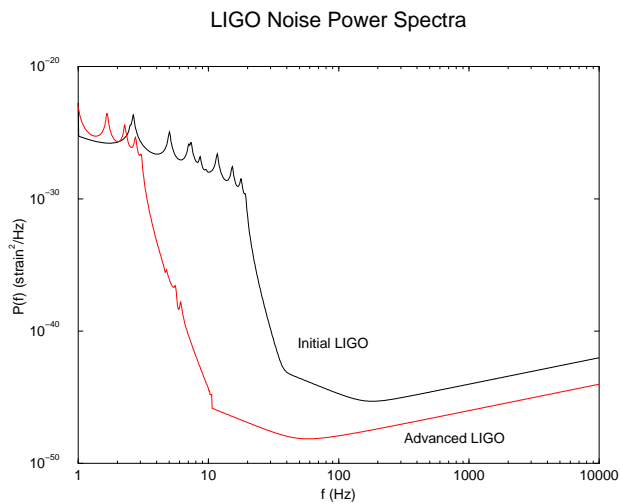,
angle=-90,width=3.4in,bbllx=25pt,bblly=50pt,bburx=590pt,bbury=740pt}}
\caption{\label{f:noise_ligo}
A log-log plot of the predicted noise power spectra for the initial and 
advanced LIGO detectors.   
The data for these noise power spectra were taken from 
the published design goals \protect\cite{science92}.}
\end{center}
\end{figure}

To evaluate the variance $\sigma^2$, we will make use of the 
assumption that the noise intrinsic to the detectors are much larger in 
magnitude than the gravitational strain.
Then 
\widetext%
\begin{eqnarray}
\sigma^2
&:=&\langle S^2\rangle-\langle S\rangle^2\approx\langle S^2\rangle
\label{e:var_1}\\
&=&\int_{-T/2}^{T/2}dt\int_{-T/2}^{T/2}dt'\ 
\langle s_1(t)s_2(t)s_1(t')s_2(t')\rangle
\label{e:var_1a}\\
&\approx&\int_{-T/2}^{T/2}dt\int_{-T/2}^{T/2}dt'\ 
\langle n_1(t)n_2(t)n_1(t')n_2(t')\rangle\\
&=&\int_{-T/2}^{T/2}dt\int_{-T/2}^{T/2}dt'\ 
\langle n_1(t)n_1(t')\rangle\langle n_2(t)n_2(t')\rangle\ ,
\label{e:var}
\end{eqnarray}
\narrowtext\noindent%
where we used the statistical independence of $n_1(t)$ and $n_2(t)$
to obtain the last line.
By definition,%
\footnote{Equation~(\ref{e:P_i(f)}) can also be written in the
frequency domain:
\begin{displaymath}
\langle\tilde n_i^*(f)\tilde n_i(f')\rangle=
{1\over 2}\delta(f-f')\ P_i(|f|)\ .
\end{displaymath}
See the discussion surrounding Eq.~(\ref{e:n_i(f)n_i(f')}) for more
details.}
\begin{equation}
\langle n_i(t)n_i(t')\rangle=:{1\over 2}\int_{-\infty}^\infty df\ 
e^{i2\pi f(t-t')}\ P_i(|f|)\ ,
\label{e:P_i(f)}
\end{equation}
where $P_i(|f|)$ is the (one-sided) {\em noise power spectrum\,} of the
$i$th detector ($i=1,2$).
$P_i(|f|)$ is a real, non-negative function, defined with a factor of 
$1/2$ to agree with the standard (one-sided) definition used by 
instrument-builders.
It satisfies%
\footnote{Unlike $\sigma_h^2$ and $\sigma^2$, $\sigma_h^2$ and 
$\sigma_{n_i}^2$ have the same dimensions (${\rm strain}^2$).}
\begin{equation}
\sigma_{n_i}^2:=\langle n_i^2(t)\rangle =\int_0^\infty df\ P_i(f)\ ,
\end{equation}
so the total noise power is the integral of $P_i(f)$ over all 
{\em positive\,} frequencies $f$ from $0$ to $\infty$, not from 
$-\infty$ to $\infty$. 
(Hence the reason for the name {\em one-sided}.)
Graphs of the predicted noise power spectra for the initial and 
advanced LIGO detectors are shown in Fig.~\ref{f:noise_ligo}.
Graphs of the predicted noise power spectra for the other major
interferometers (i.e., VIRGO, GEO-600, and TAMA-300) 
and for the ``enhanced'' LIGO detectors are shown in 
Figs.~\ref{f:noise_virgo}-\ref{f:enhanced} in Sec.~\ref{subsec:nps}.

Inserting (\ref{e:P_i(f)}) into (\ref{e:var}) yields
\widetext%
\begin{eqnarray}
\sigma^2 
&\approx&{1\over 4}\int_{-T/2}^{T/2}dt\int_{-T/2}^{T/2}dt'
\int_{-\infty}^\infty df\int_{-\infty}^\infty df'
\nonumber\\ 
&&\quad\quad\times\ e^{i2\pi f(t-t')}\ e^{-i2\pi f'(t-t')}\ 
P_1(|f|)P_2(|f'|)\ ,
\end{eqnarray}
\narrowtext\noindent%
where we used the reality of $n_2(t)$ and $P_2(|f|)$ to produce
the minus sign in the power of the second exponential.
If we integrate this expression over $t$ and $t'$, we find
\begin{equation}
\sigma^2\approx
{1 \over 4}\int_{-\infty}^\infty df\int_{-\infty}^\infty df'\
\delta_T^2(f-f') P_1(|f|) P_2(|f'|)\ ,
\end{equation}
where
\begin{equation} 
\delta_T(f):=\int_{-T/2}^{T/2}dt\ e^{-i2\pi ft}
= {\sin(\pi f T)\over\pi f}
\label{e:delta_T(f)}
\end{equation}
is a finite-time approximation to the Dirac delta function
$\delta(f)$. 
In the limit $T\to\infty$, $\delta_T(f)$ reduces to $\delta(f)$,
but for a finite observation time $T$, one has $\delta_T(0)=T$.  
Since in practice the observation time $T$ will be large enough so that 
$\delta_T(f-f')$ is sharply peaked over a region in $f-f'$ whose size 
$\approx 1/T$ is very small compared to the scale very on which the 
functions $P_1(|f|)$ and $P_2(|f|)$ are varying,%
\footnote{Typically, an observation time $T$ will be on the 
order of months (i.e., $10^7$ seconds), while the noise power spectra
$P_i(|f|)$ vary on a scale of greater than a few Hz.}
we can replace one of the 
finite-time delta functions $\delta_T(f-f')$ by an ordinary Dirac 
delta function, and evaluate the other at $f=f'$. 
Doing this yields
\begin{equation} 
\sigma^2\approx{T \over 4}
\int_{-\infty}^\infty df\ P_1(|f|)P_2(|f|)\ ,
\label{e:sigma^2}
\end{equation}
which is the second of our desired results.

Using (\ref{e:mu}) and (\ref{e:sigma^2}), we can now form the
signal-to-noise ratio%
\footnote{Warning!!  
The signal processing strategy described above, leading to 
Eq.~(\ref{e:snr}), is {\em not\,} 
optimal even for the case of coincident and coaligned detectors.
The optimal signal processing strategy, which is described in
Sec.~\ref{subsec:optimal_filtering}, leads to the signal-to-noise
ratio given by Eq.~(\ref{e:SNR_large_noise}).
Setting $\gamma(f)=1$ in Eq.~(\ref{e:SNR_large_noise}) yields an
expression for the optimally-filtered signal-to-noise ratio for the
case of coincident and coaligned detectors.}
\widetext%
\begin{equation}
{\rm SNR}:={\mu\over\sigma}\approx
{{3H_0^2\over 10\pi^2}\ \sqrt{T}\ 
{\int_{-\infty}^\infty df\ |f|^{-3}\ \Omega_{\rm gw}(|f|)\over
\left[\int_{-\infty}^\infty df \> P_1(|f|) P_2(|f|)\right]^{1/2}}}\ .
\label{e:snr}
\end{equation}
\narrowtext\noindent%
The multiplicative factor of $\sqrt{T}$ means that we can 
{\em always\,} exceed any prescribed value of the signal-to-noise
ratio by correlating the outputs of two gravity-wave detectors for a 
long enough period of time.%
\footnote{This assumes that there is no systematic source of correlated
detector noise.
In Sec.~\ref{subsec:correlated_detector_noise}, we discuss the
limits that correlated detector noise impose.}
We will have more to say about signal detection, parameter estimation,
and sensitivity levels for stochastic background searches in 
Sec.~\ref{sec:detection_etc}.

\subsection{Overlap reduction function}
\label{subsec:overlap}

To provide a rigorous treatment of the signal analysis for a stochastic
background of gravitational radiation, we must take into account the fact 
that the two gravity-wave detectors will not necessarily be either coincident 
or coaligned.
There will be a reduction in sensitivity due to: 
(i) the separation time delay between the two detectors, and
(ii) the non-parallel alignment of the detector arms.
These two effects imply that $h_1(t)$ and $h_2(t)$ are no longer equal;
the overlap between the gravitational strains in the two detectors is only 
partial.
Statistically, these effects are most apparent in the frequency domain.

The {\em overlap reduction function\,} $\gamma(f)$, first calculated in 
closed-form by Flanagan \cite{flan}, quantifies these two effects.
This is a dimensionless function of frequency $f$, which is determined
by the relative positions and orientations of a pair of detectors.  
Explicitly,
\widetext%
\begin{equation}
\gamma(f):={5\over 8\pi}\sum_A\int_{S^2}d\hat\Omega\
e^{i2\pi f\hat\Omega\cdot\Delta \vec x/c}\ 
F_1^A(\hat\Omega)F_2^A(\hat\Omega)\ ,
\label{e:gamma(f)}
\end{equation}
\narrowtext\noindent%
where $\hat \Omega$ is a unit vector specifying a direction on 
the two-sphere, $\Delta\vec x:=\vec x_1-\vec x_2$ is the separation 
vector between the central stations of the two detector sites, and 
\begin{equation}
F_i^A(\hat\Omega):=e_{ab}^A(\hat\Omega)\ d_i^{ab}:=
e_{ab}^A(\hat\Omega)\ {1\over 2}\left(\hat X_i^a\hat X_i^b-
\hat Y_i^a\hat Y_i^b\right)\ 
\label{e:F_i^A}
\end{equation}
is the response of the $i$th detector $(i=1,2)$ to the $A=+,\times$ 
polarization.  
(See also Eq.~(\ref{e:F^A}).)
The symmetric, tracefree tensor $d_i^{ab}$ specifies the orientation
of the two arms of the $i$th detector.
The overlap reduction function $\gamma(f)$ equals unity for coincident
and coaligned detectors.
It decreases below unity when the detectors are shifted apart 
(so there is a phase shift between the signals in the 
two detectors), or  rotated out of coalignment (so the detectors are 
sensitive to different polarizations).
In Sec.~\ref{subsec:optimal_filtering}, 
we will see that $\gamma(f)$ arises naturally when evaluating
the expectation value of the product of the gravitational 
strains at two different detectors when they are driven by an isotropic 
and unpolarized stochastic background of gravitational radiation.

To get a better feeling for the meaning of $\gamma(f)$, let us look at 
each term in Eq.~(\ref{e:gamma(f)}) separately:
(i) The overall normalization factor $5/8\pi$ is chosen so that for a pair 
of coincident and coaligned detectors $\gamma(f)=1$ for all frequencies $f$.
(ii) The sum over polarizations $A$ is appropriate for an unpolarized
stochastic background.
(iii) The integral over the two-sphere is an isotropic average over all 
directions $\hat\Omega$ of the incoming radiation.
(iv) The exponential phase factor is the phase shift arising from the time 
delay between the two detectors for radiation arriving along the direction 
$\hat \Omega$.  
In the limit $f\rightarrow 0$, this phase shift also goes to zero, and the 
two detectors become effectively coincident.
(v) The quantity 
$\sum_A F_1^A(\hat\Omega)F_2^A(\hat\Omega)$
is the sum of products of the responses of the two detectors to the $+$ 
and $\times$ polarization waves.
For coaligned detectors, $F_1^A(\hat\Omega)=F_2^A(\hat\Omega)$ and the 
integral of this quantity over the two-sphere equals the inverse of the 
overall normalization factor.
(See Eq.~(\ref{e:8pi5}).)

\begin{figure}[htb!]
\begin{center}
{\epsfig{file=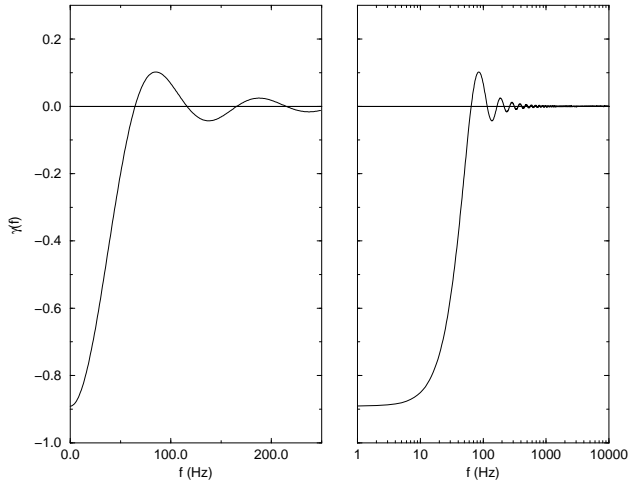,
angle=-90,width=3.4in,bbllx=25pt,bblly=25pt,bburx=600pt,bbury=740pt}}
\caption{\label{f:overlap}
The overlap reduction function $\gamma(f)$ for the Hanford, WA and 
Livingston, LA LIGO detector pair.
(The horizontal axis of the left-hand graph is linear, while that of the 
right-hand graph is $\rm log_{10}$.)  
The overlap reduction function has its first zero at 64 Hz, as explained
in the text.  
It falls off rapidly at higher frequencies.}
\end{center}
\end{figure}

Figure~\ref{f:overlap} shows a graph of the overlap reduction function
$\gamma(f)$ for the Hanford, WA and Livingston, LA LIGO detector pair.%
\footnote{Figures~\ref{f:LIGO-WA_overlap}-\ref{f:TAMA-300_overlap} in
Sec.~\ref{subsec:orf} show graphs of the overlap reduction functions for 
different detector pairs.}
Note that the overlap reduction function for the LIGO detector pair
is {\em negative\,} as $f \rightarrow 0$.  
This is because the arm orientations of the two LIGO detectors are 
not parallel to one another, but are rotated by $90^\circ$.  
If, for example, the Livingston, LA detector arms were rotated by 
$90^\circ$ in the clockwise direction, only the overall sign of $\gamma(f)$
would change.
Note also that the magnitude of $\gamma(0)$ is not unity, because the 
planes of the Hanford, WA and Livingston, LA detectors are not identical.%
\footnote{The two LIGO detectors are separated by an angle of $27.2^\circ$ 
as seen from the center of the earth.}
Thus, the arms of the two detectors are not exactly parallel, and
$|\gamma(0)|=0.89$, which is less than $1$.

\begin{figure}[htb!]
\begin{center}
{\epsfig{file=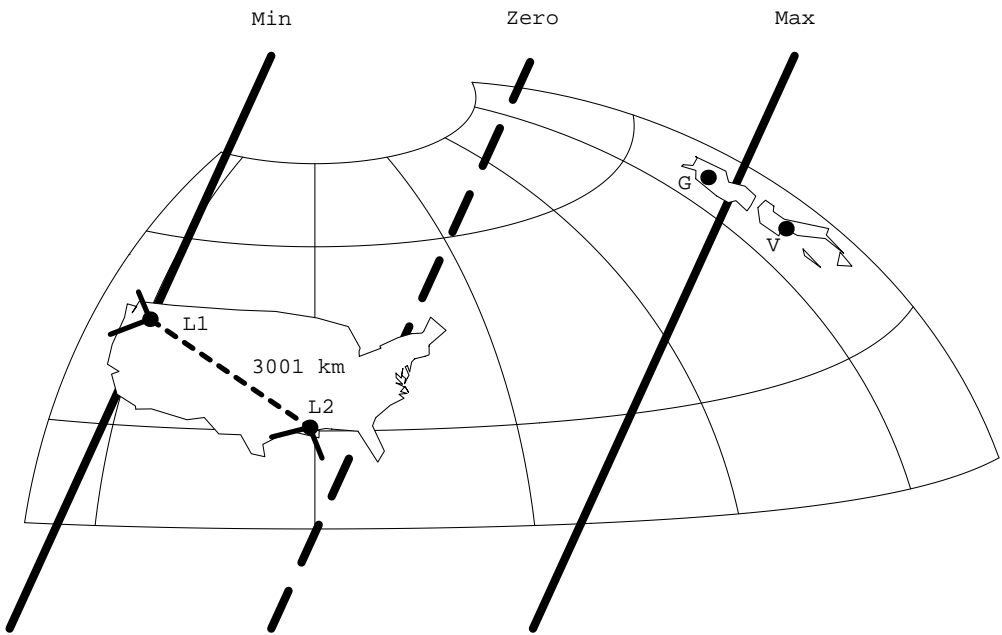,width=5.5in,bbllx=2.5cm,bblly=8.7cm,bburx=19.2cm,
bbury=19.2cm}}
\caption{\label{f:earth} 
The surface of the earth ($15^\circ<{\rm latitude}< 75^\circ$, 
$-130^\circ <{\rm longitude}<20^\circ$) including the LIGO detectors 
in Hanford, WA (L1) and Livingston, LA (L2), the VIRGO detector (V) 
in Pisa, Italy, and the GEO-600 (G) detector in Hanover, Germany.  
The perpendicular arms of the LIGO detectors are also illustrated (though 
not to scale).  
A plane gravitational wave passing by the earth is indicated by successive 
minimum, zero, and maximum of the wave.  
As this wave passes by the pair of LIGO detectors, it excites the two in 
coincidence at the moment shown, since both detectors are driven negative 
by the wave.  
During the time when the zero is between L1 and L2, the two detectors 
respond in anti-coincidence.  
Provided that the wavelength of the incident gravitational wave is larger 
than twice the separation ($d=3001$ km) between the detectors, the two 
detectors are driven in coincidence more of the time than in anti-coincidence.}
\end{center}
\end{figure}

From Fig.~\ref{f:overlap}, one also sees that the overlap reduction 
function for the two LIGO detectors has its first zero at 64 Hz.
This can be explained by the fact that a gravitational plane 
wave passing by the earth excites a pair of detectors in coincidence 
when the positive (or negative) amplitude part of the wave is passing 
by both detectors at the same time; it excites the two detectors in 
anti-coincidence when the positive (or negative) amplitude part of the 
wave is passing by one detector, and the negative (or positive) amplitude 
part of the wave is passing by the other detector.
(See Fig.~\ref{f:earth}.)
Provided that the wavelength of the incident gravitational wave is 
larger than twice the distance between the two detectors, the detectors 
will be driven in coincidence (on the average).  
For the case of the LIGO detector pair, this means that the Hanford, WA 
and Livingston, LA detectors will be driven in coincidence (on the 
average) by an isotropic and unpolarized stochastic background of 
gravitational radiation having a frequency of less than $f=c/(2d)=50$ Hz. 
The actual frequency of the zero ($f= 64$ Hz) is slightly larger than this, 
since $\gamma(f)$ is a sum of three spherical Bessel functions, 
which does not vanish at exactly 50 Hz.

\mediumtext%
\narrowtext\noindent%

In Appendix B of Ref.~\cite{flan}, Flanagan outlines a derivation of a 
closed-form expression for the overlap reduction function $\gamma(f)$.
The resulting expression applies to {\em any\,} pair of gravity-wave 
detectors, including interferometers with nonperpendicular arms and/or 
arbitrary orientations.
This is a useful result, because, e.g., the arms of the 
GEO-600 detector are separated by $94.33^\circ$.
Below we give a more detailed version of the derivation that appears in
Ref.~\cite{flan}, and correct a typographical error that appears in 
Eq.~(B6) of that paper.

We take, as our starting point for the derivation, 
the integral expression (\ref{e:gamma(f)}) for $\gamma(f)$.
To simplify the notation in what follows, we also define
\begin{equation}
\Delta\vec x:=d\ \hat s\quad{\rm and}\quad
\alpha:={2\pi fd\over c}\ ,
\end{equation}
where $\hat s$ is a unit vector that points in the direction connecting 
the two detectors, and $d$ is the distance between the two detectors.
In terms of these quantities, we can write
\begin{equation}
\gamma(f)=d_1^{ab}d_2^{cd}\ \Gamma_{abcd}(\alpha,\hat s)\ ,
\label{e:gamma_Gamma}
\end{equation}
where
\widetext%
\begin{equation}
\Gamma_{abcd}(\alpha,\hat s):=
{5\over 8\pi}\sum_A\int_{S^2}d\hat\Omega\ 
e^{i\alpha\hat\Omega\cdot\hat s}\ 
e_{ab}^A(\hat\Omega)e_{cd}^A(\hat\Omega)\ .
\label{e:Gamma}
\end{equation}
\narrowtext\noindent%
$\Gamma_{abcd}(\alpha,\hat s)$ is a tensor which is symmetric under 
the interchanges $a\leftrightarrow b$, $c\leftrightarrow d$,
$ab\leftrightarrow cd$.
It is also tracefree with respect to the $ab$ and $cd$ index pairs.

To evaluate $\Gamma_{abcd}(\alpha,\hat s)$, we begin by writing down 
the most general tensor constructed from $\delta_{ab}$ and $s_a$ that 
has the above-mentioned symmetry properties:
\widetext%
\begin{eqnarray}
\Gamma_{abcd}(\alpha,\hat s)&=&A(\alpha)\ \delta_{ab}\delta_{cd}
+B(\alpha)\ (\delta_{ac}\delta_{bd}+\delta_{bc}\delta_{ad})
+C(\alpha)\ (\delta_{ab}s_c s_d + \delta_{cd}s_a s_b)
\nonumber\\
&&+ D(\alpha)\ (\delta_{ac}s_b s_d+\delta_{ad}s_b s_c
+\delta_{bc}s_a s_d+\delta_{bd}s_a s_c)
+E(\alpha)\ s_a s_b s_c s_d\ .
\label{e:Gamma_exp}
\end{eqnarray}
\narrowtext\noindent%
We then contract (\ref{e:Gamma_exp}) with
$\delta^{ab}\delta^{cd}$, 
$(\delta^{ac}\delta^{bd}+\delta^{bc}\delta^{ad})$, $\cdots$,
$s^a s^b s^c s^d$ to obtain a linear system of equations for the functions 
$A,B,\cdots,E$:
%
\begin{equation}
\left[
\begin{array}{rrrrr}
9  & 6  & 6  & 4  & 1\\
6  & 24 & 4  & 16 & 2\\
6  & 4  & 8  & 8  & 2\\
4  & 16 & 8  & 24 & 4\\
1  & 2  & 2  & 4  & 1
\end{array}
\right]\
\left[
\begin{array}{c}
A\\
B\\
C\\
D\\
E
\end{array}
\right](\alpha)
=
\left[
\begin{array}{c}
p\\
q\\
r\\
s\\
t
\end{array}
\right](\alpha)\ ,
\label{e:pqrst}
\end{equation}
where
\widetext%
\begin{eqnarray}
p(\alpha)&:=&\Gamma_{abcd}(\alpha,\hat s)\ 
\delta^{ab}\delta^{cd}\ ,\nonumber\\
q(\alpha)&:=&\Gamma_{abcd}(\alpha,\hat s)\ 
(\delta^{ac}\delta^{bd}+\delta^{bc}\delta^{ad})\ ,\nonumber\\
r(\alpha)&:=&\Gamma_{abcd}(\alpha,\hat s)\
(\delta^{ab}s^c s^d + \delta^{cd}s^a s^b)\ ,\\
s(\alpha)&:=&\Gamma_{abcd}(\alpha,\hat s)\ 
(\delta^{ac}s^b s^d+\delta^{ad}s^b s^c
+\delta^{bc}s^a s^d+\delta^{bd}s^a s^c)\ ,\nonumber\\
t(\alpha)&:=&\Gamma_{abcd}(\alpha,\hat s)\ 
s^a s^b s^c s^d\ .\nonumber
\end{eqnarray}
\narrowtext\noindent%
From Eq.~(\ref{e:Gamma}), we see that the functions
$p,q,\cdots,t$ 
are scalar integrals that involve contractions of the spin-two
polarization tensors $e_{ab}^A(\hat\Omega)$.

To evaluate these integrals, we choose (without loss of generality) a 
coordinate system where the unit vector $\hat s$ coincides with unit 
vector $\hat z$.
Then 
\begin{equation}
\hat\Omega\cdot\hat s=\cos\theta\ ,\quad
\hat m\cdot\hat s=0\ ,\quad
\hat n\cdot\hat s=-\sin\theta\ ,
\end{equation}
and
\begin{eqnarray}
p(\alpha)&=&0\ ,\nonumber\\
q(\alpha)&=&10\int_{-1}^1 dx\ e^{i\alpha x}=20\ j_0(\alpha)\ ,\nonumber\\
r(\alpha)&=&0\ ,\\
s(\alpha)&=&10\int_{-1}^1 dx\ e^{i\alpha x}\ (1-x^2)
={40\over\alpha}\ j_1(\alpha)\ ,\nonumber\\
t(\alpha)&=&{5\over 4}\int_{-1}^1 dx\ e^{i\alpha x}\ (1-x^2)^2
={20\over\alpha^2}\ j_2(\alpha)\ ,\nonumber
\end{eqnarray}
where $j_0(\alpha)$, $j_1(\alpha)$, and $j_2(\alpha)$ are the
standard spherical Bessel functions:
\begin{eqnarray}
j_0(\alpha)&=&{\sin\alpha\over\alpha} ,\nonumber\\ 
j_1(\alpha)&=&{\sin\alpha\over\alpha^2}-{\cos\alpha\over\alpha}\ ,\\ 
j_2(\alpha)&=&3\ {\sin\alpha\over\alpha^3}-3\ {\cos\alpha\over\alpha^2}
-{\sin\alpha\over\alpha}\ .\nonumber
\end{eqnarray}
Note that $p(\alpha)=0$ and $r(\alpha)=0$ are immediate consequences 
of the tracefree property of $\Gamma_{abcd}(\alpha,\hat s)$.

The above linear system of equations (\ref{e:pqrst}) 
can be inverted for the functions
$A,B,\cdots,E$.
The results are:
\begin{equation}
\left[ 
\begin{array}{c}
A\\
B\\
C\\
D\\
E
\end{array}
\right](\alpha)
={1\over 2\alpha^{2}}
\left[
\begin{array}{rrr}
-5\alpha^2  &   10\alpha   &    5\\
 5\alpha^2  &  -10\alpha   &    5\\
 5\alpha^2  &  -10\alpha   &  -25\\
-5\alpha^2  &   20\alpha   &  -25\\
 5\alpha^2  &  -50\alpha   &  175\\
\end{array}
\right]\
\left[
\begin{array}{c}
j_0\\
j_1\\
j_2
\end{array}
\right](\alpha)\ .
\label{e:ABCDE}
\end{equation}
Finally, to obtain an expression for the overlap reduction function 
$\gamma(f)$, we substitute (\ref{e:Gamma_exp}) into (\ref{e:gamma_Gamma}).
Since $d_i^{ab}$ $(i=1,2)$ is tracefree, it follows that
\widetext%
\begin{equation}
\gamma(f)=2B(\alpha)\ d_1^{ab}d_{2ab}
+4D(\alpha)\ d_1^{ab}d_{2a}{}^c s_b s_c
+E(\alpha)\ d_1^{ab}d_2^{cd}s_a s_b s_c s_d\ .
\label{e:gamma1}
\end{equation}
Substituting the expressions for the functions $B,D,E$ given by 
(\ref{e:ABCDE})
into (\ref{e:gamma1}) yields
\begin{equation}
\gamma(f)=\rho_1(\alpha)\ d_1^{ab}d_{2ab}
+\rho_2(\alpha)\ d_1^{ab}d_{2a}{}^c s_b s_c
+\rho_3(\alpha)\ d_1^{ab}d_2^{cd}s_a s_b s_c s_d\ ,
\end{equation}
\narrowtext\noindent%
where
\begin{equation}
\left[ 
\begin{array}{c}
\rho_1\\
\rho_2\\
\rho_3
\end{array}
\right](\alpha)
=
{1\over 2\alpha^2}
\left[
\begin{array}{rrr}
 10\alpha^2 & -20\alpha   & 10\\
-20\alpha^2 &  80\alpha   & -100\\
  5\alpha^2 & -50\alpha   & 175
\end{array}
\right]\
\left[
\begin{array}{c}
j_0\\
j_1\\
j_2
\end{array}
\right](\alpha)\ .
\label{e:rho1rho2rho3}
\end{equation}
This is the desired result.

Note: In Eq.~(B6) of Ref.~\cite{flan}, the factor multiplying 
$j_1(\alpha)$ in $\rho_1(\alpha)$ is $-2/\alpha$.
As shown in Eq.~(\ref{e:rho1rho2rho3}), 
this factor should equal $-10/\alpha$.

\subsection{Optimal filtering}
\label{subsec:optimal_filtering}

Using the techniques developed in the previous two subsections, we are 
now in a position to give a rigorous derivation of the optimal signal 
processing required for the detection of a stochastic background of 
gravitational radiation.
We start by writing the cross-correlation signal $S$ between the outputs
of the two detectors in the following form:
\begin{equation} 
S:=\int_{-T/2}^{T/2} dt\ \int_{-T/2}^{T/2} dt'\
s_1(t) s_2(t') Q(t,t')\ ,
\label{e:Sdef}
\end{equation}
where as before
\begin{eqnarray}
s_1(t)&:=&h_1(t)+n_1(t)\ ,
\label{e:s_1(t)}\\
s_2(t)&:=&h_2(t)+n_2(t)\ ,
\label{e:s_2(t)}
\end{eqnarray}
but now $Q(t,t')$ is a filter function, which is not necessarily equal 
to $\delta(t-t')$ as we assumed in
Sec.~\ref{subsec:coincident_coaligned}.
Because we are assuming in this section that the statistical properties 
of the stochastic gravity-wave background and noise intrinsic to the 
detectors are both stationary, the best choice of filter function 
$Q(t,t')$ can depend only upon the time difference $\Delta t:=t - t'$.
The goal is to find the {\em optimal choice\,} of filter function 
$Q(t-t'):=Q(t,t')$ in a rigorous way.

\begin{figure}[htb!]
\begin{center}
{\epsfig{file=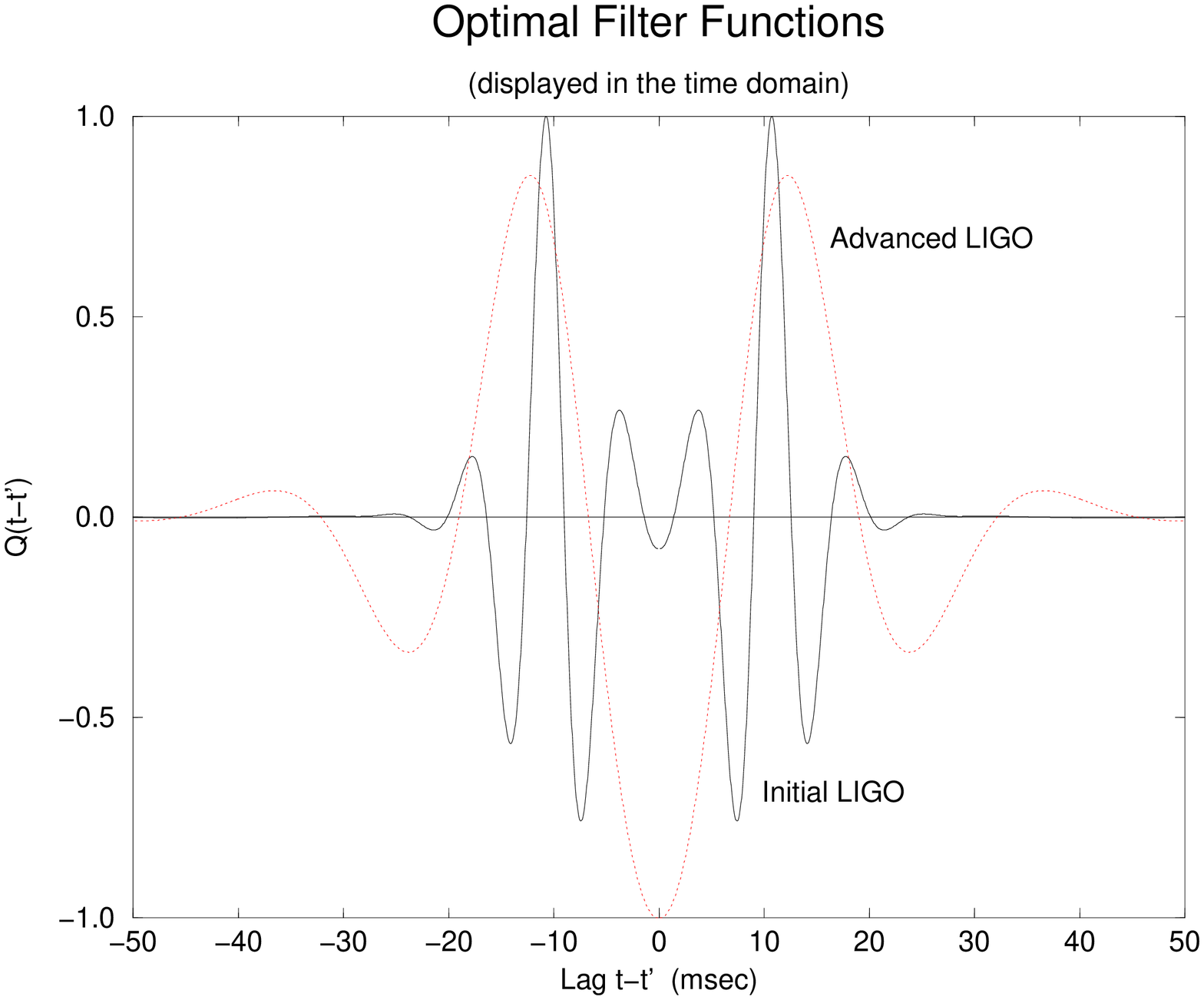,
angle=-90,width=3.4in,bbllx=25pt,bblly=50pt,bburx=590pt,bbury=740pt}}
\caption{\label{f:filters_time}
Optimal filter functions $Q(t-t')$ for the initial and advanced LIGO 
detector pairs, 
for a stochastic background having a constant frequency spectrum
$\Omega_{\rm gw}(f) =\Omega_0$.
Both filters are normalized to have maximum magnitude equal to unity.}
\end{center}
\end{figure}

The optimal choice of filter function $Q(t-t')$ will depend upon the 
locations and orientations of the detectors, as well as on the spectrum 
of the stochastic gravity-wave background and the noise power spectra of 
the detectors.  
It falls off rapidly to zero for time delays $\Delta t=t-t'$ whose 
magnitude is large compared to the light travel time $d/c$ between the 
two sites.%
\footnote{$d/c=10^{-2}$ sec for the LIGO detector pair.}
(See Fig.~\ref{f:filters_time}, which is located at the end of this 
section.)  
Since a typical observation time $T$ will be $\gg d/c$, 
we are justified in changing the limits on one of the integrations in 
(\ref{e:Sdef}) to obtain
\begin{equation} 
S=\int_{-T/2}^{T/2} dt\ \int_{-\infty}^{\infty} dt'\ 
s_1(t) s_2(t') Q(t-t')\ .
\label{e:Stime}
\end{equation}
This change of limits simplifies the mathematical analysis that follows.

We can also write Eq.~(\ref{e:Stime}) in the frequency domain.
Using the convention
\begin{equation}
\tilde g(f):= \int_{-\infty}^{\infty}dt\ 
e^{-i2 \pi f t}\ g(t)
\label{e:g(f)}
\end{equation}
for the Fourier transform of $g(t)$, it follows that
\begin{equation} 
S = \int_{-\infty}^{\infty}df\ \int_{-\infty}^{\infty} df'\ 
\delta_T(f-f') \tilde s_1^*(f) \tilde s_2(f') \tilde Q(f')\ ,
\label{e:Sfreq}
\end{equation}
where $\tilde s_1(f)$, $\tilde s_2(f)$, and $\tilde Q(f)$ are the 
Fourier transforms of $s_1(t)$, $s_2(t)$, and $Q(t-t')$, and
$\delta_T(f-f')$ is the finite-time approximation to the Dirac delta
function $\delta(f-f')$ defined by (\ref{e:delta_T(f)}).
Note also that for a real $Q(t-t')$, $\tilde Q(-f)=\tilde Q{}^*(f)$.

The optimal choice of filter function also depends on the quantity
that we want to maximize.
As we shall see in Sec.~\ref{sec:detection_etc},
in the context of stochastic background searches, it is natural to
maximize the signal-to-noise ratio 
\begin{equation}
{\rm SNR}:={\mu\over\sigma}\ ,
\label{e:SNR_display}
\end{equation}
where $\mu$ and $\sigma^2$ are the mean value and variance of the 
cross-correlation signal $S$ defined by (\ref{e:mu_label}) and
(\ref{e:sigma2_label}).
The techniques that we will use to evaluate $\mu$ 
and $\sigma^2$ are very similar to those that we used in 
Sec.~\ref{subsec:coincident_coaligned} 
for the case of coincident and coaligned detectors.

The calculation of the mean value $\mu$ is straightforward.
Since we are assuming that the noise intrinsic to the two detectors are 
statistically independent of each other and of the gravitational strains, 
it follows immediately from (\ref{e:Sfreq}) that
\widetext%
\begin{equation}
\mu:=\langle S\rangle
=\int_{-\infty}^{\infty} df\
\int_{-\infty}^{\infty} df'\ \delta_T(f-f')
\langle\tilde h_1^*(f)\tilde h_2(f')\rangle \tilde Q(f')\ .
\label{e:mudef}
\end{equation}
\narrowtext\noindent%
To calculate the expectation value
$\langle\tilde h_1^*(f)\tilde h_2(f')\rangle$, we again make use of the 
plane wave expansion (\ref{e:h_ab}) and the expectation value (\ref{e:hAhA'}).
Since 
\widetext%
\begin{equation}
\tilde h_i(f)=\sum_A\int_{S^2} d\hat\Omega\ h_A(f,\hat\Omega)\ 
e^{-i2\pi f\hat\Omega\cdot\vec x_i/c}\ F_i^A(\hat\Omega)\ ,
\end{equation}
where $i=1,2$ labels the two detectors, we find
\begin{eqnarray}
\langle \tilde h_1^*(f) \tilde h_2(f') \rangle&=&
\sum_A\sum_{A'} \int_{S^2} d\hat\Omega \int_{S^2}\ d\hat\Omega'\ 
\langle h_A^*(f,\hat\Omega) h_{A'}(f',\hat\Omega')\rangle 
\nonumber\\
&&\quad\quad\times\
e^{i2 \pi f  \hat\Omega \cdot\vec x_1/c}\
e^{-i2\pi f' \hat\Omega'\cdot\vec x_2/c}\ 
F_1^A(\hat\Omega)F_2^{A'}(\hat\Omega') \\
&=&{3 H_0^2 \over 32\pi^3}\ \delta(f-f')\ |f|^{-3}\ \Omega_{\rm gw}(|f|)
\nonumber\\
&&\quad\quad\times
\sum_{A}\int_{S^2}d\hat\Omega\ e^{i2\pi f\hat\Omega\cdot\Delta\vec x/c}\  
F_1^A(\hat\Omega) F_2^{A}(\hat\Omega)\\
&=&{3 H_0^2 \over  20 \pi^2}\ \delta(f-f')\ |f|^{-3}\ \Omega_{\rm gw}(|f|) 
\gamma(|f|)\ ,
\label{e:h_1(f)h_2(f')}
\end{eqnarray}
\narrowtext\noindent%
where we used (\ref{e:hAhA'}) to obtain the second equality and the 
definition (\ref{e:gamma(f)}) of the overlap reduction function to obtain 
the third.
Substituting (\ref{e:h_1(f)h_2(f')}) into (\ref{e:mudef}) yields
\begin{equation}
\mu={3 H_0^2 \over 20 \pi^2}\ T\
\int_{-\infty}^{\infty} df\ |f|^{-3}\ \Omega_{\rm gw} (|f|)\ 
\gamma(|f|)\tilde Q(f)\ .
\label{e:mu_final}
\end{equation}
The factor of $T$ on the RHS arises from evaluating $\delta_T(0)$.

Before calculating the variance $\sigma^2$, it is worthwhile to make a
slight digression and study in more detail the expectation value 
(\ref{e:h_1(f)h_2(f')}) derived above.
It turns out that this equation has an important physical implication.
In terms of the time domain variables $h_1(t)$ and $h_2(t')$, 
Eq.~(\ref{e:h_1(f)h_2(f')}) can be rewritten as
\begin{equation}
\langle h_1(t) h_2(t')\rangle=
\int_{-\infty}^\infty df\ e^{i2\pi f(t-t')}\ H_{12}(f)\ ,
\label{e:h_1(t)h_2(t')}
\end{equation}
where
\begin{equation}
H_{12}(f)={3H_0^2\over 20\pi^2}\ |f|^{-3}\ \Omega_{\rm gw}(|f|)\ 
\gamma(|f|)\ .
\label{e:H_12(f)}
\end{equation}
In other words, $H_{12}(f)$ is just the Fourier transform of the 
cross-correlation of the gravitational strains $h_1(t)$ and $h_2(t')$ 
at the two detector sites.
Moreover, since the noise intrinsic to the
two detectors are statistically independent of one another and of the 
gravitational strains, it follows that
\begin{equation}
\langle h_1(t) h_2(t')\rangle=\langle s_1(t) s_2(t')\rangle\ .
\end{equation}
Thus, $H_{12}(f)$ is the Fourier transform of the cross-correlation of the
outputs of the two detectors.
But this correlation is something that we can {\em measure\,}
(or at least estimate) given enough data.%
\footnote{For example, suppose we measure the outputs of the two detectors 
for a total observation time of one year. 
To estimate the cross-correlation $\langle s_1(t) s_2(t')\rangle$,
we simply form the products of $s_1(t)$ and $s_2(t')$ for all $t$ and 
$t'$ having the same $\Delta t=t-t'$ (e.g., 1 msec), and then average the 
results.
We then repeat this procedure for $\Delta t=2$ msec, 3 msec, etc.}
This in turn implies that we can measure (or at least estimate) 
$\Omega_{\rm gw}(|f|)$.
Explicitly, given the measured values of $\langle s_1(t) s_2(t')\rangle$,
we take their Fourier transform (with respect $t-t'$), multiply by
$|f|^3$, and divide by ${3H_0^2\over 20\pi^2}\ \gamma(|f|)$, to determine
$\Omega_{\rm gw}(|f|)$.
This will yield a good approximation to the real stochastic 
gravity-wave spectrum provided that the noise intrinsic to the detectors 
is not too large, or equivalently, if we measure the detector outputs for 
a long enough period of time $T$.
Also, the approximation of $\Omega_{\rm gw}(|f|)$ will be best for frequencies
$1/T<f<c/d$ (where $d/c$ is the light travel time between the two detectors):
For $f<1/T$, the Fourier transform lacks sufficient time domain data 
to provide useful information;
for $f>c/d$, the overlap reduction function quickly approaches zero, so the
division by $\gamma(f)$ makes $\Omega_{\rm gw}(|f|)$ ill-behaved.
 
Let us return now to the calculation of the signal-to-noise ratio
${\rm SNR}:=\mu/\sigma$.
To calculate the variance $\sigma^2$, we assume (as in 
Sec.~\ref{subsec:coincident_coaligned}) that the noise intrinsic to the 
two detectors are much larger in magnitude than the stochastic 
gravity-wave background.
Then
\widetext%
\begin{eqnarray}
\sigma^2
&:=&\langle S^2\rangle -\langle S\rangle^2\approx\langle S^2\rangle\\
&\approx&
\int_{-\infty}^\infty df\ \int_{-\infty}^\infty df'\
\int_{-\infty}^\infty dk\ \int_{-\infty}^\infty dk'\
\nonumber\\
&&\quad\quad\times\delta_T(f-f')\delta_T(k-k')\ 
\langle\tilde n_1^*(f)\tilde n_2(f')\tilde n_1^*(k)\tilde n_2(k')\rangle\ 
\tilde Q(f')\tilde Q(k')\\
&\approx&
\int_{-\infty}^\infty df\ \int_{-\infty}^\infty df'\
\int_{-\infty}^\infty dk\ \int_{-\infty}^\infty dk'\
\nonumber\\
&&\quad\quad\times\delta_T(f-f')\delta_T(k-k')\ 
\langle\tilde n_1^*(f)\tilde n_1(-k)\rangle
\langle\tilde n_2^*(-f')\tilde n_2(k')\rangle\ 
\tilde Q(f')\tilde Q(k')\ ,
\label{e:sigma^2def}\end{eqnarray}
\narrowtext\noindent%
where we used the statistical independence and reality of $n_1(t)$ and 
$n_2(t)$ to obtain the last line.

In Sec.~\ref{subsec:coincident_coaligned}, we defined the noise power 
spectrum $P_i(|f|)$ in terms of the expectation value 
$\langle n_i(t)n_i(t')\rangle$ of the time domain random variables $n_i(t)$.
(See Eq.~(\ref{e:P_i(f)}).)
An analogous expression holds in the frequency domain as well.
Using definition (\ref{e:g(f)}) for the Fourier transform 
$\tilde n_i(f)$ and definition (\ref{e:P_i(f)}) for the noise power spectrum
$P_i(|f|)$, it follows that
\begin{equation}
\langle\tilde n_i^*(f)\tilde n_i(f')\rangle=
{1\over 2}\delta(f-f')\ P_i(|f|)\ .
\label{e:n_i(f)n_i(f')}
\end{equation}
Substituting this result into (\ref{e:sigma^2def}) then yields
\widetext%
\begin{eqnarray}
\sigma^2
&\approx&{1 \over 4} \int_{-\infty}^\infty df\ \int_{-\infty}^\infty df'\
\delta_T^2(f-f') P_1(|f|) P_2(|f'|) \tilde Q (f)\tilde Q^*(f')\\
&\approx&{T\over 4}\int_{-\infty}^\infty df\ P_1(|f|) P_2(|f|) 
|\tilde Q (f)|^2\ ,
\end{eqnarray}
\narrowtext\noindent%
where we replaced one of the finite-time delta functions $\delta_T(f-f')$ 
by an ordinary Dirac delta function, and evaluated the other at $f=f'$ to 
obtain the last line.

To summarize:
\begin{eqnarray}
\mu&=&{3 H_0^2 \over 20 \pi^2}\ T\
\int_{-\infty}^{\infty} df\ |f|^{-3}\ \Omega_{\rm gw} (|f|)
\gamma(|f|)\tilde Q(f)\ ,
\label{e:mu2}\\
\sigma^2&\approx&{T\over 4}\int_{-\infty}^\infty df\ P_1(|f|) P_2(|f|) 
|\tilde Q (f)|^2\ .
\label{e:sigma^22}
\end{eqnarray}

The problem now is to find the filter function $\tilde Q(f)$ that maximizes 
the signal-to-noise ratio (\ref{e:SNR_display}), with $\mu$ and $\sigma^2$
as given above.
This turns out to be remarkably simple if we first introduce an {\em inner 
product\,} $(A,B)$ for any pair of complex functions $A(f)$ and $B(f)$.
The inner product of $A(f)$ and $B(f)$ is a complex number defined by
\begin{equation}
(A,B):=\int_{-\infty}^{\infty} df \ A^*(f) B(f) P_1(|f|) P_2(|f|)\ .
\label{e:inner_product}
\end{equation}
Since $P_i(|f|)>0$, it follows that $(A,A)\ge 0$, and $(A,A)=0$ if and only 
if $A(f)=0$. 
In addition, $(A,B)=(B,A)^*$ and $(A,B+\lambda C)=(A,B)+\lambda(A,C)$ 
for any complex number $\lambda$.
Thus, $(A,B)$ is a {\em positive-definite\,} inner product.
It satisfies all of the properties of an ordinary dot product of vectors 
in three-dimensional Euclidean space.

In terms of this inner product, the mean value $\mu$ and variance
$\sigma^2$ can be written as
\begin{eqnarray}
\mu&=&{3 H_0^2 \over 20 \pi^2}\ T\ 
\left(\tilde Q, {\gamma(|f|) \Omega_{\rm gw}(|f|) \over |f|^3
P_1(|f|) P_2(|f|)}\right)\\
\sigma^2&\approx& {T\over 4} (\tilde Q, \tilde Q)\ .
\end{eqnarray}
The problem is to choose $\tilde Q(f)$ so that it maximizes the 
signal-to-noise ratio (\ref{e:SNR_display}), or equivalently, the squared 
signal-to-noise ratio
\begin{equation} 
{\rm SNR}^2={\mu^2\over\sigma^2}\approx
\left({3 H_0^2 \over 10 \pi^2 }\right)^2 T\ 
{\left(\tilde Q,{\gamma(|f|)\Omega_{\rm gw}(|f|) 
\over |f|^3 P_1(|f|) P_2(|f|)}\right)^2
\over (\tilde Q,\tilde Q)}\ .
\label{e:SNR^2}
\end{equation}
But this is trivial!  
For suppose we are given a fixed three-dimensional vector $\vec A$, and are 
asked to find the three-dimensional vector $\vec Q$ that maximizes the ratio 
$(\vec Q\cdot\vec A)^2/\vec Q\cdot\vec Q$.  
Since this ratio is proportional to the squared cosine of the angle between 
the two vectors, it is maximized by choosing $\vec Q$ to point in the same
direction as $\vec A$.
The problem of maximizing (\ref{e:SNR^2}) is identical.
The solution is
\begin{equation} 
\tilde Q(f) = \lambda 
{\gamma(|f|) \Omega_{\rm gw}(|f|) \over |f|^3 P_1(|f|)
P_2(|f|)}\ ,
\label{e:optimal}
\end{equation}
where $\lambda$ is a (real) overall normalization constant.

\begin{figure}[htb!]
\begin{center}
{\epsfig{file=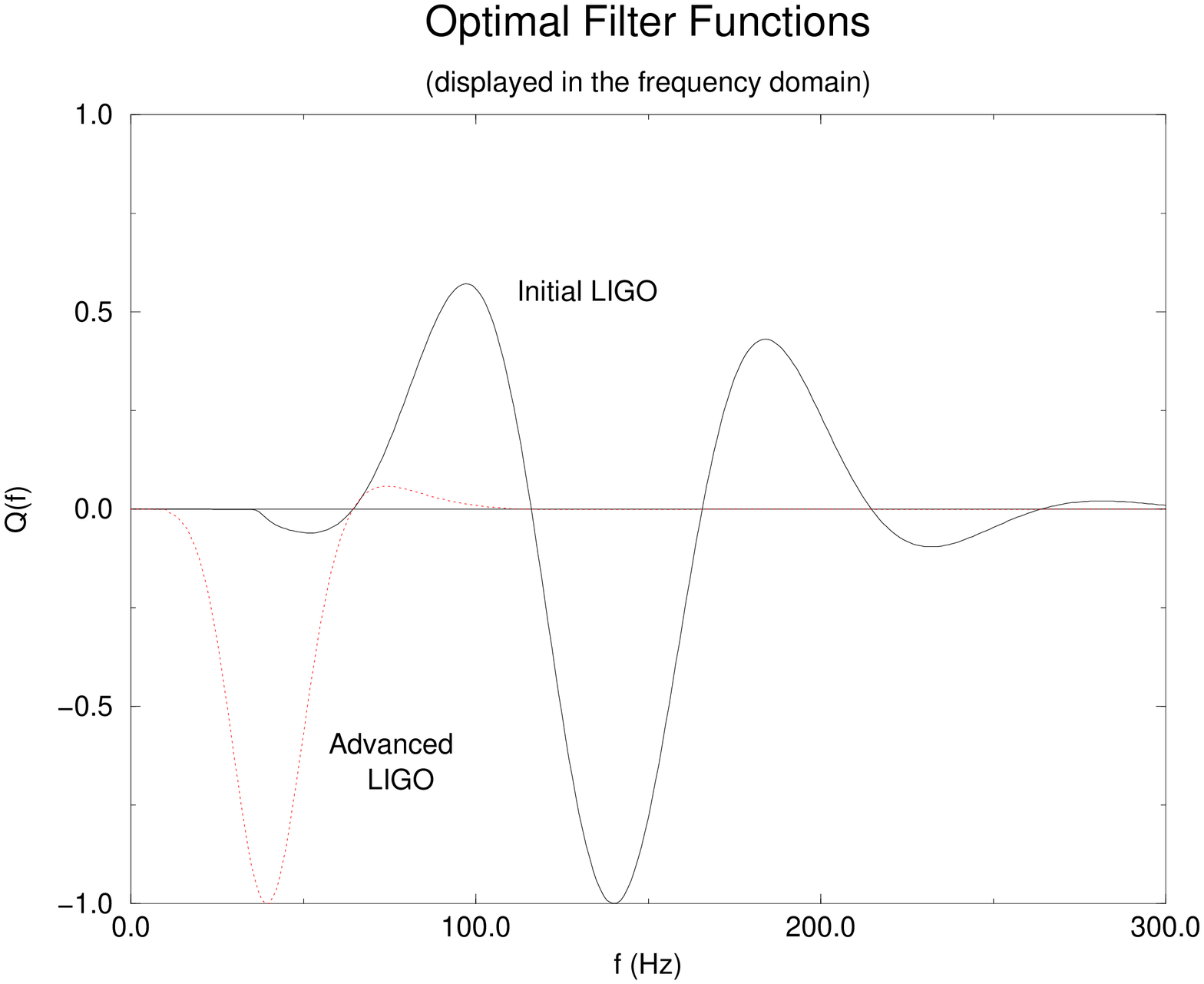,
angle=-90,width=3.4in,bbllx=25pt,bblly=50pt,bburx=590pt,bbury=740pt}}
\caption{\label{f:filters_freq}
Optimal filter functions $\tilde Q(f)$ for the initial and advanced LIGO 
detector pairs, 
for a stochastic background having a constant frequency spectrum
$\Omega_{\rm gw}(f) =\Omega_0$.
Both filters are normalized to have maximum magnitude equal to unity.}
\end{center}
\end{figure}

One of the curious things about expression (\ref{e:optimal}) for the optimal
filter $\tilde Q(f)$ is that it depends upon the spectrum $\Omega_{\rm gw}(f)$
of the stochastic gravity-wave background. 
This is a function that we do not know a~priori.%
\footnote{See, however, the discussion surrounding 
Eqs.~(\ref{e:h_1(t)h_2(t')}) and (\ref{e:H_12(f)}).}
In practice this means that we cannot use a single optimal filter 
when performing the data analysis; 
we will need to use a {\em set} of such filters.  
For example, within the bandwidth of interest for the ground-based
interferometers, it is reasonable to assume that the spectrum is given by 
a power-law $\Omega_{\rm gw}(f)=\Omega_\alpha f^\alpha$ 
(where $\Omega_\alpha={\rm constant}$).%
\footnote{The $\alpha$ in $\Omega_\alpha$ and $f^\alpha$ is just a number; 
it is not an index label.}
We could then construct a set of optimal filters $\tilde Q_\alpha(f)$
(say, for $\alpha=-4,-7/2,\cdots,7/2,4$) with the overall
normalization constants $\lambda_\alpha$ chosen so that
\begin{equation}
\mu=\Omega_\alpha\ T\ .
\end{equation}
With this choice of normalization, the optimal filter functions 
$\tilde Q_\alpha(f)$ are {\em completely} specified by the exponent 
$\alpha$, the overlap reduction function, and the noise power spectra of
the two detectors.
We would then analyze the outputs of the two detectors for each of these 
filters separately.
Figure~\ref{f:filters_freq} shows the optimal filter functions 
(displayed in the frequency domain) for both the initial and advanced 
LIGO detector pairs, for a stochastic background
having a constant frequency spectrum $\Omega_{\rm gw}(f)=\Omega_0$ 
(i.e., $\alpha=0$).%
\footnote
{Figures~\ref{f:LIGO-WA_LIGO-LA_optimal}-\ref{f:GEO-600_TAMA-300_optimal}
in Sec.~\ref{subsec:off} show the analogous optimal filter 
functions for different detector pairs.}
Figure~\ref{f:filters_time} shows these same optimal filter functions
displayed in the time domain---i.e., as a function of the {\em lag\,}
$t-t'$.

Having found the optimal choice of filter function $\tilde Q(f)$, it is now
straightforward to calculate the signal-to-noise ratio for a given pair of 
detectors.  
Substituting (\ref{e:optimal}) into (\ref{e:SNR^2}) and taking the square root
gives 
\begin{equation}
{\rm SNR}\approx{3H_0^2\over 10\pi^2}\ \sqrt{T}\ 
\left[\int_{-\infty}^\infty df\
{\gamma^2 (|f|) \Omega_{\rm gw}^2(|f|) \over f^6 P_1(|f|) P_2(|f|)}
\right]^{1/2}\ .
\label{e:SNR_large_noise}
\end{equation}
We will use this result in later sections to calculate
signal-to-noise ratios and sensitivity levels for different detector pairs, 
assuming that the stochastic gravity-wave background has a constant 
frequency spectrum
$\Omega_{\rm gw}(f)=\Omega_0$.
Tables~\ref{t:snr}-\ref{t:omega_min_4d} in Sec.~\ref{subsec:snrs} 
contain the results of these calculations.

\section{Detection, estimation, and sensitivity levels}
\label{sec:detection_etc}

Once the detectors have gone ``on-line'' and are generating data that 
needs to be analyzed, we will be confronted with the following questions:
(i) ``How do we decide, from the experimental data, if we've detected a 
stochastic gravity-wave signal?''
(ii) ``Assuming that a stochastic gravity-wave signal is present, 
how do we estimate its strength?''
(iii) ``Assuming that a stochastic gravity-wave signal is present, 
what is the minimum value of $\Omega_0$ required to detect it 95\% of 
the time?''
In this section, we answer these questions, using a {\em frequentist\,} 
approach to the theory of probability and statistics.%
\footnote{There are actually {\em two\,} approaches that one can take
when analyzing data: the frequentist (or classical frequency 
probability) approach, which is
adopted in this paper, and the Bayesian (or subjective probability) 
approach, which is adopted in Refs.~\cite{lsf1,lsf2}.
Although we will not describe the similarities and differences of 
these two approaches in any detail in this 
paper, it is important to emphasize that the frequentist and Bayesian 
approaches are {\em inequivalent\,} methods of analyzing data.
Frequentists and Bayesians ask {\em different\,} questions about data and 
hypotheses, and consequently obtain different answers and draw different
conclusions.
In fact, there are certain questions that one can ask and answer in the 
Bayesian approach, that are ill-defined for a frequentist.
Interested readers should see Refs.~\cite{lsf1,lsf2} 
for a detailed discussion of the frequentist and Bayesian approaches 
applied to gravitational-wave data analysis with multiple detectors.}

\subsection{Statistical considerations}
\label{subsec:statistical_considerations}

When performing a search for a stochastic background of gravitational
radiation, it is convenient
to break the data set (which might be hours, days, or weeks 
in length) into shorter stretches of $N$ points, which we can then FFT and 
correlate with data from other detectors.
Depending on the choice of $N$ and on the sampling rate of the detector,
these shorter segments of data will typically last on the order of 
seconds.%
\footnote{For example, for $N=65536=2^{16}$ and a sampling rate of 20 kHz, 
$T=3.2768$ sec.
For the same $N$ and a sampling rate of 
$16.384\ {\rm kHz}=2^{14}\ {\rm Hz}$, $T=4.0$ sec.
Since the relevant bandwith for stochastic background detection is 
$f<300\ {\rm Hz}$, the data stream could, in principle, be decimated 
to sampling rates which are substantially smaller---i.e., 1024 Hz.}
From Eq.~(\ref{e:Sfreq}), we see that in a measurement over a single
observation period $T\approx 4$ sec, the signal $S$ is a sum (over $f$ 
and $f'$) of approximately 400 statistically independent
random variables (products of the Fourier amplitudes of the signals).
This is because $\tilde s_1(f)$ and $\tilde s_2(f')$ are correlated only
when $|f-f'| < 1/T \approx .25$ Hz, and the bandwidth over which the integral 
in (\ref{e:Sfreq}) gets its major contribution is $\sim 100$ Hz wide.  
Thus, by virtue of the central limit theorem, $S$ is well-approximated by
a Gaussian random variable, provided we are not too far away from the mode
of the distribution.
Equivalently, the values of $S$ in a set of measurements over 
statistically independent time intervals (each of length $T$) are normally 
distributed.
The mean value of this distribution is $\mu:=\langle S\rangle$, and the
variance is $\sigma^2:=\langle S^2 \rangle-\langle S\rangle^2$.

Let ${\bf s}:=(S_1,S_2,\cdots,S_n)$
be a set of such measurements over statistically
independent time intervals, each of length $T$.%
\footnote{In order that the measurements be statistically independent, 
the time intervals should be non-overlapping, and $T$ should be $\gg$ 
the light travel time $d/c$ between the two detectors.
For the LIGO detector pair, this corresponds to $T\gg 10^{-2}$ sec, 
which is satisfied for $T\approx 4$ sec.}
We can think of these measurements as $n$ independent samples drawn
from a normal distribution having mean $\mu$ and variance $\sigma^2$.
The set $\bf s$ represents the outcome of a single experiment.
From these samples, we can construct the {\em sample\,} mean
\begin{equation}
\hat\mu:={1\over n}\sum_{i=1}^n S_i
\label{e:samplemean}
\end{equation}
and the {\em sample\,} variance
\begin{equation}
\hat\sigma^2:={1\over n-1}\sum_{i=1}^n(S_i-\hat\mu)^2\ .
\label{e:samplevar}
\end{equation}
Given the values of these estimators $\hat\mu$ and $\hat\sigma^2$, 
we would like to decide, in some reliable way, whether or not we
have detected a stochastic gravity-wave signal.

To make such a decision, we will apply a standard theorem from the
theory of of probability and statistics 
(see, e.g., p.178 of Ref.~\cite{mf}).
The theorem states that if $\hat\mu$ is the sample mean of a set of 
$n$ independent samples drawn from a normal distribution having mean
$\mu$ and variance $\sigma^2$, then
\begin{equation}
t={\hat\mu-\mu\over\hat\sigma/\sqrt{n}}
\label{e:t_distribution}
\end{equation}
is the value of a random variable having Student's $t$-distribution 
with parameter $\nu=n-1$.
This is the classic Student's $t$-test.
It is used to compare the means of two normal distributions that
have the same variance, when the variance $\sigma^2$ is unknown.

Since tables of Student's $t$-distribution and its associated
cumulative probability distribution function can be found in most 
handbooks on statistics
(see, e.g., \cite{handbook}), we could do all of the remaining 
calculations in this section in terms of the $t$-distribution.
The drawback to this approach, however, is that the $t$-distribution
depends on the parameter $\nu=n-1$.
This means that all of our results would depend on the number of 
observations $n$ that constitute a single experiment.
For stochastic background searches, this undesirable feature
can be avoided by choosing $n$ large enough so that the $t$-distribution 
and standard normal distribution are virtually indistinguishable.
Since a typical total observation time will be on the order 
of months or years ($\sim 10^7$ seconds),
while $T$ (the duration of a single observation period) is typically on
the order of seconds, it is no problem to choose $n\sim 10^3$ or more.%
\footnote{In fact, even for $n=30$, 
the $2\sigma$ values for the $t$-distribution and standard normal 
distribution differ by less than 5\%.
As $n\rightarrow\infty$, this difference goes to zero.}
For such large $n$, Eq.~(\ref{e:t_distribution}) can be rewritten as
\begin{equation}
z\approx{\hat\mu-\mu\over\hat\sigma/\sqrt{n}}\ ,
\label{e:z_distribution}
\end{equation}
where $z$ is the value of a random variable having the standard normal 
distribution---i.e., $z$ is a Gaussian random variable having zero 
mean and unit variance.
Note that the approximation becomes a strict equality 
%
%
if $\hat\sigma$ is replaced by its expected value $\sigma$.
This is because a linear combination of $n$ Gaussian random variables
(e.g., $\hat\mu$) is also a Gaussian random variable, independent of $n$.   

In the calculations that follow, we will want to assign probabilities
to different events.
From a frequentist point of view, this means that we should perform 
(or imagine performing) some fixed experiment many, many times.
The probability of an event is then defined as the frequency of
occurence of that event, in the limit of an infinite number of repeated, 
independent experiments.

\subsection{Signal detection}
\label{subsec:signal_detection}

In order to decide whether or not we've detected a stochastic 
gravity-wave signal, we need a rule that, given a set of measured 
data, will select for us one of two alternative hypotheses:
\begin{description}
\item[$H_0$:]
A stochastic gravity-wave signal is absent.
\item[$H_1$:]
A stochastic gravity-wave signal is present, characterized by 
some fixed, but {\em unknown\,}, mean value $\mu>0$.
\end{description}
Moreover, we would like this rule to be ``optimal'' with respect to 
some chosen set of criteria.
This method of decision making, or ``hypothesis testing'' as it is
more formally called, is a well-studied branch of frequentist statistics.
As such, we will not go into any of the details here.
After making the appropriate definitions, we will simply state the rule 
that we adopt for our stochastic background searches and explain 
in what sense it is optimal.
Interested readers should see Ref.~\cite{helstrom} for a much more
thorough discussion of the statistical theory of signal detection.

To begin, let us note that the two hypotheses $H_0$ and $H_1$ 
are exhaustive and mutually exclusive---i.e., a stochastic gravity-wave
signal is either absent or present.
And, if present, it will be characterized by some fixed mean value $\mu$,
which is proportional to $\Omega_0$ for a stochastic background of 
gravitational radiation having a constant frequency spectrum
$\Omega_{\rm gw}(f)=\Omega_0$.
The only requirement is that $\mu>0$.
$H_0$ is a {\em simple\,} hypothesis, since it does not depend on any
unknown parameters.
$H_1$ is a {\em complex\,} (or composite) hypothesis, since it depends
on a range of the unknown parameter $\mu$.
Explicitly,
\begin{equation}
H_1=\bigcup_{n=0}^\infty\ H_{nd\mu,d\mu}\ ,
\end{equation}
where $H_{\mu,d\mu}$ is the hypothesis that a 
stochastic gravity-wave signal is present, characterized by a fixed 
mean value lying in the range $(\mu,\mu+d\mu]$.

As before, let
${\bf s}:=(S_1,S_2,\cdots,S_n)$
be a set of $n$ statistically independent measurements of the
cross-correlation signal $S$.
Due to the noise intrinsic to the detectors and to errors inherent
to the measurement process, the outcome of an experiment ${\bf s}$
is a random variable.
It is described statistically by the probability density functions: 
\begin{description}
\item[$p({\bf s}|0)$:]
Probability density function for the outcome of an experiment to be
$\bf s$, given that a stochastic gravity-wave signal is absent.
\item[$p({\bf s}|\mu)$:]
Probability density function for the outcome of an experiment to be
$\bf s$, given that a stochastic gravity-wave signal is present, 
characterized by the fixed mean value $\mu>0$.
\end{description}
From (\ref{e:z_distribution}), it follows that 
\begin{eqnarray}
p({\bf s}|0)&=&(2\pi\hat\sigma^2)^{-n/2}\ 
\exp\left[\ -\sum_{i=1}^n {S_i^2\over 2\hat\sigma^2}\ \right]\ ,
\label{e:p(s;0)}\\
p({\bf s}|\mu)&=&(2\pi\hat\sigma^2)^{-n/2}\ 
\exp\left[\ -\sum_{i=1}^n {(S_i-\mu)^2\over 2\hat\sigma^2}\ \right]\ .
\label{e:p(s;u)}
\end{eqnarray}
The fact that we can use the sample variance $\hat\sigma^2$
(instead of the true variance $\sigma^2$) on the right-hand-sides 
of (\ref{e:p(s;0)}) and (\ref{e:p(s;u)}) follows from the large $n$ 
approximation that we used to obtain Eq.~(\ref{e:z_distribution}).

A decision rule that, given the outcome of an experiment, selects 
for us either $H_0$ or $H_1$ is equivalent to a division of the space 
of all possible experimental outcomes into two disjoint regions $R_0$ 
and $R_1$:
If ${\bf s}\in R_0$, then $H_0$ is chosen;
if ${\bf s}\in R_1$, then $H_1$ is chosen.
The success and failure of such a rule is characterized by two types of 
errors:
A {\em type I\,} (or {\em false alarm\,}) error occurs when the decision rule
chooses $H_1$ when $H_0$ is really true.
A {\em type II\,} (or {\em false dismissal\,}) error occurs when the decision
rule chooses $H_0$ when $H_1$ is really true.
In terms of $p({\bf s}|0)$, $p({\bf s}|\mu)$, $R_0$, and $R_1$, we have
\begin{eqnarray}
\alpha\equiv&&\hbox{\rm False alarm rate}:=
\int_{R_1}d{\bf s}\ p({\bf s}|0)
\label{e:false_alarm}\\
\beta(\mu)\equiv&&\hbox{\rm False dismissal rate}:=
\int_{R_0}d{\bf s}\ p({\bf s}|\mu)
\label{e:false_dismissal}
\end{eqnarray}
Note that, for the complex hypothesis $H_1$,
the false dismissal rate is actually a function of the mean value
$\mu>0$.
Note also that $1-\alpha$ is the fraction of experimental outcomes
that the decision rule correctly identifies the {\em absence\,} 
(not presence!!) of a stochastic gravity-wave signal.
If we want to talk about detection, we should evaluate
\begin{equation}
\gamma(\mu)\equiv\hbox{\rm Detection rate}:=1-\beta(\mu)\ ,
\end{equation}
which is the fraction of experimental outcomes that the decision rule 
correctly identifies the presence of a stochastic gravity-wave signal, 
characterized by the fixed mean value $\mu>0$.%
\footnote{The detection rate $\gamma(\mu)$ should not be confused with 
the overlap reduction function $\gamma(f)$, which was defined in
Sec.~\ref{subsec:overlap}.}

\begin{figure}[htb!]
\begin{center}
{\epsfig{file=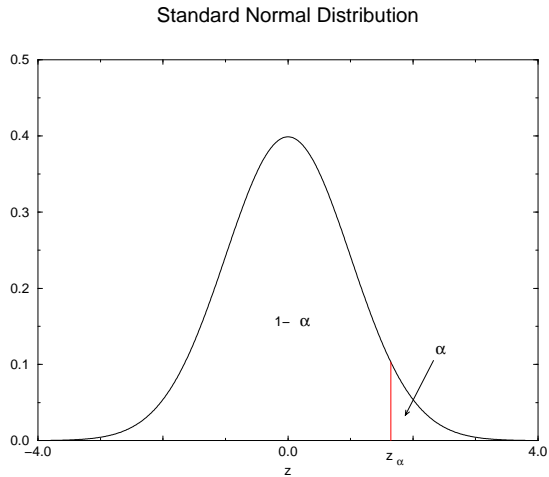,
angle=-90,width=3.4in,bbllx=25pt,bblly=50pt,bburx=590pt,bbury=740pt}}
\caption{\label{f:normal1}
The standard normal probability distribution for a Gaussian random 
variable having zero mean and unit variance.
$z_{\alpha}$ is that value of $z$ for which the area under the standard 
normal distribution to its right is equal to $\alpha$.
Typical values for $1-\alpha$ are 0.90, 0.95, and 0.99.
The corresponding values for $z_{\alpha}$ are
$z_{0.10}=1.28$, $z_{0.05}=1.65$, and $z_{0.01}=2.33$.
These are the threshold values appropriate for the one-sided test 
described in the text.}
\end{center}
\end{figure}

\begin{figure}[htb!]
\begin{center}
{\epsfig{file=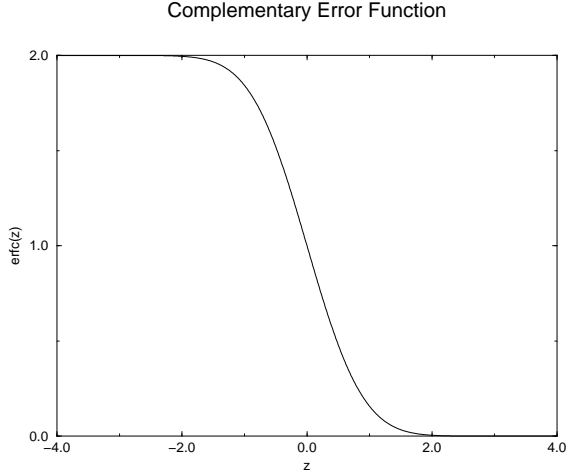,
angle=-90,width=3.4in,bbllx=25pt,bblly=50pt,bburx=590pt,bbury=740pt}}
\caption{\label{f:erfc}
The complementary error function.}
\end{center}
\end{figure}

In order for the decision rule to choose the regions $R_0$ and $R_1$ 
in an ``optimal'' way, we must first select some set of criteria 
with respect to which ``optimal'' can be defined.
For stochastic background searches, where one does not know a~priori 
the ``costs'' that one should associate with false alarm and false 
dismissal errors, it is reasonable to choose a decision rule that 
minimizes the false dismissal rate $\beta(\mu)$ for a fixed value of 
the false alarm rate $\alpha$.
Equivalently, one chooses a decision rule that 
maximizes the probability of detecting a stochastic gravity-wave signal, 
while keeping the false alarm rate fixed.
This decision criterion is known in the literature as the 
{\em Neyman-Pearson\,} criterion.
In general, for a complex hypothesis, the Neyman-Pearson criterion 
yields regions $R_0$ and $R_1$ that depend on the unknown parameter $\mu$.
But for this case, where $H_1$ involves {\em all\,} parameter values 
$\mu>0$, $R_0$ and $R_1$ are actually independent of $\mu$.%
\footnote{See pp.152-153 of Ref.~\cite{helstrom} for more details.}
Without going into details here, let us simply state the result:
Namely, in the context of stochastic background searches as described 
above, the Neyman-Pearson criterion is satisfied if, given the outcome of an 
experiment $\bf s$,
we form the estimators $\hat\mu$ and $\hat\sigma^2$ according to
Eqs.~(\ref{e:samplemean}) and (\ref{e:samplevar}), and then 
\begin{equation}
\hbox{\rm 
choose $H_0$ if $\hat\mu < z_\alpha\ \hat\sigma/\sqrt{n}$, and
choose $H_1$ if $\hat\mu \ge z_\alpha\ \hat\sigma/\sqrt{n}$\ .}
\label{e:mu_rule}
\end{equation}
Here  $z_{\alpha}$ is that value of the random variable $z$ 
for which the area under the standard normal distribution to its right 
is equal to $\alpha$.
(See Fig.~\ref{f:normal1}.)
In terms of the {\em complementary\,} error function
\begin{equation}
{\rm erfc}(z):={2\over\sqrt{\pi}}\int_z^\infty dx\ e^{-x^2}
\end{equation}
(see Fig.~\ref{f:erfc}),
\begin{equation}
z_\alpha=\sqrt{2}\ {\rm erfc}^{-1}(2\alpha)\ ,
\end{equation}
where ${\rm erfc}^{-1}$ denotes the inverse of {\rm erfc}.
Since $\hat\mu$ and $\hat\sigma$ are functions of the outcome of the
experiment $\bf s$, the inequalities
$\hat\mu < z_\alpha\ \hat\sigma/\sqrt{n}$
and $\hat\mu \ge z_\alpha\ \hat\sigma/\sqrt{n}$ define $R_0$ and $R_1$.
Note also that the decision rule can be restated as
\begin{equation}
\hbox{\rm 
choose $H_0$ if $\sqrt{n}\ \widehat{\rm SNR} < z_\alpha$, and
choose $H_1$ if $\sqrt{n}\ \widehat{\rm SNR} \ge z_\alpha$\ ,}
\label{e:snr_rule}
\end{equation}
where
\begin{equation}
\sqrt{n}\ \widehat{\rm SNR}:=\sqrt{n}\ {\hat\mu\over\hat\sigma}
\end{equation}
is the measured signal-to-noise ratio after $n$ observation periods. 
The fact that the signal-to-noise ratio enters the above inequalities
is one of the main reasons why we paid so much attention to evaluating 
it in Sec.~\ref{sec:detection}.

Given the above decision rule for signal detection, we can now 
calculate the false alarm and false dismissal rates
defined by Eqs.~(\ref{e:false_alarm}) and (\ref{e:false_dismissal}).
First, for the false alarm rate, one finds
\begin{eqnarray}
\alpha&=&{\rm Prob}(\hat\mu\ge z_{\alpha}\ \hat\sigma/\sqrt{n}|\mu=0)\\
&=&{\rm Prob}(z\ge z_{\alpha})\\
&=&{1\over 2}\ {\rm erfc}\left({z_{\alpha}\over\sqrt{2}}\right)\ ,
\label{e:alpha_eqn}
\end{eqnarray}
where we used Eq.~(\ref{e:z_distribution}) with $\mu=0$ to obtain 
the second equality.
Thus, our two uses of the symbol $\alpha$ (for the false alarm
rate and for the area under the standard normal distribution
to the right of $z_{\alpha}$) are consistent.

\begin{figure}[htb!]
\begin{center}
{\epsfig{file=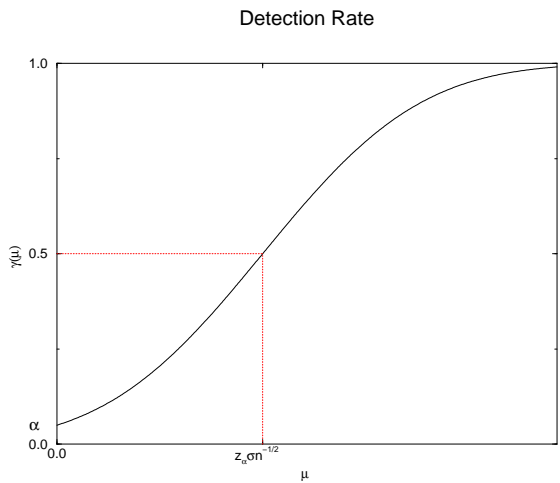,
angle=-90,width=3.4in,bbllx=25pt,bblly=50pt,bburx=590pt,bbury=740pt}}
\caption{\label{f:detection_rate}
The detection rate $\gamma(\mu):=1-\beta(\mu)$ plotted as a function
of the mean value $\mu$.}
\end{center}
\end{figure}

Second, for the false dismissal rate,
\begin{eqnarray}
\beta(\mu)&=&{\rm Prob}(\hat\mu<z_{\alpha}\ \hat\sigma/\sqrt{n}| 
\mu>0\ {\rm fixed})\\
&=&{\rm Prob}(z<z_{\alpha}-\sqrt{n}\ \mu/\hat\sigma)\\
&=&1-{1\over 2}\ {\rm erfc}\left({z_{\alpha}-
\sqrt{n}\ \mu/\hat\sigma\over\sqrt{2}}\right)\ ,
\label{e:beta_eqn}
\end{eqnarray}
where we again used Eq.~(\ref{e:z_distribution}) (but this time 
with $\mu>0$) to obtain the second equality.
A graph of the detection rate
$\gamma(\mu):=1-\beta(\mu)$ is shown in Fig.~\ref{f:detection_rate}.
From this graph, we see
that the detection rate $\gamma(\mu)$ approaches $\alpha$ as 
$\mu\rightarrow 0$; it approaches 1 as $\mu\rightarrow\infty$.
Also, the detection rate equals 0.50 for 
$\mu=z_{\alpha}\ \hat\sigma/\sqrt{n}$ (or, equivalently, for
$\sqrt{n}\ \mu/\hat\sigma=z_{\alpha}$).
Thus, if we replace $\hat\sigma$ with its expected value $\sigma$, we see
that the detection rate is only 50\% for a signal having a theoretical 
signal-to-noise ratio after $n$ observation periods equal to $z_\alpha$.

It is also interesting to note that the Neyman-Pearson detection 
criterion, when applied to stochastic background searches, is 
equivalent to the {\em maximum-likelihood} detection criterion 
\cite{helstrom}.
In other words, we will obtain the same decision rule given above if 
we first construct the likelihood ratio
\begin{equation}
\Lambda({\bf s}|\mu):={p({\bf s}|\mu)\over p({\bf s}|0)}\ ;
\end{equation}
maximize $\Lambda({\bf s}|\mu)$ with respect to variations of the 
parameter $\mu>0$:
\begin{equation}
\Lambda_{\rm max}({\bf s}):=\max_{\mu>0}\ \Lambda({\bf s}|\mu)\ ;
\end{equation}
and then divide the space of all possible experimental outcomes 
$\bf s$ into two regions $R_0$ and $R_1$ according the the rule
\begin{equation}
\hbox{\rm 
choose $H_0$ if $\Lambda_{\rm max}({\bf s})<\Lambda_0$, and
choose $H_1$ if $\Lambda_{\rm max}({\bf s})\ge\Lambda_0$\ ,}
\label{e:lamba_rule}
\end{equation}
where $\Lambda_0$ is chosen so that the false alarm rate equals $\alpha$.
Explicitly, from Eqs.~(\ref{e:p(s;0)}) and (\ref{e:p(s;u)}), it follows 
that 
\widetext%
\begin{equation}
\Lambda({\bf s}|\mu)=\exp\left[\ {n\mu\over\hat\sigma^2}
\left({1\over n}\sum_{i=1}^n S_i-{1\over 2}\mu\right)\ \right]\ 
=\exp\left[\ {n\mu\over\hat\sigma^2}
\left(\hat\mu-{1\over 2}\mu\right)\ \right]\ .
\end{equation}
\narrowtext\noindent%
$\Lambda({\bf s}|\mu)$ is maximized when $\mu$ equals the sample 
mean $\hat\mu$:%
\footnote{The sample mean $\hat\mu$ is said to be the
{\em maximum-likelihood estimator\,} for this problem.}
\begin{equation}
\Lambda_{\rm max}({\bf s})
=\Lambda({\bf s}|\hat\mu)
=\exp\left[\ {1\over 2}{n\hat\mu^2\over\hat\sigma^2}\ \right]\ .
\end{equation}
The decision surface $\Lambda_0$ is obtained by
setting $\hat\mu$ equal to the threshold value 
$z_\alpha\ \hat\sigma/\sqrt{n}$:
\begin{equation}
\Lambda_0=\Lambda_{\rm max}(z_\alpha\ \hat\sigma/\sqrt{n})
=\exp\left[\ {1\over 2}z_\alpha^2\ \right]\ .
\end{equation}

\subsection{Parameter estimation}
\label{subsec:parameter_estimation}

\begin{figure}[htb!]
\begin{center}
{\epsfig{file=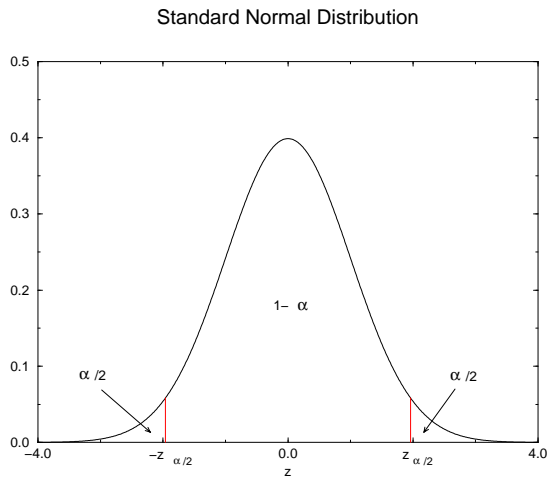,
angle=-90,width=3.4in,bbllx=25pt,bblly=50pt,bburx=590pt,bbury=740pt}}
\caption{\label{f:normal2}
The standard normal probability distribution for a Gaussian random 
variable having zero mean and unit variance.
$z_{\alpha/2}$ is that value of $z$ 
for which the area under the standard normal distribution to its right is 
equal to $\alpha/2$.
The area under the standard normal distribution between 
$-z_{\alpha/2}$ and $z_{\alpha/2}$ is thus $1-\alpha$.
Typical values for $1-\alpha$ are 0.90, 0.95, and 0.99.
The corresponding values for $z_{\alpha/2}$ are
$z_{0.05}=1.65$, $z_{0.025}=1.96$, and $z_{0.005}=2.58$.}
\end{center}
\end{figure}

Assuming that a stochastic gravity-wave signal is present, 
characterized by some fixed, but unknown, mean value $\mu>0$, 
parameter estimation attempts to answer the question:
``What is the value of $\mu$?''
It does this by first constructing the sample mean $\hat\mu$ and sample 
variance $\hat\sigma^2$ of a set ${\bf s}:=(S_1,S_2,\cdots,S_n)$ of 
$n$ statistically independent measurements of the cross-correlation 
signal $S$, as described in Sec.~\ref{subsec:statistical_considerations}.
From Eq.~(\ref{e:z_distribution}), we then know that
\begin{equation}
z\approx{\hat\mu-\mu\over\hat\sigma/\sqrt{n}}\ 
\end{equation}
is the value of a Gaussian random variable having zero mean and unit
variance.
Thus, if $z_{\alpha/2}$ is that value of the random variable $z$ for 
which the area under the standard normal distribution to its right is 
equal to $\alpha/2$ (see Fig.~\ref{f:normal2}), 
then $(1-\alpha)\cdot 100\%$ of the time
\begin{equation}
-z_{\alpha/2}<z<z_{\alpha/2}\ ,
\end{equation}
or, equivalently,
\begin{equation}
\hat\mu - z_{\alpha/2}\ \hat\sigma/\sqrt{n} < \mu <
\hat\mu + z_{\alpha/2}\ \hat\sigma/\sqrt{n}\ .
\end{equation}
Said another way, in an ensemble of observations of the same stochastic
background, a fraction $1-\alpha$ of the intervals
\begin{equation}
I_\alpha:=
\left[\ \hat\mu - z_{\alpha/2}\ \hat\sigma/\sqrt{n},\ 
\hat\mu + z_{\alpha/2}\ \hat\sigma/\sqrt{n}\ \right]\ ,
\label{e:I_alpha}
\end{equation}
constructed from the measured data, 
will contain the value of the true mean $\mu$.
Equivalently, $\alpha$ is the fraction of intervals $I_\alpha$ that 
{\em fail\,} to contain  the value of the true mean $\mu$.
Of course, given the outcome of a single experiment $\bf s$, the interval 
$I_\alpha$ either contains or does not contain the value of $\mu$.
And the value of $\mu$, if it is contained in $I_\alpha$, need not be 
any closer to the center of the interval than to either of its edges.
Thus, the confidence that one associates with the above estimation
procedure is not equivalent to our degree of belief that the true mean 
$\mu$ lies within a given interval.
(This is a Bayesian interpretation of probability.)
Rather, it is the fraction of experimental outcomes that our estimation 
procedure will produce an interval that contains the true mean $\mu$, in 
the limit of an infinite number of repeated, independent experiments.

\subsection{Sensitivity levels}
\label{subsec:sensitivity_levels}

Let us assume once again that a stochastic gravity-wave signal is present, 
characterized by some fixed, but unknown, mean value $\mu>0$.
Then it is reasonable to ask:
``What is the minimum value of $\mu$ required so that our decision
rule correctly identifies the presence of a signal at least
$\gamma\cdot 100\%$ of the time?''

The answer to this question can be obtained by applying the results
of Sec.~\ref{subsec:signal_detection}.
Namely, we require that the detection rate $\gamma(\mu):=1-\beta(\mu)$ 
be greater than or equal to the desired rate $\gamma$, and then solve 
the resulting inequality for $\mu$.
Explicitly, from Eq.~(\ref{e:beta_eqn}), it follows that
\begin{equation}
1-\beta(\mu)={1\over 2}\ {\rm erfc}\left({z_{\alpha}
-\sqrt{n}\ \mu/\hat\sigma\over\sqrt{2}}\right)\ge\gamma\ ,
\end{equation}
or, equivalently,
\begin{equation}
z_{\alpha}-\sqrt{n}\ \mu/\sigma\le\sqrt{2}\ 
{\rm erfc}^{-1}(2\gamma)\ ,
\end{equation}
where we replaced the estimator $\hat\sigma$ by its expected value
$\sigma$ to obtain the LHS of the above equation.
Thus,
\begin{eqnarray}
\mu&&\ge{\sigma\over\sqrt{n}}\ \left(z_{\alpha}-
\sqrt{2}\ {\rm erfc}^{-1}(2\gamma)\right)\\
&&={\sigma\over\sqrt{n}}\ 
\sqrt{2}\left({\rm erfc}^{-1}(2\alpha)-{\rm erfc}^{-1}(2\gamma)\right)\ .
\label{e:mu_sensitivity}
\end{eqnarray}
Equivalently,
\begin{equation}
\sqrt{n}\ {\rm SNR}\ge 
\sqrt{2}\left({\rm erfc}^{-1}(2\alpha)-{\rm erfc}^{-1}(2\gamma)\right)\ ,
\label{e:SNR_sensitivity}
\end{equation}
where $\sqrt{n}\ {\rm SNR}:=\sqrt{n}\ \mu/\sigma$ 
is the theoretical signal-to-noise ratio after $n$ observation periods.
Note that the right-hand-sides of Eqs.~(\ref{e:mu_sensitivity})
and (\ref{e:SNR_sensitivity}) depend on both $\alpha$ and $\gamma$.%
\footnote{The dependence on $\alpha$ is via the threshold value 
$z_\alpha$, which defines the decision rule (\ref{e:snr_rule}).}
This means that to calculate the minimum value of $\mu$ (or the 
minimum signal-to-noise ratio), we must specify the desired detection 
rate $\gamma$ in addition to the false alarm rate $\alpha$.
In the past (see, e.g., Refs.~\cite{flan,leshouches}), people have
only specified the false alarm rate $\alpha$.
It seems that they have mistakenly assumed that the probability of 
correctly identifying the presence of a signal was $1-\alpha$.
But, as mentioned in Sec.~\ref{subsec:signal_detection}, 
$1-\alpha$ is the probability that the
decision rule correctly identifies the {\em absence\,} of a signal.
To talk about detection (and the associated minimum value of $\mu$), 
one needs to include the additional parameter $\gamma$.

For stochastic background searches, we can rewrite 
Eq.~(\ref{e:mu_sensitivity}) in terms of $\Omega_0$, $\gamma(f)$, and 
the noise power spectra $P_1(f)$ and $P_2(f)$ of the two detectors:
\widetext%
\begin{equation}
\Omega_0\ge{1\over\sqrt{T_{\rm tot}}}\ {10\pi^2\over 3 H_0^2}\ 
\left[\int_{-\infty}^\infty df\ 
{\gamma^2(|f|)\over f^6 P_1(|f|) P_2(|f|)}\right]^{-1/2}\ 
\sqrt{2}\left({\rm erfc}^{-1}(2\alpha)-{\rm erfc}^{-1}(2\gamma)\right)\ ,
\label{e:Omega_sensitivity}
\end{equation}
\narrowtext\noindent%
where $T_{\rm tot}:=n T$ is the duration of a single experiment 
$\bf s$.
This result follows from Eq.~(\ref{e:SNR_large_noise}) for a stochastic 
background of gravitational radiation having a constant frequency spectrum 
$\Omega_{\rm gw}(f)=\Omega_0$.
Let us now evaluate the minimum value of $\Omega_0$ for 4 months of 
observation (i.e., $T_{\rm tot}=10^7$\ sec), for a false alarm rate 
$\alpha=0.05$, and for a detection rate $\gamma=0.95$.  
Denoting the solution by $\Omega_0^{95\%,5\%}$, we find%
\footnote{There is nothing special about the choice $\alpha=0.05$ and
$\gamma=0.95$.
For example, $\alpha$ and $\gamma$ need not sum to one.
We could equally well have chosen $\alpha=0.10$ and $\gamma=0.95$, and
calculated a different minimum value $\Omega_0^{95\%,10\%}$.}
\begin{itemize}
\item[(i)]
 $\Omega_0^{95\%,5\%}\ h_{100}^2=5.74\times 10^{-6}\ $\ 
for the initial LIGO detector pair;
\item[(ii)] 
$\Omega_0^{95\%,5\%}\ h_{100}^2=5.68\times 10^{-11}\ $\ 
for the advanced LIGO detector pair.
\end{itemize}
Note, however, that these values {\em disagree} with those quoted in the
literature (see, e.g., Refs~\cite{flan,leshouches}).
In Refs.~\cite{flan,leshouches}, the minimum value of $\Omega_0$ is 
determined by the equation
\widetext%
\begin{equation}
\Omega_0={1\over\sqrt{T_{\rm tot}}}
\ {10\pi^2\over 3 H_0^2}\ 
\left[\int_{-\infty}^\infty df\ 
{\gamma^2(|f|)\over f^6 P_1(|f|) P_2(|f|)}\right]^{-1/2}\ 
\sqrt{2}\ {\rm erfc}^{-1}(\alpha)\ .
\label{e:literature}
\end{equation}
\narrowtext\noindent%
There are two mistakes:
First, the argument of the inverse complementary error function 
is $\alpha$ instead of $2\alpha$.
But this is appropriate for a {\em two-sided\,} test, which would be 
correct if the mean value $\mu$ could be either positive or negative.
Second, and more importantly, there is no term proportional to 
${\rm erfc}^{-1}(2\gamma)$.
As mentioned above, it seems that the authors mistakenly assumed that 
the probability of correctly identifying the presence of a signal was 
$1-\alpha$.
Even the $\alpha=0.10$, $\gamma=0.90$ value of 
Eq.~(\ref{e:Omega_sensitivity}) does not agree with the values
of $\Omega_0^{90\%}$ quoted in Refs.~\cite{flan,leshouches}.
Thus, by $\Omega_0^{90\%}$ the authors do not mean the minimum value 
of $\Omega_0$ for a false alarm rate equal to 10\% and a detection rate
equal to 90\%.

Alternatively, the absence of the term ${\rm erfc}^{-1}(2\gamma)$ 
from Eq.~(\ref{e:literature}) is equivalent to calculating the 
minimum value of $\Omega_0$ for a detection rate $\gamma=0.50$.
This is because ${\rm erfc}^{-1}(1)=0$.
Thus, the values of $\Omega_0^{90\%}$ quoted in 
Refs.~\cite{flan,leshouches} 
are for a false alarm rate equal to 5\% and a detection rate equal 
to 50\%---not 90\% as they claim.
This is why the minimum values of $\Omega_0$ quoted in those papers
are smaller than those found here.

Table \ref{t:snr} in Sec.~\ref{subsec:snrs} 
contains theoretical signal-to-noise ratios after 4 months of 
observation, for a stochastic background having a constant 
frequency spectrum $\Omega_{\rm gw}(f)=\Omega_0=6\times 10^6\ h_{100}^{-2}$.

Table \ref{t:omega_min} in Sec.~\ref{subsec:snrs} 
contains minimum values of $\Omega_0\ h_{100}^2$ 
for 4 months of observation, 
for a false alarm rate equal to 5\%, and for a detection rate equal
to 95\%.

\subsection{Summary}
\label{subsec:summary}

We started this section by asking a series of questions.
To conclude, we summarize the answers obtained above, 
and address a couple of other related issues.

\begin{itemize}
\item[(i)]
{\em ``How do we decide, from the experimental data, if we've 
detected a stochastic gravity-wave signal?''}
\end{itemize}

Answer:
We compare the measured signal-to-noise ratio after $n$ observation periods 
to the threshold value $z_\alpha$.
If $\sqrt{n}\ \widehat{\rm SNR}<z_\alpha$, we conclude that a stochastic
gravity-wave signal is absent.
If $\sqrt{n}\ \widehat{\rm SNR}\ge z_\alpha$, we conclude that a stochastic
gravity-wave signal is present, characterized by some fixed, but unknown, 
mean value $\mu>0$.

Note, however, that we can never conclude, with 100\% confidence, 
that a stochastic gravity-wave signal is absent or present.
Our decision rule leads us to infer one of these two possibilities, 
but the rule is not perfect.
The false alarm rate $\alpha$ and false dismissal rates $\beta(\mu)$
(defined by Eqs.~(\ref{e:false_alarm}) and (\ref{e:false_dismissal}))
are the error rates associated with the rule.
Thus, claims about the absence or presence of a stochastic 
gravity-wave signal should always be made with these error rates in mind.

Moreover, one has to be very careful about trying to define 
``termination'' criteria.
For example, it would be misleading to try to terminate an 
experiment by correlating the outputs of two gravity-wave detectors 
until the measured signal-to-noise ratio for the 
total observation period exceeds some threshold value $z_\alpha$.
One can show that the false alarm rate associated with such a rule
is 100\%!
In other words, the conclusion drawn from such an experiment would 
{\em always\,} be that a stochastic gravity-wave signal is present, 
even in the absence of a signal.
Noise intrinsic to the detectors and errors inherent to the 
measurement process are sufficient to guarantee that the measured
signal-to-noise ratio for the total observation period will eventually
exceed any $z_\alpha$.

If, however, the value of the threshold level increases with
observation time in an appropriate manner, then one {\em can\,} define 
termination criteria that have false alarm rates less than 100\%.
A famous theorem of probability and statistics, called 
{\em The Law of the Iterated Logarithm\,} \cite{feller}, states that 
if $S_1, S_2,\cdots $ are statistically independent and identically
distributed random variables with zero mean and finite variance
$\sigma^2$, then%
\footnote{The $\limsup_{n\rightarrow\infty}$ 
of a sequence $x_1,x_2,\cdots$ is defined as follows:
Let $a_m$ equal the least upper bound of the subsequence 
$x_m,x_{m+1},\cdots$\ .
(Note that $a_1\ge a_2\ge\cdots$\ .)
Then 
$\limsup_{n\rightarrow\infty} x_n:=\lim_{m\rightarrow\infty}a_m$.}
\begin{equation}
{\rm Prob}\left(\limsup_{n\rightarrow\infty}
{S_1+S_2+\cdots+S_n\over\sigma\sqrt{2n \log(\log(n))}}=1\right)
=1\ .
\end{equation}
This means that if $\lambda<1$, there is unit probability that
\begin{equation}
{S_1+S_2+\cdots+S_n\over\sigma\sqrt{2n \log(\log(n))}}>\lambda
\label{e:lil}
\end{equation}
for infinitely many $n$.
If $\lambda>1$, there is unit probability that (\ref{e:lil}) holds 
for only finitely many $n$.
In terms of the measured signal-to-noise ratio after $n$
observation periods, Eq.~(\ref{e:lil}) can be rewritten as 
\begin{equation}
\sqrt{n}\ \widehat{\rm SNR}>\lambda\ \sqrt{2\log(\log(n))}\ ,
\end{equation}
where we have replaced $\hat\sigma$ in the definition of
$\widehat{\rm SNR}$ by its expected value $\sigma$.
Thus, if we want to define a termination criterion with
a false alarm rate less than 100\%, we should compare 
$\sqrt{n}\ \widehat{\rm SNR}$ with threshold levels $t_n$ 
satisfying
\begin{equation}
\lim_{n\rightarrow\infty} {t_n\over\sqrt{\log(\log(n))}}=\infty\ .
\label{e:t_n}
\end{equation}
This is consistent with the claim made in the previous paragraph
that constant threshold levels $t_n=z_\alpha$ will always be
exceeded for some $n$.
Unfortunately, we have not been able to write down a simple analytic 
expression for the false alarm rate, for arbitrary threshold levels 
$t_n$ satisfying (\ref{e:t_n}).
The probability that $\sqrt{n}\ \widehat{\rm SNR}\ge t_n$ for
some $n$ involves integrals of products of the Gaussian
probability density function with the complementary error function.

\begin{itemize}
\item[(ii)]
{\em ``Assuming that a stochastic gravity-wave signal is 
present, how do we estimate its strength?''}
\end{itemize}

Answer:
Assuming that a stochastic gravity-wave signal is present, 
characterized by some fixed, but unknown, mean value $\mu>0$, 
we estimate $\mu$ by constructing the interval 
\begin{equation}
I_\alpha:=
\left[\ \hat\mu - z_{\alpha/2}\ \hat\sigma/\sqrt{n},\ 
\hat\mu + z_{\alpha/2}\ \hat\sigma/\sqrt{n}\ \right]\ .
\end{equation}
In an ensemble of observations of the same stochastic background, 
$1-\alpha$ is the fraction of intervals $I_\alpha$ constructed in this 
way that contain the value of the true mean $\mu$.

We should emphasize that parameter estimation {\em assumes\,} the 
{\em presence\,} of a signal.
First, as mentioned above, we can never be 100\% certain that a 
stochastic gravity-wave signal is present.
Second, it does not make any sense to try to estimate the parameters 
of something that we assume does not exist.

\begin{itemize}
\item[(iii)]
{\em ``Assuming that a stochastic gravity-wave signal is present, 
what is the minimum value of $\Omega_0$ required to detect it 
$\gamma\cdot 100\%$ of time?''}
\end{itemize}

Answer:
Assuming that a stochastic gravity-wave signal is present, 
characterized by some fixed, but unknown, mean value $\mu>0$, 
the minimum value of 
$\Omega_0$, for an observation time $T_{\rm tot}$, false alarm rate 
$\alpha$, and detection rate $\gamma$, is given by
\widetext%
\begin{equation}
\Omega_0
={1\over\sqrt{T_{\rm tot}}}\ {10\pi^2\over 3 H_0^2}\ 
\left[\int_{-\infty}^\infty df\ 
{\gamma^2(|f|)\over f^6 P_1(|f|) P_2(|f|)}\right]^{-1/2}\ 
\sqrt{2}\left({\rm erfc}^{-1}(2\alpha)-{\rm erfc}^{-1}(2\gamma)\right)\ .
\label{e:upper_limit1}
\end{equation}
\narrowtext\noindent%
For fixed $\alpha$ and $\gamma$, 
the factor of $T_{\rm tot}^{-1/2}$ implies that the minimum 
value of $\Omega_0$ decreases with increasing observation time.
Thus, the {\em sensitivity\,} improves as the total observation 
time increases.
This means that as $T_{\rm tot}$ increases we can put a tighter 
``upper limit'' on the values of $\Omega_0$ for stochastic 
gravity-wave signals that  we will falsely dismiss more than 
$(1-\gamma)\cdot 100\%$ of the time.

Alternatively, for a fixed $\Omega_0$ and a fixed false alarm 
rate $\alpha$, the factor of $T_{\rm tot}^{-1/2}$ implies that the false 
dismissmal rate $1-\gamma$, for a stochastic gravity-wave signal having 
a strength equal to $\Omega_0$, decreases with increasing observation time.
For example, suppose that a particular theory---like cosmic 
strings---predicts a value of $\Omega_0=10^{-7}$.
Moreover, suppose that, in successive years of observation, we fail 
to detect the presence of a signal at this level of sensitivity.
Then we can still say that the probability of our falsely dismissing 
a stochastic gravity-wave signal having a strength equal to 
$\Omega_0=10^{-7}$ has decreased over the course of the observation.

The above upper limit on stochastic background signal strengths 
(defined by Eq.~(\ref{e:upper_limit1})) is
different from that obtained by setting the measured signal-to-noise
ratio for the total observation period equal to the threshold value
$z_\alpha$:
\widetext%
\begin{equation}
\Omega_0
={1\over\sqrt{T_{\rm tot}}}\ {10\pi^2\over 3 H_0^2}\ 
\left[\int_{-\infty}^\infty df\ 
{\gamma^2(|f|)\over f^6 P_1(|f|) P_2(|f|)}\right]^{-1/2}\ 
\sqrt{2}\ {\rm erfc}^{-1}(2\alpha)\ .
\label{e:upper_limit2}
\end{equation}
\narrowtext\noindent%
This alternative upper limit is the maximum value of the test statistic 
leading to the conclusion that a stochastic gravity-wave signal is absent.
The decision rule---by its very construction---will never allow one to 
conclude that a stochastic gravity-wave signal is present with an 
$\Omega_0$ less than this upper limit.

The above two definitions of upper limit agree when the detection 
rate $\gamma=0.50$.
This is because ${\rm erfc}^{-1}(1)=0$.
This means that, if a stochastic gravity-wave signal is present, 
with an $\Omega_0$ less than or equal to this maximum value,
then there is more than a 50\% chance of falsely dismissing it.
Note also that if we change the argument of the inverse complementary
error function from $\alpha$ to $2\alpha$ in 
Eq.~(\ref{e:literature}), then Eqs.~(\ref{e:literature}) and
(\ref{e:upper_limit2}) agree.
Thus, the minimum values of $\Omega_0$ quoted in 
Refs.~\cite{flan,leshouches} can be interpreted as either the 
$\gamma=0.50$, $\alpha=0.05$ upper limit defined by 
Eq.~(\ref{e:upper_limit1}), or the $\alpha=0.05$ upper limit defined by 
Eq.~(\ref{e:upper_limit2}).

\section{Complications}
\label{sec:complications}

In Sec.~\ref{sec:detection}, we discussed optimal filtering under the 
assumptions that the noise intrinsic to the 
detectors were: (i) stationary, (ii) Gaussian, (iii) statistically
independent of one another and of the stochastic gravity-wave background, 
and (iv) much larger in magnitude than the stochastic 
gravity-wave background.
In this section, we describe the modifications that are necessary when 
some of these assumptions are removed.%
\footnote{We will {\em always\,} assume that the noise intrinsic to a
detector is Gaussian and statistically independent of the gravitational 
strains.}
We will also describe how one can correlate the outputs of 4 or more 
detectors, and how one can combine data from multiple detector pairs to 
increase the sensitivity of a stochastic background search.

\subsection{Signal-to-noise ratios for arbitrarily large stochastic 
backgrounds}
\label{subsec:arbitrary_signal_strengths}

In Sec.~\ref{subsec:optimal_filtering}, 
we calculated the signal-to-noise ratio for a stochastic background of
gravitational radiation, assuming that the noise intrinsic 
to the detectors were much larger in magnitude than the gravitational 
strains.
Although this assumption is most likely valid, there are at least two
reasons why we want to consider stochastic background signals whose
magnitudes are comparable to (or larger than) the noise intrinsic to the 
detectors:
(i) Computer simulations for stochastic background searches allow one to
``dial-in'' arbitrarily large signal strengths.
Thus, in order to compare theoretical predictions with the results of
computer simulations, we need to be able to analyze the large signal
case.
(ii) Future generations of gravity-wave detectors might have intrinsic
detector noise levels comparable to the level of a real stochastic background
signal.
For this case, optimal filtering of large signal data will also be
necessary.

To begin, let us recall the main results of 
Sec.~\ref{subsec:optimal_filtering}.
Under the assumption that the noise intrinsic to the detectors were much
larger than the gravitational strains, the calculation of the variance
$\sigma^2:=\langle S^2\rangle-\langle S\rangle^2$ simplified considerably.
For the large noise case, $\sigma^2$ was dominated by the pure detector 
noise contribution:
\begin{equation} 
\sigma^2\approx{T \over 4}
\int_{-\infty}^\infty df \> P_1(|f|) P_2(|f|)|\tilde Q(f)|^2\ .
\label{e:sigma^2Va1}
\end{equation}
This is Eq.~(\ref{e:sigma^22}) from Sec.~\ref{subsec:optimal_filtering}.

If, however,
the magnitude of the stochastic background signal is comparable to
the noise intrinsic to the detectors, the calculation 
of the signal-to-noise ratio ${\rm SNR}:=\mu/\sigma$ is more involved.
Although the mean value $\mu$ is independent of the relative size of the 
stochastic background signal and the detector noise,%
\footnote{See the discussion surrounding
Eqs.~(\ref{e:mudef})-(\ref{e:mu_final}) in 
Sec.~\ref{subsec:optimal_filtering}.}
\begin{equation}
\mu={3 H_0^2 \over 20 \pi^2}\ T\
\int_{-\infty}^{\infty} df\ |f|^{-3}\ \Omega_{\rm gw}(|f|)\ 
\gamma(|f|)\tilde Q(f)\ , 
\label{e:mu_new}
\end{equation}
the variance $\sigma^2$ is not.
Explicitly,
\begin{equation}
\sigma^2={T\over 4}\int_{-\infty}^\infty df\ 
|\tilde Q(f)|^2\ R(f)\ ,
\label{e:sigma^2_new}
\end{equation}
where
\widetext%
\begin{eqnarray}
R(f):=\bigg[&&\ P_1(|f|)P_2(|f|)
+\left({3 H_0^2\over 10\pi^2}\right)
{\Omega_{\rm gw}(|f|)\over |f|^3}\ \Big(P_1(|f|)+P_2(|f|)\Big)
\nonumber\\
&&+\left({3 H_0^2\over 10\pi^2}\right)^2
{\Omega^2_{\rm gw}(|f|)\over f^6}\ \left(1+\gamma^2(|f|)\right)\ 
\bigg]\ .
\label{e:R(f)}
\end{eqnarray}
\narrowtext\noindent%
Note the additional terms that contribute to the variance.
Roughly speaking, they can be thought of as two ``signal+noise'' 
cross-terms and a ``pure signal'' variance term.%
\footnote{These are the terms proportional to $\Omega_{\rm gw}(|f|)$ 
and $\Omega^2_{\rm gw}(|f|)$, respectively.}
When $\Omega_{\rm gw}(|f|)$ is negligible compared to the detector noise 
power spectra $P_i(|f|)$ $(i=1,2)$, Eqs.~(\ref{e:sigma^2_new}) and 
(\ref{e:R(f)}) for $\sigma^2$ reduce to the ``pure noise'' variance 
term (\ref{e:sigma^2Va1}) as they should.

At this stage of the analysis, the filter function $\tilde Q(f)$ in 
(\ref{e:mu_new}) and (\ref{e:sigma^2_new}) is arbitrary.
For the case of large detector noise, we were able to make an {\em optimal\,} 
choice for $\tilde Q(f)$, which maximized the signal-to-noise ratio.
This was facilitated by introducing an inner product
\begin{equation}
(A,B):=\int_{-\infty}^\infty df \ A^*(f)B(f)P_1(|f|)P_2(|f|)\ ,
\label{e:(A,B)_old}
\end{equation}
and writing $\mu$ and $\sigma^2$ in terms of this inner product.
The squared signal-to-noise ratio then took the form
\begin{equation}
{\rm SNR}^2=
\left({3 H_0^2 \over 10 \pi^2 }\right)^2 T\ 
{\left(\tilde Q,{\gamma(|f|)\Omega_{\rm gw}(|f|) 
\over |f|^3 P_1(|f|) P_2(|f|)}\right)^2
\over (\tilde Q,\tilde Q)}\ ,
\label{e:snr^2_old}
\end{equation}
which was maximized by choosing
\begin{equation}
\tilde Q(f) = \lambda 
{\gamma(|f|) \Omega_{\rm gw}(|f|) \over |f|^3 P_1(|f|)
P_2(|f|)}\ .
\label{e:Q(f)_old}
\end{equation}
Although $\tilde Q(f)$ depended on $\Omega_{\rm gw}(f)$, we could 
construct a set of optimal filters $\tilde Q_\alpha(f)$ for 
stochastic backgrounds having power-law spectra
$\Omega_{\rm gw}(f)=\Omega_\alpha f^\alpha$ 
(where $\Omega_\alpha={\rm constant}$).
The proportionality constants $\Omega_\alpha$ could always be absorbed 
into the normalization constants $\lambda_\alpha$.
The resulting set of optimal filters were then {\em completely\,} 
specified by the exponent $\alpha$, the overlap reduction function, and 
the noise power spectra of the two detectors.

For the case where the stochastic background is comparable to 
the noise intrinsic to the detectors, we can try to do something 
similar.
We can define a {\em new\,} inner product
\begin{equation}
(A,B):=\int_{-\infty}^\infty df\ A^*(f)B(f) R(f)\ ,
\label{e:(A,B)_new}
\end{equation}
where $R(f)$ is given by (\ref{e:R(f)}).
Although this inner product is more complicated in form than the 
original inner product (\ref{e:(A,B)_old}), it is still positive-definite,
since $R(f)$ is real and positive.
In terms of (\ref{e:(A,B)_new}),
\begin{equation}
\mu={3 H_0^2 \over 20 \pi^2}\ T\ (\tilde Q,A)\quad{\rm and}\quad
\sigma^2={T\over 4}\ (\tilde Q, \tilde Q)\ ,
\end{equation}
where
\begin{equation}
A(f):=|f|^{-3}\ \Omega_{\rm gw}(|f|)\gamma(|f|) R^{-1}(f)\ .
\label{e:A(f)}
\end{equation}
The squared signal-to-noise ratio is thus
\begin{equation}
{\rm SNR}^2=
\left({3 H_0^2 \over 10 \pi^2 }\right)^2 T\ 
{(\tilde Q,A)^2\over (\tilde Q,\tilde Q)}\ ,
\label{e:snr^2_new}
\end{equation}
which has the same form as (\ref{e:snr^2_old}), although with a much 
more complicated expression for $A(f)$.
But the same argument for maximizing the squared signal-to-noise ratio
still goes through.
The optimal choice of filter function is 
\begin{equation}
\tilde Q(f)=\lambda A(f)\ ,
\label{e:Q(f)_new}
\end{equation}
where $\lambda$ is a (real) overall normalization constant.

But this is where the similarity with the large detector noise case ends, 
and where the complications start to arise.
The main problem is that the optimal filter function $\tilde Q(f)$ has a 
complicated functional dependence on the stochastic gravity-wave spectrum 
$\Omega_{\rm gw}(f)$.
For the large detector noise case, this problem did not exist.
As mentioned earlier,
we could always consider stochastic backgrounds having power-law spectra
$\Omega_{\rm gw}(f)=\Omega_\alpha f^\alpha$,
and then construct a set of optimal filters $\tilde Q_\alpha(f)$ 
labeled by the different values of $\alpha$.
But for the optimal filter function $\tilde Q(f)$ given by 
(\ref{e:Q(f)_new}), (\ref{e:A(f)}), and (\ref{e:R(f)}), 
this idea of constructing a set of filters labeled by only the power-law 
exponents $\alpha$ fails.
The proportionality constants $\Omega_\alpha$ {\em cannot\,}
be absorbed into the normalization constants $\lambda_\alpha$.
For example, if we consider a stochastic background having a constant
frequency spectrum $\Omega_{\rm gw}(f)=\Omega_0$ 
(i.e., $\alpha=0$), then
\widetext%
\begin{eqnarray}
\tilde Q(f)=\lambda\Omega_0|f|^{-3}\ \gamma(|f|)\ \bigg[&&\ P_1(|f|)P_2(|f|)
+\left({3 H_0^2\over 10\pi^2}\right)
{\Omega_0\over |f|^3}\ \Big(P_1(|f|)+P_2(|f|)\Big)
\nonumber\\
&&+\left({3 H_0^2\over 10\pi^2}\right)^2
{\Omega_0^2\over f^6}\ \left(1+\gamma^2(|f|)\right)\ 
\bigg]^{-1}\ .
\end{eqnarray}
\narrowtext\noindent%
Although the factor of $\Omega_0$ in the numerator {\em can\,} be absorbed 
into the normalization constant $\lambda$, the factors of $\Omega_0$ and 
$\Omega_0^2$ in the denominator cannot.
In other words, if we want to construct a set of optimal filters for 
arbitrarily large stochastic backgrounds having power-law spectra
$\Omega_{\rm gw}(f)=\Omega_\alpha f^\alpha$, we need 
to specify the proportionality constants $\Omega_\alpha$ in addition to 
the exponents $\alpha$.
The ``space of optimal filters'' thus becomes a much larger set, 
parametrized by $(\alpha,\Omega_\alpha)$.
Although in principle this poses no problem, in practice it requires
a more sophisticated search algorithm; 
the detector outputs will have to be analyzed for each of the filters 
$\tilde Q_{(\alpha,\Omega_\alpha)}(f)$ separately.

\subsection{Nonstationary detector noise}
\label{subsec:nonstationary_detector_noise}

It is not at all uncommon for the power spectra of the noise intrinsic
to the detectors to change over the course of time.
There will be periods of time when the detectors are relatively ``quiet,'' 
and other periods of time when the detectors are relatively ``noisy.''%
\footnote{Although the frequency of these more ``noisy'' periods should
decrease as the detectors are gradually improved over the course of months
or years, variations in the detector noise power spectra will still
{\em inevitably\,} occur.}
These changes in the power spectra will, in turn, lead to measurements 
whose statistical properties also change with time.
For example, during the quiet periods of detector operation, 
the measurements
${}^{(1)}\!S_1,{}^{(1)}\!S_2,\cdots,{}^{(1)}\!S_{n_1}$ will have an
associated variance ${}^{(1)}\!\sigma^2$ that will be smaller 
than the variance ${}^{(2)}\!\sigma^2$ associated with the measurements
${}^{(2)}\!S_1,{}^{(2)}\!S_2,\cdots,{}^{(2)}\!S_{n_2}$ taken during the 
noisy periods.
Moreover, if the optimal filter function is normalized 
(by an appropriate choice of $\lambda$ in Eq.~(\ref{e:optimal})) 
so that the theoretical mean $\mu$ is equal to some fixed value (e.g., 
$\mu=\Omega_\alpha\ T$ for a stochastic background having a power-law
spectrum $\Omega_{\rm gw}(f)=\Omega_\alpha f^\alpha$), then 
the quiet periods will have a correspondingly larger signal-to-noise ratio.
Thus, a natural question that arises in this context is:
``Can one combine the different sets of measurements, corresponding to 
the quiet and noisy periods of detector operation, so as to maximize the 
overall signal-to-noise ratio?''
The answer to this question is ``yes,'' and the proof is sketched below.

In order to handle the most general case, let us consider $m$ different 
sets of measurements
\begin{equation}
{}^{(i)}\!S_1,\ {}^{(i)}\!S_2,\ \cdots\ ,\ {}^{(i)}\!S_{n_i}\ 
\label{e:meas}
\end{equation}
corresponding to $m$ different levels of detector noise, or $m$ different
periods of detector operation.
(Here $i=1,2,\cdots,m$.)
Each of these measurements is taken over an identical time interval of
length $T$.
These measurements can be thought of as realizations of $m$ random 
variables ${}^{(i)}\!S$, each having the {\em same\,} theoretical mean 
\begin{equation}
{}^{(i)}\!\mu:=\langle{}^{(i)}\!S\rangle=:\mu\ ,
\end{equation}
but {\em different\,} theoretical variances 
\begin{equation}
{}^{(i)}\!\sigma^2:=\langle{}^{(i)}\!S^2\rangle-
\langle {}^{(i)}\!S\rangle^2\ .
\end{equation}
The equality of the mean values follows because we assume identical 
normalization conventions on the optimal 
filters---e.g., $\mu=\Omega_\alpha\ T$ for a power-law spectrum 
$\Omega_{\rm gw}(f)=\Omega_\alpha f^\alpha$.
We also assume that {\em all\,} of the measurements
${}^{(1)}\!S_1,{}^{(1)}\!S_2,\cdots,{}^{(m)}\!S_{n_m}$
are statistically independent of one another.%
\footnote{As mentioned in Sec.~\ref{subsec:statistical_considerations},
this means that the measurements are taken over distinct, non-overlapping
periods of operation, with $T\gg$ the light travel 
time $d/c$ between the two detectors.}

For each set of measurements, we can construct the
sample mean (or estimator)
\begin{equation}
{}^{(i)}\!\hat\mu:={1\over n_i}
\sum_{j=1}^{n_i} {}^{(i)}\!S_j\ .
\end{equation}
Viewed as a random variable in its own right, ${}^{(i)}\!\hat\mu$ 
has mean value 
\begin{equation}
\mu_i:=\langle{}^{(i)}\!\hat\mu\rangle=\mu\ ,
\end{equation}
and variance
\begin{equation}
\sigma_i^2:=\langle{}^{(i)}\!\hat\mu^2\rangle
-\langle{}^{(i)}\!\hat\mu\rangle^2
={{}^{(i)}\!\sigma^2\over n_i}\ .
\label{e:sigma_relationship}
\end{equation}
%

What we want to do now is combine all the measurements
${}^{(1)}\!S_1,{}^{(1)}\!S_2,\cdots,{}^{(m)}\!S_{n_m}$
(or, equivalently, combine the sample means
${}^{(1)}\!\hat\mu,{}^{(2)}\!\hat\mu,\cdots,{}^{(m)}\!\hat\mu$)
so as to maximize the overall signal-to-noise ratio.
We thus use a {\em weighted\,} average to define the estimator
\begin{equation}
\hat\mu:={\sum_{i=1}^m\lambda_i\ {}^{(i)}\hat\mu\over
\sum_{j=1}^m\lambda_j}\ ,
\label{e:w_avg}
\end{equation}
and then choose $\lambda_i>0$ to maximize the signal-to-noise ratio
of $\hat\mu$.
From (\ref{e:w_avg}), it follows that $\hat\mu$ has mean value 
\begin{equation}
\mu_{\hat\mu}:=\langle\hat\mu\rangle=\mu
\end{equation}
(which is independent of the choice of $\lambda_i$), and variance
\begin{equation}
\sigma^2_{\hat\mu}:=\langle\hat\mu^2\rangle-\langle\hat\mu\rangle^2
={\sum_{i=1}^m\lambda_i^2\ \sigma_i^2\over
\left(\sum_{j=1}^m\lambda_j\right)^2}\ .
\end{equation}
The squared signal-to-noise ratio is thus
\begin{equation}
{\rm SNR}^2_{\hat\mu}:={\mu^2_{\hat\mu}\over\sigma^2_{\hat\mu}}
=\mu^2\ {\left(\sum_{j=1}^m\lambda_j\right)^2
\over\sum_{i=1}^m\lambda_i^2\ \sigma_i^2}\ .
\label{e:snr^2_mu}
\end{equation}
%

To find the $\lambda_i$ which maximize ${\rm SNR}_{\hat\mu}^2$
(and hence which tell us how to optimally combine data from periods of 
quiet and noisy detector operation) is quite easy. 
This is because Eq.~(\ref{e:snr^2_mu}) can be written as a ratio of inner 
products, just as we were able to write the squared signal-to-noise ratio
(\ref{e:SNR^2}) in Sec.~\ref{subsec:optimal_filtering} 
as a ratio of inner products.
Explicitly,
\begin{equation}
{\rm SNR}^2_{\hat\mu}=\mu^2\ {(\lambda,\sigma^{-2})^2\over
(\lambda,\lambda)}
\label{e:snr_label}
\end{equation}
where the inner product $(\alpha,\beta)$ is defined by
\begin{equation}
(\alpha,\beta):=\sum_{i=1}^m\alpha_i^*\beta_i\ \sigma_i^2\ ,
\end{equation}
for any pair of (complex) sequences $\alpha_i$ and $\beta_i$ 
($i=1,2,\cdots,m$).
The inner product $(\alpha,\beta)$ is positive-definite, since 
$\sigma_i^2$ is real and positive, 
and it satisfies all of the properties of the ordinary dot 
product of vectors in three-dimensional Euclidean space.
Thus, choosing 
\begin{equation}
\lambda_i\propto \sigma_i^{-2}
\end{equation}
maximizes (\ref{e:snr_label}) in the same way that choosing 
$\vec A$ proportional to $\vec B$ maximizes the ratio 
$(\vec A\cdot\vec B)^2/(\vec A\cdot\vec A)$.
By averaging each set of measurements with the inverse of its associated
theoretical variance, we give more weight to signal values that are 
measured when the detectors are quiet, than to signal values that are 
measured when the detectors are noisy.
This weighting maximizes the overall signal-to-noise ratio.

For the optimal choice of weights $\lambda_i=\sigma_i^{-2}$, the 
inverse variance of the optimal estimator
\begin{equation}
\hat\mu_{\rm optimal}:=\hat\mu\big|_{\lambda_i=\sigma_i^{-2}}
\end{equation}
has a very simple form:
\begin{equation}
\sigma_{\rm optimal}^{-2}
=\sum_{i=1}^m \sigma_i^{-2}
=\sum_{i=1}^m n_i\ {}^{(i)}\!\sigma^{-2}\ .
\label{e:sigma_optimal}
\end{equation}
This result says that the variances for the optimal estimator add 
like electrical resistors in parallel.
The squared signal-to-noise ratio of the optimal estimator also
has a very simple form:
\begin{equation}
{\rm SNR}_{\rm optimal}^2=\sum_{i=1}^m {\rm SNR}_i^2
=\sum_{i=1}^m n_i\ {}^{(i)}{\rm SNR}^2\ ,
\label{e:snr_optimal}
\end{equation}
where
\begin{equation}
{\rm SNR}_i:={\mu_i\over\sigma_i}={\mu\over\sigma_i}
\end{equation}
and
\begin{equation}
{}^{(i)}{\rm SNR}:={{}^{(i)}\!\mu\over{}^{(i)}\!\sigma}
={\mu\over{}^{(i)}\!\sigma}\ .
\end{equation}
Thus, the squared signal-to-noise ratio of the optimal estimator
is simply a sum the squared signal-to-noise ratios for each
${}^{(i)}\!S$ after $n_i$ observation periods, each of length $T$.

It is instructive to compare the results for the optimal estimator 
$\hat\mu_{\rm optimal}$ with those for the ``naive'' estimator
\begin{equation}
\hat\mu_{\rm naive}:={1\over n_{\rm tot}}\sum_{i=1}^m\sum_{j=1}^{n_i}\ 
{}^{(i)}\!S_j\ ,
\end{equation}
which simply averages all of the signal estimates, paying no
attention to the different variances ${}^{(i)}\!\sigma^2$ associated 
with the ${}^{(i)}\!S$.
The naive estimator $\hat\mu_{\rm naive}$ corresponds to 
Eq.~(\ref{e:w_avg}) with 
$\lambda_i=n_i$ and $n_{\rm tot}:=\sum_{i=1}^m n_i$.
One can show that 
\widetext%
\begin{equation}
\sigma_{\rm naive}^2
={1\over n_{\rm tot}^2}\sum_{i=1}^m n_i^2\ \sigma_i^2
={1\over n_{\rm tot}^2}\sum_{i=1}^m n_i\ {}^{(i)}\!\sigma^2
\label{e:sigma_naive}
\end{equation}
and
\begin{equation}
{\rm SNR}_{\rm naive}^{-2}
={1\over n_{\rm tot}^2}\sum_{i=1}^m n_i^2\ {\rm SNR}_i^{-2}
={1\over n_{\rm tot}^2}\sum_{i=1}^m n_i\ {}^{(i)}{\rm SNR}^{-2}\ .
\label{e:snr_naive}
\end{equation}
\narrowtext\noindent%
Note that in addition to the factors of $1/n_{\rm tot}^2$, 
the signs of the exponents for the variances and signal-to-noise ratios 
in Eqs.~(\ref{e:sigma_naive}) and (\ref{e:snr_naive})
are opposite to those in 
Eqs.~(\ref{e:sigma_optimal}) and (\ref{e:snr_optimal}).

To give a numerical example, suppose we have just two different periods of
detector operation, with the second period twice as noisy as the 
first---i.e., ${}^{(2)}\!\sigma=2\ {}^{(1)}\!\sigma$.
Since $\sigma_i^2={}^{(i)}\!\sigma^2/n_i$, it follows that if there
are {\em four\,} times as many measurements during the noisy period
(i.e., $n_2=4 n_1$), then $\sigma_1^2=\sigma_2^2$.
Intuition suggests that the optimal way of combining the measurements 
for this case is to weight the estimators ${}^{(1)}\!\hat\mu$ and
${}^{(2)}\!\hat\mu$ equally (i.e., $\lambda_1=\lambda_2=\sigma_1^{-2}$).
This agrees with the above mathematical analysis, and yields a
squared signal-to-noise ratio for the optimal estimator equal to  
\begin{equation}
{\rm SNR}_{\rm optimal}^2=2n_1\ {}^{(1)}{\rm SNR}^2
=2n_1\ {\mu^2\over{}^{(1)}\!\sigma^2}\ .
\end{equation}
For the naive estimator,
\begin{equation}
{\rm SNR}_{\rm naive}^2={25\over 17}n_1\ {}^{(1)}{\rm SNR}^2
={25\over 17}n_1\ {\mu^2\over{}^{(1)}\!\sigma^2}\ .
\end{equation}
Thus, for this particular example, the signal-to-noise ratio for the
optimal estimator is $\sqrt{34/25}=1.17$ times larger than that for the
naive estimator.

\subsection{Multiple detector pairs}
\label{subsec:multiple_detector_pairs}

Combining measurements from multiple detector pairs in order to increase
the sensitivity of a stochastic background search is {\em identical\,}
to combining different sets of measurements corresponding to quiet and
noisy periods of detector operation in order to maximize the overall 
signal-to-noise ratio.
The only difference between the two is one of interpretation and notation.
In Sec.~\ref{subsec:nonstationary_detector_noise},
\begin{equation}
{}^{(i)}\!S_1,\ {}^{(i)}\!S_2,\ \cdots\ ,\ {}^{(i)}\!S_{n_i}\ 
\end{equation}
denoted $n_i$ different measurements, each of length $T$, of the 
optimally-filtered cross-correlation signal ${}^{(i)}\!S$
when the level of detector noise was characterized by the variance
\begin{equation}
{}^{(i)}\!\sigma_i^2:=\langle{}^{(i)}\!S^2\rangle
-\langle{}^{(i)}\!S\rangle^2\ .
\end{equation}
In this section, for multiple detector pairs, 
\begin{equation}
{}^{(ij)}\!S_1,\ {}^{(ij)}\!S_2,\ \cdots\ ,\ {}^{(ij)}\!S_{n_{ij}}
\end{equation} 
denote $n_{ij}$ different measurements, again each of length $T$,
of the optimally-filtered cross-correlation signal ${}^{(ij)}\!S$
between the $i$th and $j$th detectors.%
\footnote{In Sec.~\ref{subsec:nonstationary_detector_noise},
$i=1,2,\cdots,m$\ labeled $m$ different levels of detector noise, or
$m$ different periods of detector operation.
In this section, $i,j=1,2,\cdots,l$\ label $l$
different detectors, and $m:=l(l-1)/2$ is the number of different 
detector pairs.}
As usual,
\begin{equation}
{}^{(ij)}\!\sigma_i^2:=\langle{}^{(ij)}\!S^2\rangle
-\langle{}^{(ij)}\!S\rangle^2
\end{equation}
denotes the variance of the cross-correlation signal ${}^{(ij)}\!S$.

Provided that each cross-correlation signal measurement occurs over the 
same time interval $T$ (which should be $\gg$
the light travel time between any pair of detectors), and that
the optimal filter functions for each detector pair have identical 
normalizations
(e.g., $\lambda$ in Eq.~(\ref{e:optimal}) is chosen so that
${}^{(ij)}\!\mu=\langle{}^{(ij)}\!S\rangle=\Omega_\alpha\ T$
for a stochastic gravity-wave background having a 
power-law spectrum $\Omega_{\rm gw}(f)=\Omega_\alpha f^\alpha$),
the mathematical analysis of 
Sec.~\ref{subsec:nonstationary_detector_noise} goes through unchanged.%
\footnote{Although cross-correlation signals
${}^{(ij)}\!S$ and ${}^{(kl)}\!S$ taken during the {\em same\,} time 
interval $T$ {\em are\,} correlated, i.e.,
\begin{displaymath} 
{\rm cov}\{{}^{(ij)}\!S,{}^{(kl)}\!S\}:=
\langle{}^{(ij)}\!S{}^{(kl)}\!S \rangle-
\langle{}^{(ij)}\!S\rangle \langle {}^{(kl)}\!S \rangle
\ne 0\ ,
\end{displaymath}
the variance-covariance matrix 
$C_{(ij)(kl)}:={\rm cov}\{{}^{(ij)}\!S,{}^{(kl)}\!S\}$ is dominated by
the diagonal terms $C_{(ij)(ij)}={\rm cov}\{{}^{(ij)}\!S,{}^{(ij)}\!S\}
=:{}^{(ij)}\!\sigma^2$ in the large noise approximation.
Thus, in practice, one can treat {\em all\,} of the measurements
${}^{(12)}\!S_1,{}^{(12)}\!S_2,\ \cdots\ ,\ {}^{(l-1,l)}\!S_{n_{l-1,l}}$,
as effectively uncorrelated.
This is because the detector noises (which are statistically independent 
of one another) are the only contributors to ${}^{(ij)}\!\sigma^2$ in this
approximation.}
The optimal way of combining the estimators
\begin{equation}
{}^{(ij)}\!\hat\mu:={1\over n_{ij}}\sum_{k=1}^{n_{ij}}{}^{(ij)}\!S_k
\end{equation}
for each detector pair is
\begin{equation}
\hat\mu:={\sum_{i=1}^l\sum_{j<i}^l\lambda_{ij}\ {}^{(ij)}\!\hat\mu\over
\sum_{i=1}^l\sum_{j<i}^l\lambda_{ij}}\ ,
\end{equation}
where%
\footnote{For completeness, we note that the optimal way of combining 
{\em correlated} random variables $x_1,\cdots,x_m$ is given by 
$\hat\mu:=\sum_{i=1}^m\lambda_i x_i/\sum_{j=1}^m\lambda_j$, where
$\lambda_i=\sum_{j=1}^m \left(C^{-1}\right)_{ij}$ and 
$\left(C^{-1}\right)_{ij}$ is the inverse of the variance-covariance
matrix $C_{ij}:={\rm cov}\{x_i,x_j\}$.
When the variance-covariance matrix is dominated by the diagonal terms 
$C_{ii}={\rm cov}\{x_i,x_i\}=:\sigma_i^2$, 
the optimal combination of data reduces (approximately) to the uncorrelated 
result $\lambda_i=\sigma_i^{-2}$.
As argued in the previous footnote, this is what happens for the 
cross-correlation signals $x_i\leftrightarrow{}^{(ij)}\!S$.}
\begin{equation}
\lambda_{ij}=\sigma_{ij}^{-2}=n_{ij}\ {}^{(ij)}\!\sigma^{-2}\ .
\end{equation}
The inverse variance for the optimal estimator 
\begin{equation}
\hat\mu_{\rm optimal}:=\hat\mu\big|_{\lambda_{ij}=\sigma_{ij}^{-2}}
\end{equation}
is given by
\begin{equation}
\sigma_{\rm optimal}^{-2}=\sum_{i=1}^l\sum_{j<i}^l n_{ij}\ 
{}^{(ij)}\!\sigma^{-2}\ .
\label{e:sigma_optimal_md}
\end{equation}
The squared signal-to-noise ratio for the optimal estimator
is given by
\begin{equation}
{\rm SNR}_{\rm optimal}^2=\sum_{i=1}^l\sum_{j<i}^l n_{ij}\ 
{}^{(ij)}{\rm SNR}^2\ .
\label{e:snr_opt_multiple}
\end{equation}
These equations should be compared with 
Eqs.~(\ref{e:sigma_optimal}) and (\ref{e:snr_optimal}) in
Sec.~\ref{subsec:nonstationary_detector_noise}.

Using Eq.~(\ref{e:sigma_optimal_md}) for the inverse variance of the 
optimal estimator, we can derive an expression for the minimum value 
of $\Omega_0$ after 4 months of observation 
(i.e., $T_{\rm tot}=10^7$ sec), for a false alarm rate equal to
5\% and a detection rate equal to 95\%, 
where we optimally combine data from different detector pairs.
In order to simplify the analysis, we will assume that $T_{\rm tot}=nT$, 
where $n:=n_{ij}$ is the same for all pairs of detectors.

For a single pair of detectors, the minimum value of $\Omega_0$
is given by
\begin{equation}
{}^{(ij)}\Omega_0^{95\%,5\%}={2\cdot 1.65\over 10^7\ {\rm sec}}\sqrt{n}\ 
{}^{(ij)}\!\sigma\ ,
\label{e:Omega^ij}
\end{equation}
where
\begin{equation}
{}^{(ij)}\!\sigma^2\approx T\left({10\pi^2\over 3 H_0^2}\right)^2
\left[\int_{-\infty}^{\infty} 
df\ {\gamma^2(|f|)\over f^6 P_i(|f|) P_j(|f|)}\right]^{-1}\ .
\end{equation}
The above expression for ${}^{(ij)}\!\sigma^2$ follows from 
Eqs.~(\ref{e:sigma^22}) and (\ref{e:optimal}), 
together with the normalization condition
$\mu=\Omega_0\ T$ for a stochastic background having a constant 
frequency spectrum $\Omega_{\rm gw}(f)=\Omega_0$.
The factor of $2\cdot 1.65$ in Eq.~(\ref{e:Omega^ij}) comes from 
the choice of the false alarm and detection rates.
(See, e.g., Eq.~(\ref{e:Omega_sensitivity}).)
Table~\ref{t:omega_min} in Sec.~\ref{subsec:snrs} contains values of
${}^{(ij)}\Omega_0^{95\%,5\%}\ h_{100}^2$ for different detector pairs.

For the optimal combination of data from multiple detector pairs, 
we have
\begin{equation}
\left(\Omega_0^{95\%,5\%}\big|_{\rm optimal}\right)^{-2}=
\left({2\cdot 1.65\over 10^7\ {\rm sec}}\right)^{-2}n^{-1}
\sum_{i=1}^l\sum_{j<i}^l {}^{(ij)}\!\sigma^{-2}\ .
\end{equation}
This can also be written as
\begin{equation}
\left(\Omega_0^{95\%,5\%}\big|_{\rm optimal}\right)^{-2}=
\sum_{i=1}^l\sum_{j<i}^l \left({}^{(ij)}\Omega_0^{95\%,5\%}\right)^{-2}\ .
\label{e:omega_min_multiple}
\end{equation}
Tables~\ref{t:triples}-\ref{t:quadruples} in Sec.~\ref{subsec:snrs}
contain values of $\Omega_0^{95\%,5\%}\big|_{\rm optimal}$ for the
optimal combination of data from multiple detector pairs, for 
all possible triples and quadruples of the five major interferometers.

\subsection{4-detector correlation}
\label{subsec:4-detector}

Rather than combine data from multiple detector pairs as described in
the previous section, one can directly correlate the outputs of 4 detectors
(or, in general, $2N$ detectors) in a manner analogous to the single 
2-detector correlation described in Sec.~\ref{subsec:optimal_filtering}.%
\footnote{The correlation of 3 detectors (or, in general, $2N+1$ detectors) 
yields a signal that has zero mean.}
In fact, it turns out that one can write down expressions for the 
optimally-filtered squared signal-to-noise ratio and the inverse 
square of $\Omega_0^{95\%,5\%}$ for the 4-detector correlation in terms of 
a simple sum of products of the 
corresponding quantities for the individual detector pairs.
As noted above, the analysis given here can easily be generalized to the 
case of $2N$ detectors.
At the end of the section, we summarize our results by writing 
down the key equations in the general $2N$-detector form.

To begin, we define the 4-detector correlation signal $S$ to be
the integrated product
\begin{equation}
S:=
\int_{-T/2}^{T/2}dt_1\cdots\int_{-T/2}^{T/2}dt_4\ 
s_1(t_1)\cdots s_4(t_4)
\ Q(t_1,\cdots,t_4)\ , 
\label{e:S_4}
\end{equation}
where 
\begin{equation}
s_i(t):=h_i(t)+n_i(t)
\end{equation}
are the outputs of the detectors $(i=1,\cdots,4)$, and 
$Q(t_1,\cdots,t_4)$ is an arbitrary filter function, which we 
will determine shortly.
To save some writing in what follows, we will use the shorthand notation
\begin{equation}
\int_{-T/2}^{T/2}d^4t\equiv
\int_{-T/2}^{T/2}dt_1\cdots \int_{-T/2}^{T/2}dt_4\ .
\end{equation}
Throughout, we will assume that the gravitational strains $h_i(t)$
satisfy the statistical properties listed in 
Sec.~\ref{subsec:statistical_assumptions}.
We will also assume that the noise $n_i(t)$ intrinsic to the detectors
are (i) stationary, (ii) Gaussian, (iii) statistically independent of one 
another and of the gravitational strains, and (iv) much larger in 
magnitude than the gravitational strains.  
The goal is to determine the filter function $Q(t_1,\cdots,t_4)$ that
maximizes the signal-to-noise ratio ${\rm SNR}:=\mu/\sigma$ of $S$.

Let us start by calculating the variance $\sigma^2$.
Since we are assuming that the noise intrinsic to the detectors are
statistically independent of one another and of the gravitational strains, 
and that they are much larger in magnitude than the gravitational strains, 
it follows that
\widetext%
\begin{eqnarray}
\sigma^2
&:=&\langle S^2\rangle-\langle S\rangle^2\approx\langle S^2\rangle\\
&=&\int_{-T/2}^{T/2}d^4t\int_{-T/2}^{T/2}d^4t'\ 
\langle s_1(t_1)\cdots s_4(t_4)s_1(t_1')\cdots s_4(t_4')\rangle\ 
Q(t_1,\cdots,t_4)Q(t_1',\cdots,t_4')\\
&\approx&\int_{-T/2}^{T/2}d^4t\int_{-T/2}^{T/2}d^4t'\ 
\langle n_1(t_1)\cdots n_4(t_4)n_1(t_1')\cdots n_4(t_4')\rangle\ 
Q(t_1,\cdots,t_4)Q(t_1',\cdots,t_4')\\
&=&\int_{-T/2}^{T/2}d^4t\int_{-T/2}^{T/2}d^4t'\
\langle n_1(t_1)n_1(t_1')\rangle\cdots\langle n_4(t_4)n_4(t_4')\rangle\ 
Q(t_1,\cdots,t_4)Q(t_1',\cdots,t_4')\ .
\label{e:n_products}
\end{eqnarray}
(See Eqs.~(\ref{e:var_1})-(\ref{e:var}).)
Using the definition (\ref{e:P_i(f)}) of the noise power spectra $P_i(|f|)$,
Eq.~(\ref{e:n_products}) can be rewritten as
\begin{eqnarray}
\sigma^2\approx
&&\left({1\over 2}\right)^4\int_{-\infty}^\infty d^4f\ 
P_1(|f_1|)\cdots P_4(|f_4|)\nonumber\\
&&\times\int_{-T/2}^{T/2} d^4t\ 
e^{i2\pi f_1t_1}\cdots e^{i2\pi f_4t_4}\ 
Q(t_1,\cdots,t_4)\nonumber\\
&&\times\int_{-T/2}^{T/2} d^4t'\ 
e^{-i2\pi f_1t_1'}\cdots e^{-i2\pi f_4t_4'}\ 
Q(t_1',\cdots,t_4')\ ,
\end{eqnarray}
where
\begin{equation}
\int_{-\infty}^{\infty}d^4f\equiv
\int_{-\infty}^{\infty}df_1\cdots \int_{-\infty}^{\infty}df_4\ .
\end{equation}
If we further define the (finite-time) Fourier transform
\begin{equation}
\tilde Q(f_1,\cdots,f_4):=
\int_{-T/2}^{T/2} d^4t\ 
e^{-i2\pi f_1t_1}\cdots e^{-i2\pi f_4t_4}\ 
Q(t_1,\cdots,t_4)\ ,
\label{e:Q_ft}
\end{equation}
which has as its inverse (for $-T/2<t_1,\cdots,t_4<T/2$)
\begin{equation}
Q(t_1,\cdots,t_4)=
\int_{-\infty}^{\infty} d^4f\ 
e^{i2\pi f_1t_1}\cdots e^{i2\pi f_4t_4}\ 
\tilde Q(f_1,\cdots,f_4)\ ,
\label{e:Q_ift}
\end{equation}
we obtain
\begin{equation}
\sigma^2\approx
\left({1\over 2}\right)^4\int_{-\infty}^\infty d^4f\ 
P_1(|f_1|)\cdots P_4(|f_4|)\ |\tilde Q(f_1,\cdots,f_4)|^2\ .
\end{equation}
This can be written in an even more convenient form if we define an
inner product
\begin{equation}
(A,B):=
\int_{-\infty}^\infty d^4f\
P_1(|f_1|)\cdots P_4(|f_4|)\ A^*(f_1,\cdots,f_4) B(f_1,\cdots,f_4)\ ,
\label{e:ip_4}
\end{equation}
\narrowtext\noindent%
where $A(f_1,\cdots,f_4)$ and $B(f_1,\cdots,f_4)$ are any two
complex-valued functions of four variables.
(See Eq.~(\ref{e:inner_product}).)
Note that the inner product is positive-definite since $P_i(|f|)>0$.
Using (\ref{e:ip_4}) it follows that
\begin{equation}
\sigma^2\approx
\left({1\over 2}\right)^4\left(\tilde Q,\tilde Q\right)\ .
\label{e:sigma^2_ip}
\end{equation}

To calculate the mean value $\mu$, we proceed in a similar manner.
Since the noise intrinsic to the detectors are statistically independent
of one another and of the gravitational strains, it follows that
\begin{eqnarray}
\mu
&:=&\langle S\rangle\\
&=&\int_{-T/2}^{T/2} d^4t\ 
\langle s_1(t_1)\cdots s_4(t_4)\rangle\ Q(t_1,\cdots,t_4)\\
&=&\int_{-T/2}^{T/2} d^4t\ 
\langle h_1(t_1)\cdots h_4(t_4)\rangle\ Q(t_1,\cdots,t_4)\ .
\end{eqnarray}
This can be expanded to 
\widetext%
\begin{equation}
\mu=\int_{-T/2}^{T/2} d^4t\ 
\Big(\langle h_1(t_1)h_2(t_2)\rangle\langle h_3(t_3)h_4(t_4)\rangle
+\langle 13\rangle \langle 24\rangle 
+\langle 14\rangle \langle 23\rangle\Big)\ Q(t_1,\cdots,t_4)
\label{e:mu_41}
\end{equation}
by using the ``factorization'' property
\begin{equation}
\langle x_1 x_2 x_3 x_4\rangle =
\langle x_1 x_2\rangle \langle x_3 x_4\rangle+ 
\langle x_1 x_3\rangle \langle x_2 x_4\rangle+ 
\langle x_1 x_4\rangle \langle x_2 x_3\rangle 
\end{equation}
\narrowtext\noindent%
for Gaussian random variables $x_1,x_2,x_3,x_4$ each having zero mean.%
\footnote{In Eq.~(\ref{e:mu_41}), $\langle 13\rangle\langle 24\rangle$ is 
used as a shorthand notation for 
$\langle h_1(t_1)h_3(t_3)\rangle\langle h_2(t_2)h_4(t_4)\rangle$, etc.}

To express $\mu$ in terms of the inner product (\ref{e:ip_4}), we first
use (\ref{e:Q_ift}) to rewrite $Q(t_1,\cdots,t_4)$ in terms of its Fourier
transform $\tilde Q(f_1,\cdots,f_4)$:
\widetext%
\begin{eqnarray}
\mu=
&&\int_{-T/2}^{T/2} d^4t\ 
\Big(\langle h_1(t_1)h_2(t_2)\rangle\langle h_3(t_3)h_4(t_4)\rangle
+\langle 13\rangle \langle 24\rangle 
+\langle 14\rangle \langle 23\rangle\Big)\nonumber\\
&&\times\int_{-\infty}^{\infty} d^4f\ 
e^{i2\pi f_1t_1}\cdots e^{i2\pi f_4t_4}\ 
\tilde Q(f_1,\cdots,f_4)\ .
\end{eqnarray}
Then by interchanging the order of integrations and rearranging terms
in the integrand, we see that
\begin{eqnarray}
\mu=
&&\int_{-\infty}^{\infty} d^4f\int_{-T/2}^{T/2} d^4t\ 
e^{i2\pi f_1t_1}\cdots e^{i2\pi f_4t_4}\nonumber\\
&&\times\Big(\langle h_1(t_1)h_2(t_2)\rangle\langle h_3(t_3)h_4(t_4)\rangle
+\langle 13\rangle \langle 24\rangle 
+\langle 14\rangle \langle 23\rangle\Big)\ 
\tilde Q(f_1,\cdots,f_4)\ ,
\end{eqnarray}
or, equivalently,
\begin{equation}
\mu=\left(A,\tilde Q\right)\ ,
\end{equation}
where 
\begin{eqnarray}
A(f_1,\cdots,f_4):=
&&{1\over P_1(|f_1|)\cdots P_4(|f_4|)}\ 
\int_{-T/2}^{T/2} d^4t\ 
e^{-i2\pi f_1t_1}\cdots e^{-i2\pi f_4t_4}\nonumber\\
&&\times\Big(\langle h_1(t_1)h_2(t_2)\rangle\langle h_3(t_3)h_4(t_4)\rangle
+\langle 13\rangle \langle 24\rangle 
+\langle 14\rangle \langle 23\rangle\Big)\ .
\label{e:A1}
\end{eqnarray}
\narrowtext\noindent%

To simplify this expression for $A(f_1,\cdots,f_4)$, we expand the
expectation values $\langle h_i(t_i)h_j(t_j)\rangle$ as
\begin{equation}
\langle h_i(t_i)h_j(t_j)\rangle=
\int_{-\infty}^\infty df\ e^{i2\pi f(t_i-t_j)}\ 
H_{ij}(f)\ ,
\label{e:ev_time}
\end{equation}
where
\begin{equation}
H_{ij}(f)={3H_0^2\over 20\pi^2}\ 
|f|^{-3}\ \Omega_{\rm gw}(|f|)\ \gamma_{ij}(|f|)\ ,
\end{equation}
and $\gamma_{ij}(f)$ denotes the overlap reduction function between
the $ij$ detector pair.
(See Eqs.~(\ref{e:h_1(t)h_2(t')}) and ({\ref{e:H_12(f)}).)
Substituting Eq.~(\ref{e:ev_time}) into (\ref{e:A1}) yields
\widetext%
\begin{eqnarray}
A(f_1,\cdots,f_4):=
&&{1\over P_1(|f_1|)\cdots P_4(|f_4|)}\ 
\int_{-\infty}^\infty df\int_{-\infty}^\infty df'\ 
H_{12}(f)H_{34}(f')\nonumber\\
&&\times\int_{-T/2}^{T/2} d^4t\ 
e^{-i2\pi f_1t_1}\cdots e^{-i2\pi f_4t_4}\ 
e^{i2\pi f(t_1-t_2)}e^{i2\pi f'(t_3-t_4)}\nonumber\\
&&+13,24+14,23\ ,
\label{e:A2}
\end{eqnarray}
\narrowtext\noindent%
where $13,24$ and $14,23$ denote the analogous terms with the appropriate 
interchange of detector indices $1,\cdots,4$.
Since
\begin{equation}
\delta_T(f)=\int_{-T/2}^{T/2}dt\ e^{i2\pi ft}
\end{equation}
(see Eq.~\ref{e:delta_T(f)}), we can explicitly integrate over the time 
variables $t_1,\cdots,t_4$:
\widetext%
\begin{eqnarray}
A(f_1,\cdots,f_4):=
&&{1\over P_1(|f_1|)\cdots P_4(|f_4|)}\ 
\int_{-\infty}^\infty df\int_{-\infty}^\infty df'\ 
H_{12}(f)H_{34}(f')\nonumber\\
&&\times\delta_T(f_1-f)\delta_T(f_2+f)\delta_T(f_3-f')\delta_T(f_4+f')
\nonumber\\
&&+13,24+14,23\ .
\label{e:A3}
\end{eqnarray}
\narrowtext\noindent%
Also, since the noise power spectra $P_i(|f_i|)$ and filter function
$\tilde Q(f_1,\cdots,f_4)$ are not expected to vary much over the support,
$1/T$, of $\delta_T(f_1-f)$ and $\delta_T(f_3-f')$, we are justified in 
approximating these two finite-time delta functions by ordinary Dirac delta 
functions $\delta(f_1-f)$ and $\delta(f_3-f')$.
This approximation allows us to eliminate the integrations over $f$ and $f'$, 
yielding
\widetext
\begin{eqnarray}
A(f_1,\cdots,f_4)\approx
&&{H_{12}(f_1)\over P_1(|f_1|)P_2(|f_2|)}\delta_T(f_1+f_2)\ 
{H_{34}(f_3)\over P_3(|f_3|)P_4(|f_4|)}\delta_T(f_3+f_4)\nonumber\\ 
&&+13,24+14,23\ .
\label{e:A4}
\end{eqnarray}
\narrowtext\noindent%

Given the above expressions for $\sigma^2$, $\mu$, and $A(f_1,\cdots,f_4)$,
it is now a simple matter to evaluate the squared signal-to-noise ratio
of the 4-detector correlation $S$, and to determine the filter function
$\tilde Q(f_1,\cdots,f_4)$ that maximizes this ratio.
In terms of the inner product (\ref{e:ip_4}), we have
\begin{equation}
{\rm SNR}^2:={\mu^2\over\sigma^2}\approx 2^4\ {\left(A,\tilde Q\right)^2
\over\left(\tilde Q,\tilde Q\right)}\ .
\end{equation}
As we saw already in Secs.~\ref{subsec:optimal_filtering} and 
\ref{subsec:arbitrary_signal_strengths}, such a 
ratio of inner products is maximized by choosing
\begin{equation}
\tilde Q(f_1,\cdots,f_4)=\lambda\ A(f_1,\cdots,f_4)\ ,
\end{equation}
where $\lambda$ is an arbitrary (real) overall normalization constant.
For this choice of $\tilde Q(f_1,\cdots,f_4)$, the value of the squared
signal-to-noise ratio is given by
\begin{equation}
{\rm SNR}_{\rm optimal}^2\approx 2^4\ (A,A)\ .
\label{e:snr_opt_4}
\end{equation}

To find an explicit expression for ${\rm SNR}^2_{\rm optimal}$, we
substitute (\ref{e:A4}) into the RHS of (\ref{e:snr_opt_4}) and expand
the product of $A(f_1,\cdots,f_4)$ with itself.
This leads to nine different terms: 
three diagonal (i.e., squared) terms and six off-diagonal terms.
The diagonal terms are given by
\widetext%
\begin{eqnarray}
\hbox{{\rm diagonal terms}}=
&&2^4\ \int_{-\infty}^\infty d^4f\ 
{H_{12}^2(f_1)\over P_1(|f_1|)P_2(|f_2|)}\delta_T^2(f_1+f_2)\ 
{H_{34}^2(f_3)\over P_3(|f_3|)P_4(|f_4|)}\delta_T^2(f_3+f_4)\nonumber\\ 
&&+13,24+14,23\ .
\label{e:diag1}
\end{eqnarray}
A typical off-diagonal term is given by
\begin{eqnarray}
\hbox{{\rm off-diagonal term}}=
&&2^4\ \int_{-\infty}^\infty d^4f\ 
{1\over P_1(|f_1|)\cdots P_4(|f_4|)}\ 
H_{12}(f_1)H_{34}(f_3)H_{13}(f_1)H_{24}(f_2)\nonumber\\
&&\times\delta_T(f_1+f_2)\delta_T(f_3+f_4)\delta_T(f_1+f_3)\delta_T(f_2+f_4)\ .
\label{e:off_diag1}
\end{eqnarray}
\narrowtext\noindent%

Let us evaluate each of these terms separately.
First, for the diagonal terms,
by approximating one of the two finite-time delta functions in
each of the factors $\delta_T^2(f_1+f_2)$ and $\delta_T^2(f_3+f_3)$ 
in (\ref{e:diag1}) by ordinary Dirac delta functions, and by evaluating
the others at $f_1+f_2=0$ and $f_3+f_4=0$, we eliminate two of the 
integrations, and introduce a factor of $T^2$:
\widetext%
\begin{eqnarray}
\hbox{{\rm diagonal terms}}\approx
&&2^4\ T^2\ 
\int_{-\infty}^\infty df\ {H_{12}^2(f)\over P_1(|f|)P_2(|f|)}\ 
\int_{-\infty}^\infty df'\ {H_{34}^2(f')\over P_3(|f'|)P_4(|f'|)}\nonumber\\
&&+13,24+14,23\\
=&&\left({3H_0^2\over 10\pi^2}\right)^4\ T^2\ 
\int_{-\infty}^\infty df\ 
{\gamma_{12}^2(|f|)\Omega^2_{\rm gw}(|f|)\over f^6 P_1(|f|)P_2(|f|)}\
\int_{-\infty}^\infty df'\ 
{\gamma_{34}^2(|f'|)\Omega^2_{\rm gw}(|f'|)\over f'{}^6 P_3(|f'|)P_4(|f'|)}
\nonumber\\
&&+13,24+14,23\\
=&&{}^{(12)}{\rm SNR}^2\ {}^{(34)}{\rm SNR}^2+
{}^{(13)}{\rm SNR}^2\ {}^{(24)}{\rm SNR}^2+
{}^{(14)}{\rm SNR}^2\ {}^{(23)}{\rm SNR}^2\ ,
\end{eqnarray}
\narrowtext\noindent%
where ${}^{(ij)}{\rm SNR}^2$ denotes the squared signal-to-noise
ratios for the optimally-filtered cross-correlation signal for the
$ij$ detector pair, which we derived in Sec.~\ref{subsec:optimal_filtering}.

For the off-diagonal term (\ref{e:off_diag1}), we can again approximate
$\delta_T(f_1+f_2)$ and $\delta_T(f_3+f_4)$ by ordinary Dirac delta
functions to obtain
\widetext%
\begin{eqnarray}
\hbox{{\rm off-diagonal term}}\approx
&&2^4\ 
\int_{-\infty}^\infty df_1\int_{-\infty}^\infty df_3\ 
{1\over P_1(|f_1|)P_2(|f_1|)P_3(|f_3|)P_4(|f_3|)}\\ 
&&\times H_{12}(f_1)H_{34}(f_3)H_{13}(f_1)H_{24}(-f_1)\ \delta_T^2(f_1+f_3)\ .
\label{e:off_diag2}
\end{eqnarray}
\narrowtext\noindent%
By further approximating one of the finite-time delta functions in 
$\delta_T^2(f_1+f_3)$ by an ordinary Dirac delta function, and by 
evaluating the other at $f_1+f_3=0$, we eliminate one more integration 
and introduce a factor of $T$:
\widetext%
\begin{equation}
\hbox{{\rm off-diagonal term}}\approx
2^4\ T\ 
\int_{-\infty}^\infty df\ 
{H_{12}(f)H_{34}(f)H_{13}(f)H_{24}(f)\over 
P_1(|f|)P_2(|f|)P_3(|f|)P_4(|f|)}\ . 
\label{e:off_diag3}
\end{equation}
\narrowtext\noindent%
But since this term grow like $T$, while the diagonal terms grow like 
$T^2$, we can (for large observation times) ignore the contribution
of the off-diagonal terms to the optimal signal-to-noise ratio.
The final result is thus%
\footnote{
This expression should be compared with 
\begin{displaymath}
{\rm SNR}^2_{\rm optimal}=
{}^{(12)}{\rm SNR}^2+{}^{(13)}{\rm SNR}^2+\cdots+
{}^{(34)}{\rm SNR}^2\ ,
\end{displaymath}
which is the squared signal-to-noise ratio that we found in the 
previous section for the optimal combination of data from multiple 
detector pairs.
(See Eq.~(\ref{e:snr_opt_multiple}).)}
\widetext%
\begin{equation}
{\rm SNR}^2_{\rm optimal}\approx
{}^{(12)}{\rm SNR}^2\ {}^{(34)}{\rm SNR}^2+
{}^{(13)}{\rm SNR}^2\ {}^{(24)}{\rm SNR}^2+
{}^{(14)}{\rm SNR}^2\ {}^{(23)}{\rm SNR}^2\ .
\label{e:snr_opt_4_final}
\end{equation}
\narrowtext\noindent%
Note that the optimal signal-to-noise ratio for the 4-detector
correlation is {\em quadratic} in the signal-to-noise ratios for the 
individual $ij$ detector pairs.
This is because the 4-detector correlation signal $S$, given by
Eq.~(\ref{e:S_4}), is quartic in the detector outputs $s_i(t_i)$, 
while the cross-correlation signal values $^{(ij)}\!S$ are quadratic 
in the detector outputs.
(See, e.g., Eq.~(\ref{e:Sdef}).)
\narrowtext\noindent%

Given (\ref{e:snr_opt_4_final}), we can now ask the question:
``What is the minimum value of $\Omega_0$ required to detect a 
stochastic gravity-wave signal 95\% of the time, 
from data obtained via a 4-detector correlation experiment?''
Since
\begin{eqnarray}
S&:=&\int_{-T/2}^{T/2} d^4t\ s_1(t_1)\cdots s_4(t_4)\ Q(t_1,\cdots,t_4)\\
&=&\int_{-\infty}^{\infty} d^4f\ \tilde s_1(f_1)\cdots\tilde s_4(f_4)\ 
\tilde Q^*(f_1,\cdots,f_4)\ ,
\end{eqnarray}
is, effectively, a sum (over $f_1,\cdots,f_4$) of a large number of 
statistically independent random variables (products of the Fourier
amplitudes $\tilde s_i(f_i)$, which are correlated only when 
$|f_i-f_j|<1/T$), the central limit theorem guarantees that $S$
will be well-approximated by a Gaussian random variable.
Thus, the statistical analysis of Sec.~\ref{sec:detection_etc} 
is valid for the 4-detector correlation as well.
In particular, for a false alarm rate $\alpha=0.05$, and for a detection
rate $\gamma=0.95$, the minimum value of $\Omega_0$ for the 
4-detector correlation is determined by setting the signal-to-noise 
ratio after $n$ observation periods
equal to $2\cdot 1.65$ (see Eq.~(\ref{e:SNR_sensitivity})).
For a stochastic background having a constant frequency spectrum
$\Omega_{\rm gw}(f)=\Omega_0$ and for a total observation time 
$T_{\rm tot}:=nT$, the squared signal-to-noise ratio for the 
optimally-filtered 4-detector correlation signal $S$ can be written as
\widetext%
\begin{eqnarray}
{\rm SNR}^2_{\rm optimal}\approx
&&\Omega_0^4\ \left({3H_0^2\over 10\pi^2}\right)^4\ T_{\rm tot}^2\ 
\int_{-\infty}^{\infty}df\ 
{\gamma^2_{12}(|f|)\over f^6 P_1(|f|)P_2(|f|)}\
\int_{-\infty}^{\infty}df'\ 
{\gamma^2_{34}(|f'|)\over f'{}^6 P_3(|f'|)P_4(|f'|)}\nonumber\\
&&+13,24+14,23\ .
\end{eqnarray}
Setting ${\rm SNR}_{\rm optimal}=2\cdot 1.65$ and rearranging terms yields
\begin{eqnarray}
\left(\Omega_0^{95\%,5\%}\big|_{\rm optimal}\right)^{-4}\ 
(2\cdot 1.65)^2\approx
&&\left({3H_0^2\over 10\pi^2}\right)^4\ T_{\rm tot}^2\ 
\int_{-\infty}^{\infty}df\ 
{\gamma^2_{12}(|f|)\over f^6 P_1(|f|)P_2(|f|)}\nonumber\\
&&\times
\int_{-\infty}^{\infty}df'\ 
{\gamma^2_{34}(|f'|)\over f'{}^6 P_3(|f'|)P_4(|f'|)}
+13,24+14,23\ ,
\label{e:omega_opt_4_1}
\end{eqnarray}
which can also be written in terms of the minimum values 
${}^{(ij)}\Omega_0^{95\%,5\%}$
for the individual $ij$ detector pairs:
\begin{equation}
\left(\Omega_0^{95\%,5\%}\big|_{\rm optimal}\right)^{-2}
\approx 2\cdot 1.65\ \bigg[\ 
\left({}^{(12)}\Omega_0^{95\%,5\%}\right)^{-2}
\left({}^{(34)}\Omega_0^{95\%,5\%}\right)^{-2}
+13,24+14,23\ \bigg]^{1/2}\ .
\end{equation}
\narrowtext\noindent%

How does this minimum value of $\Omega_0$ compare with those found
in previous sections for a single 2-detector correlation, and for the 
optimal combination of data from multiple detector pairs?
First, from Eq.~(\ref{e:omega_opt_4_1}) we see immediately that
\begin{equation}
\Omega_0^{95\%,5\%}\big|_{\rm optimal}\sim T_{\rm tot}^{-1/2}\ .
\end{equation}
This dependence on the total observation time $T_{\rm tot}$ is the 
{\em same\,} as that for
${}^{(ij)}\Omega_0^{95\%,5\%}$ for a single detector pair, and for the 
optimal combination of data from multiple detector pairs:
\widetext%
\begin{equation}
\left(\Omega_0^{95\%,5\%}\big|_{\rm optimal}\right)^{-2}
=\left({}^{(12)}\Omega_0^{95\%,5\%}\right)^{-2}
+\left({}^{(13)}\Omega_0^{95\%,5\%}\right)^{-2}
+\cdots
+\left({}^{(34)}\Omega_0^{95\%,5\%}\right)^{-2}\ .
\end{equation}
\narrowtext\noindent%
(See Eq.~(\ref{e:omega_min_multiple}).)
Thus, by correlating 2, 4 (or even $2N$) detectors, 
one does {\em not\,} change the general dependence of the minimum value
of $\Omega_0$ on the total observation time $T_{\rm tot}$.
However, the numerical factors multiplying $T_{\rm tot}^{-1/2}$ 
differ from one another.
In fact, it is fairly easy to show that 
$\Omega_0^{95\%,5\%}\big|_{\rm optimal}$
for the single 4-detector correlation is {\it always greater\,} than 
that for the optimal combination of data from multiple detector pairs.
Thus, in theory, it is better to optimally combine data from multiple 
detector pairs than to optimally filter a single 4-detector correlation.
Table~\ref{t:omega_min_4d} in Sec.~\ref{subsec:snrs} lists values of
$\Omega_0^{95\%,5\%}|_{\rm optimal}\ h_{100}^2$
for the 4-detector correlations taken from the five major
interferometers: LIGO-WA, LIGO-LA, VIRGO, GEO-600, and TAMA-300.

Finally, to conclude this section, we rewrite the key equations derived 
above for the general case of $2N$ detectors:
\widetext
\begin{itemize}
\item[(i)]
The $2N$-detector correlation signal $S$ is defined by
\begin{equation}
S:=
\int_{-T/2}^{T/2}d^{2N}t\ 
s_1(t_1)\cdots s_{2N}(t_{2N})
\ Q(t_1,\cdots,t_{2N})\ . 
\end{equation}
\item[(ii)]
The ``factorization'' property for $2N$ Gaussian random variables each 
having zero mean is%
\footnote{The number of terms on the RHS of the ``factorization''
equation is given by $(2N-1)\ (2N-3)\cdots 1$. 
For $N=1,2,3,\cdots$\ , this corresponds to $1,3,15,\cdots$\ terms.}
\begin{equation}
\langle x_1 x_2 \cdots x_{2N}\rangle =
\langle x_1 x_2\rangle\langle x_3 x_4\rangle
\cdots\langle x_{2N-1}x_{2N}\rangle+ 
\hbox{{\rm\ all possible permutations}}\ .
\end{equation}
\item[(iii)]
The inner product for which
\begin{equation}
\mu=\left(A,\tilde Q\right)\quad\hbox{{\rm and}}\quad
\sigma^2=\left({1\over 2}\right)^{2N}\left(\tilde Q,\tilde Q\right)
\end{equation}
is given by 
\begin{equation}
(A,B):=
\int_{-\infty}^\infty d^{2N}f\
P_1(|f_1|)\cdots P_{2N}(|f_{2N}|)\ A^*(f_1,\cdots,f_{2N}) 
B(f_1,\cdots,f_{2N})\ ,
\label{ip_4}
\end{equation}
where $A(f_1,\cdots,f_{2N})$ and $B(f_1,\cdots,f_{2N})$ are any two 
complex-valued functions of $2N$ variables.
\item[(iv)]
The optimal filter function $Q(t_1,\cdots,t_{2N})$ is given in the frequency
domain by
\begin{equation}
\tilde Q(f_1,\cdots,f_{2N})=\lambda\ A(f_1,\cdots,f_{2N})\ ,
\end{equation}
where
\begin{eqnarray}
A(f_1,\cdots,f_{2N})=
&&{H_{12}(f_1)\over P_1(|f_1|)P_2(|f_2|)}\delta_T(f_1+f_2)\
{H_{34}(f_3)\over P_3(|f_3|)P_4(|f_4|)}\delta_T(f_3+f_4)\nonumber\\
&&\times\cdots 
{H_{2N-1,2N}(f_{2N-1})\over P_{2N-1}(|f_{2N-1}|)P_{2N}(|f_{2N}|)}
\delta_T(f_{2N-1}+f_{2N})\nonumber\\ 
&&+\hbox{{\rm\ all possible permutations}}\ .
\end{eqnarray}
\item[(v)]
The squared signal-to-noise ratio for the optimally-filtered 
$2N$-detector correlation is given by
\begin{equation}
{\rm SNR}^2_{\rm optimal}\approx
{}^{(12)}{\rm SNR}^2\ {}^{(34)}{\rm SNR}^2
\cdots{}^{(2N-1,2N)}{\rm SNR}^2
+\hbox{{\rm\ all possible permutations}}\ .
\end{equation}
\item[(vi)]
The minimum value of $\Omega_0$ required to detect a stochastic
gravity-wave signal 95\% of the time, with a false alarm rate
equal to 5\%, from data obtained via a $2N$-detector correlation 
experiment, is given by
\begin{eqnarray}
\left(\Omega_0^{95\%,5\%}\big|_{\rm optimal}\right)^{-2}=
&&(2\cdot 1.65)^{(2N-2)/N}\ \bigg[\ 
\left({}^{(12)}\Omega_0^{95\%,5\%}\right)^{-2}\ 
\left({}^{(34)}\Omega_0^{95\%,5\%}\right)^{-2}\nonumber\\
&&\times\cdots
\left({}^{(2N-1,2N)}\Omega_0^{95\%,5\%}\right)^{-2}
+\hbox{{\rm\ all possible permutations}}\ \bigg]^{1/N}\ ,
\end{eqnarray}
where ${}^{(ij)}\Omega_0^{95\%,5\%}$ are the analogous quantities for
the $ij$ detector pairs.
\end{itemize}
\narrowtext\noindent%

\subsection{Correlated detector noise}
\label{subsec:correlated_detector_noise}

We have shown how to carry out an experimental search for a stochastic
background of gravitational radiation, by correlating the 
outputs of widely-separated detectors.  Our analysis assumed that any
correlation between the two outputs arises only from a stochastic
gravity-wave background.  In this section, we address the validity
of this assumption, and look at possible sources of instrumental
contamination that give rise to a correlated signal between the
separated detectors.  Any source of correlated environmental noise or
interference in the separated detectors mimics the correlation arising
from a stochastic background, so it is important to understand the
order-of-magnitude effects of any potential sources of such correlated
noise.

This subject has already been considered in some detail, both in the the
published paper of Christensen \cite{chris} and, in more detail,
in Chapter 7 of his Ph.D. thesis  \cite{christensen_thesis}. 
In the thesis work, the following sources of correlated detector noise
are analyzed:
\begin{itemize}
\item[(i)]
Seismic noise, whose effects on initial LIGO are minimal.
\item[(ii)]
Fluctuations in the residual gas (for two interferometers sharing a
common vacuum system such as the LIGO-WA site).  The effects on
initial LIGO are minimal, and of course there is no effect for separated
detectors that do not share a common vacuum envelope.
\item[(iii)]
Acoustic noise, whose effects on LIGO are minimal.
\item[(iv)]
Cosmic ray showers, whose effects on LIGO are minimal.
\item[(v)]
Magnetic field fluctuations, whose effects on initial LIGO might
be significant.
\end{itemize}
In this section, we derive a general formalism for calculating the
effects of such correlated detector noise, and illustrate this for the
case of correlated magnetic fields, which Christensen concluded would
be the most significant source of correlated fluctuations at two widely
separated sites.

Although these ideas can be generalized to multiple site locations, for
simplicity we present only the two-site case.  Correlated noise in
two detectors can be described by the {\em cross-spectral\,} 
function $C(f)$ defined
by
\begin{equation}
\label{e:defcrossspec}
\langle n_1(t) n_2(t') \rangle = {1 \over 2} \int_{-\infty}^\infty \> df \>
e^{i2 \pi f (t-t')}\ C(f)\ .
\end{equation}
Because the LHS is real, the cross-spectrum satisfies
$C(f) = C^*(-f)$.  If this cross spectrum is non-vanishing, then
it produces a correlation between the two detectors which mimics
the effect of a stochastic gravity-wave background.

It is straightforward to determine the point at which a non-vanishing
cross spectrum $C(f)$ will significantly interfere with a stochastic
background search.   Such interference will {\it not} take place if the
correlation arising from the cross spectrum of detector noise is
significantly less than that arising from the stochastic
gravity-wave background:
\begin{equation}
\langle n_1(t) n_2(t') \rangle \ll \langle h_1(t) h_2(t') \rangle
\quad {\rm for\ } |t-t'| \lesssim d/c\ ,
\end{equation}
where $d/c$ is the light travel time between the two sites.  Making use of
Eqs.~(\ref{e:defcrossspec}) and (\ref{e:ev_time}), we see that
correlated detector noise will not impair a stochastic background
search if
\begin{equation}
{1\over 2}|C(f)|\ll {3H_0^2\over 20\pi^2}\ |f|^{-3}\ \Omega_{\rm gw}(|f|)\ 
\gamma(|f|)
\end{equation}
over the range of frequencies included in the optimal filter $Q(f)$.
For example, for the initial LIGO detectors, this range of frequencies
is from about 40 Hz to 300 Hz, and in a 4 month search, the expected
level of sensitivity is about $\Omega_0\ h_{100}^2 \sim 10^{-6}$, so the
RHS is $ \sim 3 \times 10^{-49}\ h^2_{100}\ {\rm sec}$.  Thus, in
order that correlated sources of noise do not interfere with initial
LIGO's 4-month stochastic background search, one must have
\begin{equation}
|C(f)|\ll 3 \times 10^{-49}\ h^2_{100}\ {\rm sec}\quad{\rm for}\ 
40 {\rm \ Hz\ } < f < 300 {\rm \ Hz\ }.
\end{equation}
This limit can be stated in an interesting way.  If we compare the
allowable cross spectrum $C(f)$ with the intrinsic noise power spectrum
$P(f) \sim 10^{-45} \rm \ sec$ in each detector, we see that they
differ by about four orders of magnitude.  Thus, {\it in order that
correlated noise sources do not interfere with a 4-month long
stochastic background search for initial LIGO, the correlated sources of
noise must not contribute more than 1\% of the motion of the test
masses, in the frequency range from 40 Hz to 300 Hz.}

Essentially the same limit on correlated noise can be written in another
fashion.  If we make use of Eq.~(\ref{e:mudef}), the contribution of 
correlated noise to the expected mean of the signal can be written as
\begin{equation}
\mu^2_{\rm correlated\ noise} = 
\left| 
{1 \over 2}\>  T  \int_{-\infty}^\infty \> df \>
C(f) \tilde Q (f) \right| ^2\ .
\end{equation}
Requiring that this be smaller than the magnitude of the variance 
$\sigma^2$ (see Eq.~(\ref{e:sigma^22})) leads immediately to the 
condition that the correlated noise will not interfere with a stochastic
background search over an observation time $T$ if
\begin{equation}
|C(f)|^2\ll {P_1(f) P_2(f) \over T f}\ .
\end{equation}
As before, the contribution of correlated noise to the motion of
the interferometer must be smaller than the intrinsic detector
noise motion by a factor of $(T f)^{-1/2}$.  For an observation time
$T$ of 4 months $(i.e., 10^7\ {\rm sec})$ sec and frequencies 
$f \sim 100 \rm \ Hz$, this gives the same 1\% bound as before.

One can also give a precise formula showing the effects of
correlated noise sources on the expected value $\mu$ of the signal.
Making use once again of the expected mean value of the signal
(\ref{e:mudef}), we can express the ratio of the signal mean
arising from the correlated noise to that arising from the
stochastic background as:
\widetext%
\begin{equation}
\left| { \mu_{\rm correlated\ noise}
\over\mu_{\rm stochastic\ background} } \right|
=   {10 \pi^2 \over 3 H_0^2}
\left| {\int_{-\infty}^\infty \> df \>
C(f) \tilde Q (f)
\over \int_{-\infty}^\infty \> df \>
|f|^{-3}\ \Omega_{\rm gw}(|f|) \gamma(|f|) \tilde Q (f) }
\right|\ .
\label{e:corrlimit1}
\end{equation}
\narrowtext\noindent%
This formula allows us to precisely determine the effect of any correlated
source of noise on the signal value starting from a model or measurement
of the correlation spectrum $C(f)$.
In particular, if we assume that the stochastic gravity-wave background 
has a constant frequency spectrum $\Omega_{\rm gw}(f)=\Omega_0$, then
\begin{equation}
\left| { \mu_{\rm correlated\ noise}
\over
\mu_{\rm stochastic\ background} } \right|
= { \Omega_{\rm correlated\ noise\ limit} \over \Omega_0}\ ,
\label{e:corrlimit2}
\end{equation}
where $\Omega_{\rm correlated\ noise\ limit} $ is (by definition) the 
smallest value of $\Omega_0$ that can be observed before the effects of
correlated instrument noise interfere with the measurement.

The effects of a given source of correlated noise can be modeled more
precisely.  Here, we work through one example: the effects of
correlated magnetic field fluctuations.  Christensen's work 
\cite{chris,christensen_thesis} concludes
that these are the most likely environmental source of correlated
noise between two sites.  

The LIGO interferometers use small magnets to steer and push the
optical elements (i.e., mirrors and beam splitters).  These magnets are an
integral part of the Length Sensing and Control (LSC) system. Forces
are applied to these magnets with electromagnetic coils, and they are
part of the servo loop that uses modulation techniques to lock and
monitor the path length difference between the test masses.  External
magnetic fields, present in the environment, exert forces on these
magnets thus constituting one of the different sources of instrument noise.

The external magnetic field  is of particular concern, because it
propagates at the speed of light, and therefore can give rise to
correlations between sites on the time scale $d/c$.  The magnetic field
in the laboratory consists of two parts:  a ``local" part and a
``global" part.  The local part is the magnetic field arising from the
instrumentation, wiring, and power lines within the lab; the global
part comes from Schuman resonances of the earth and the ionosphere, and
lightning strikes over the surface of the earth.  The local part of the
magnetic field will not correlate between widely-separated sites;
the global part, however, is likely to be highly-correlated.  Studies in
the Caltech 40-meter prototype lab \cite{coyne} have shown that the
spectrum of ambient magnetic fields is of order $10^{-7} \rm
\ Gauss/\sqrt{Hz}$, but most of this field is local. Christensen reports
in his thesis on a number of studies that show that the global magnetic
fields are well-modeled above 20 Hz by a power-law spectrum:
\begin{equation}
P_{\rm B}(f) = A \left( { f \over 40 {\rm \ Hz} } \right)^{-0.88}
\end{equation}
with $A \sim 1.2 \times 10^{-17} \rm \  Gauss^2/Hz$ during magnetically
noisy periods such as thunderstorms, and $A \sim 1.8 \times 10^{-19}
\rm \  Gauss^2/Hz$ during magnetically quiet periods.  Separate measurements
have shown that these fields have a coherence of order $r \sim 1/2$ in the
frequency range of interest, over widely-separated (almost antipodal!)\
points on the earth's surface.  These highly-correlated global fields
are the main concern here.

In order to reduce the effects of external magnetic fields on the test
masses and optics, the four magnets on each optic or test mass are
arranged to approximately cancel both the dipole and quadrupole parts
of the magnetic field.  The magnitude of the resulting force on a test
mass may be described by:
\begin{equation}
F = {\mu B \over \ell} \left[ \epsilon_0 + {\Delta \over \ell} \epsilon_1
+ {\Delta ^2\over \ell^2} \epsilon_2 + \cdots \right]\ .
\end{equation}
Here, $\mu$ is the magnetic dipole moment of one of the magnets, $B$ is
the magnitude of the ambient magnetic field, $\ell$ is the length scale
over which the magnetic field is varying in the vicinity of the test
masses and optics, and $\Delta$ is the separation between the magnets
on the test masses and optics.  The quantities 
$\epsilon_0,\epsilon_1,\cdots$\ are
the fractional difference between the dipole, quadrupole,$\cdots$\
moments of the different magnets (which are not perfectly matched).
For the initial LIGO detectors, the preliminary design values of these
quantities are approximately $\mu = 0.11 \rm \ Amp\cdot meters^2/c$,
$\epsilon_0=0.05$, $\ell = 6 \rm \ cm$, $\Delta= 15\rm \ cm$.  This leads
to a force
\begin{equation}
F = \kappa \> B 
\end{equation}
on the test masses, with $\kappa \sim 0.1\ {\rm dyne/Gauss}$ 
\cite{coyne}.
This force accelerates the test mass, producing an equivalent strain
\begin{equation}
\tilde n(f) \sim { \tilde F(f) \over M (2 \pi f)^2 L}\ ,
\end{equation}
where $M=10$ kg is the mass of the optic, and $L=4$ km is the
length of the arm.  This gives rise to a cross spectrum
\begin{equation}
C(f) = {r \kappa^2 \over M^2 (2 \pi f)^4 L^2}\ P_B(f)\ ,
\end{equation}
where $r$ is the coherence.  Evaluating the integral in
(\ref{e:corrlimit1}) and (\ref{e:corrlimit2}) we find that the limits
on $\Omega_0$ are 
\widetext%
\begin{equation}
\Omega_{\rm correlated\ noise\ limit} =
\cases{ 10^{-7} & during magnetically ``noisy" times; \cr
  1.5 \times 10^{-9} & during magnetically ``quiet" times.}
\end{equation}
\narrowtext\noindent%
Our conclusion is that for the initial LIGO design (where sensitivities
are on the order of $6\times 10^{-6}$ for 4 months of observation), magnetic
field induced correlations are not a concern.  However, for advanced LIGO
(where sensitivities are on the order of $6\times 10^{-11}$ for 4 months
of observation), the magnets must be eliminated from the design or they will
they will significantly constrain the measurements of (or limits placed on) 
$\Omega_0$. 

\section{Numerical Data}
\label{sec:numerical_data}

This section consists of a series of graphs and tables containing 
numerical data for the five major interferometers:
(i) Hanford, WA LIGO detector (LIGO-WA), (ii) Livingston, LA LIGO detector 
(LIGO-LA), (iii) VIRGO detector (VIRGO), (iv) GEO-600 detector (GEO-600), 
and (v) TAMA-300 detector (TAMA-300).
This data was derived from published site location/orientation information 
and detector noise power spectra design goals \cite{noise_curves}, using 
the stochastic background data analysis routines contained in GRASP 
\cite{GRASP}.
(See Sec.~\ref{sec:computer_simulation} for more information about the 
computer code that we wrote to perform these calculations.)

\subsection{Noise power spectra}
\label{subsec:nps}

Figures~\ref{f:noise_ligo_again}-\ref{f:noise_tama} show the predicted 
noise power spectra for the initial and advanced LIGO detectors, and for 
the VIRGO, GEO-600, and TAMA-300 detectors.
Figure~\ref{f:noise_all} displays all of the noise power spectra on a 
single graph.
Figure~\ref{f:enhanced} shows the predicted noise power spectra for the 
``enhanced'' LIGO detectors, 
which track the projected performance of the LIGO detector 
design over the next decade.
The data for the noise power spectra displayed in all of these figures 
were taken from the published design goals \cite{noise_curves}.

\begin{figure}[htb!]
\begin{center}
{\epsfig{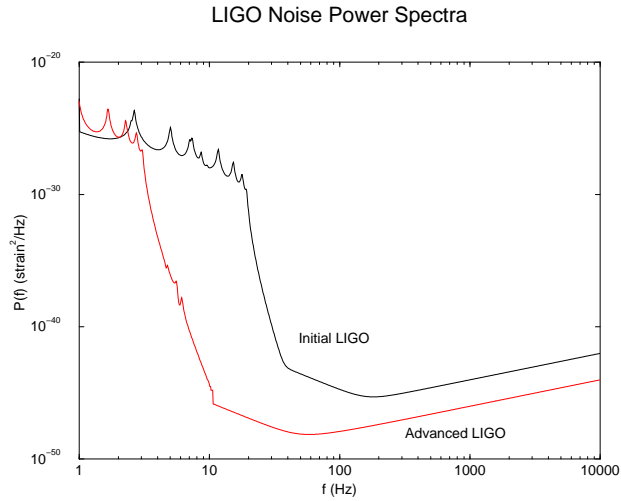}}
\caption{
\label{f:noise_ligo_again}
A log-log plot of the predicted noise power spectra for the initial and
advanced LIGO detectors.}
\end{center}
\end{figure}

\begin{figure}[htb!]
\begin{center}
{\epsfig{file=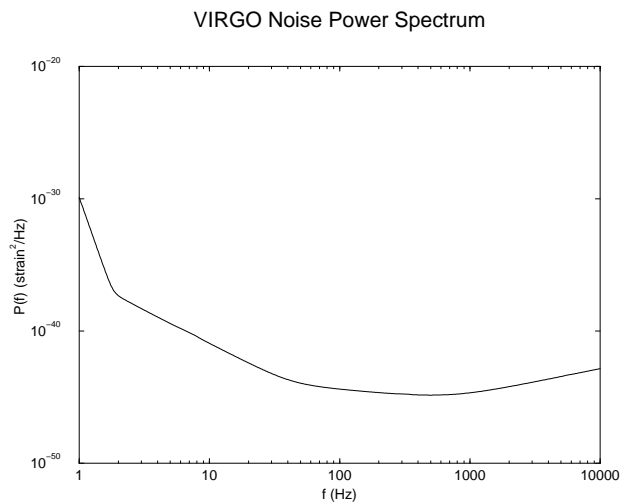,
angle=-90,width=3.4in,bbllx=25pt,bblly=50pt,bburx=590pt,bbury=740pt}}
\caption{\label{f:noise_virgo}
A log-log plot of the predicted noise power spectrum for the VIRGO
detector.}
\end{center}
\end{figure}

\begin{figure}[htb!]
\begin{center}
{\epsfig{file=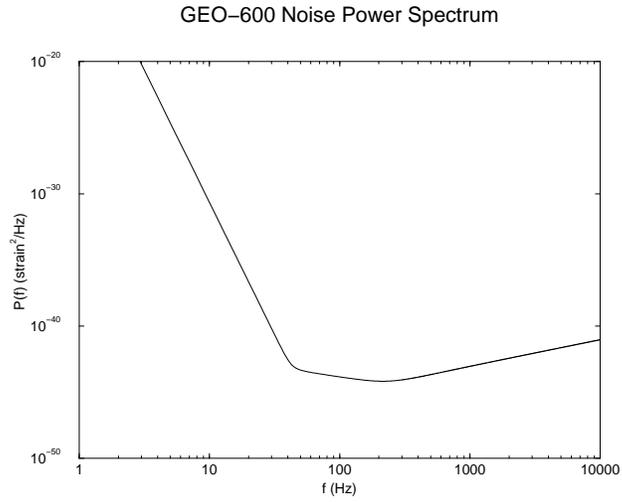,
angle=-90,width=3.4in,bbllx=25pt,bblly=50pt,bburx=590pt,bbury=740pt}}
\caption{\label{f:noise_geo}
A log-log plot of the predicted noise power spectrum for the GEO-600
detector.}
\end{center}
\end{figure}

\begin{figure}[htb!]
\begin{center}
{\epsfig{file=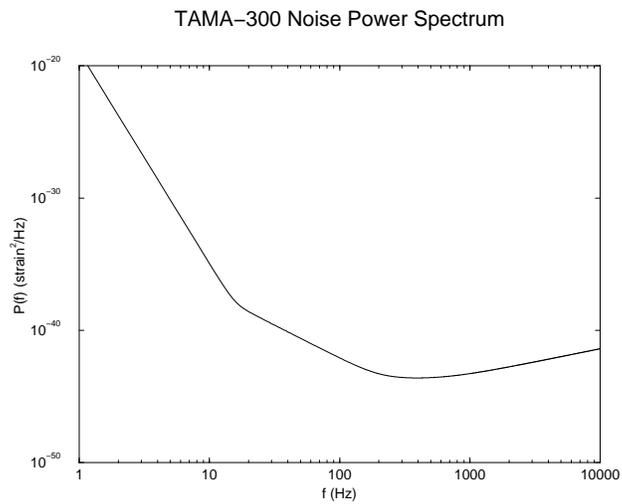,
angle=-90,width=3.4in,bbllx=25pt,bblly=50pt,bburx=590pt,bbury=740pt}}
\caption{\label{f:noise_tama}
A log-log plot of the predicted noise power spectrum for the TAMA-300
detector.}
\end{center}
\end{figure}

\begin{figure}[htb!]
\begin{center}
{\epsfig{file=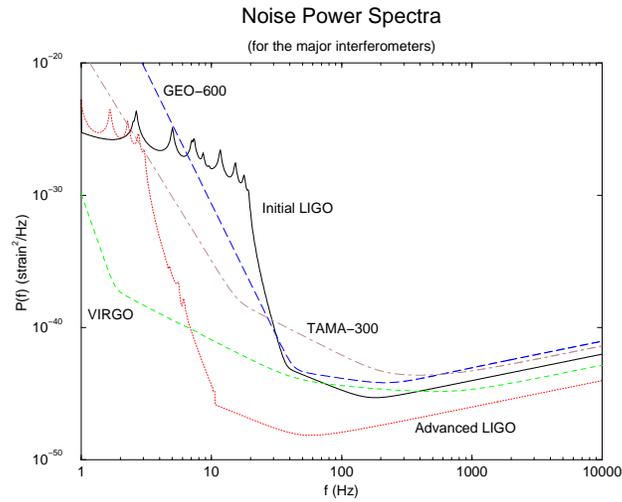,
angle=-90,width=3.4in,bbllx=25pt,bblly=50pt,bburx=590pt,bbury=740pt}}
\caption{\label{f:noise_all}
A log-log plot of the predicted noise power spectra for all the major
interferometers.}
\end{center}
\end{figure}

\begin{figure}[htb!]
\begin{center}
{\epsfig{file=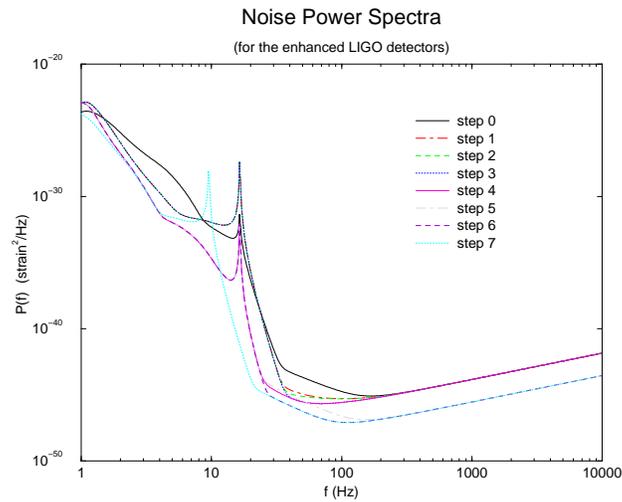,
angle=-90,width=3.4in,bbllx=25pt,bblly=50pt,bburx=590pt,bbury=740pt}}
\caption{\label{f:enhanced}
A log-log plot of the predicted noise power spectra for the ``enhanced'' 
LIGO detectors, showing the probable evolution of the detector design
over the next decade.}
\end{center}
\end{figure}

\clearpage

\subsection{Overlap reduction functions}
\label{subsec:orf}

Figures~\ref{f:LIGO-WA_overlap}-\ref{f:TAMA-300_overlap} show the
overlap reduction functions $\gamma(f)$ for different detector pairs.

\begin{figure}[htb!]
\begin{center}
{\epsfig{file=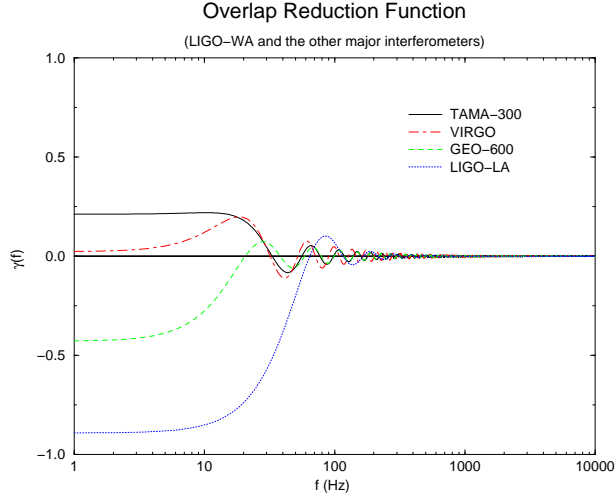,
angle=-90,width=3.4in,bbllx=25pt,bblly=50pt,bburx=590pt,bbury=740pt}}
\caption{\label{f:LIGO-WA_overlap}
The overlap reduction function $\gamma(f)$ for the LIGO-WA detector 
and the other major interferometers.
Note that the overlap reduction functions for the more distant 
detectors have their first zero at lower frequencies than those
for the more nearby detectors.}
\end{center}
\end{figure}

\begin{figure}[htb!]
\begin{center}
{\epsfig{file=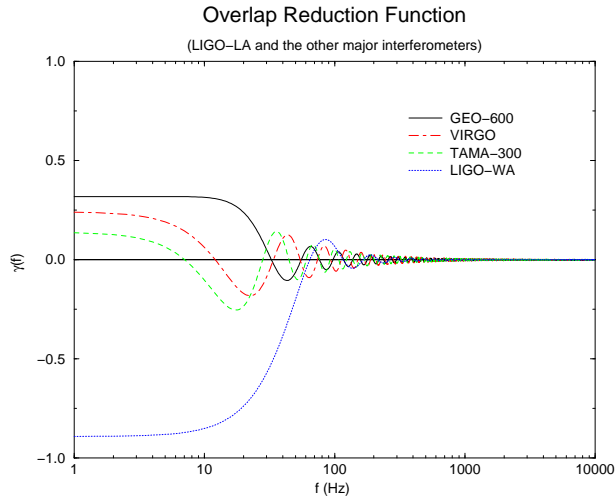,
angle=-90,width=3.4in,bbllx=25pt,bblly=50pt,bburx=590pt,bbury=740pt}}
\caption{\label{f:LIGO-LA_overlap}
The overlap reduction function $\gamma(f)$ for the LIGO-LA detector 
and the other major interferometers.}
\end{center}
\end{figure}

\begin{figure}[htb!]
\begin{center}
{\epsfig{file=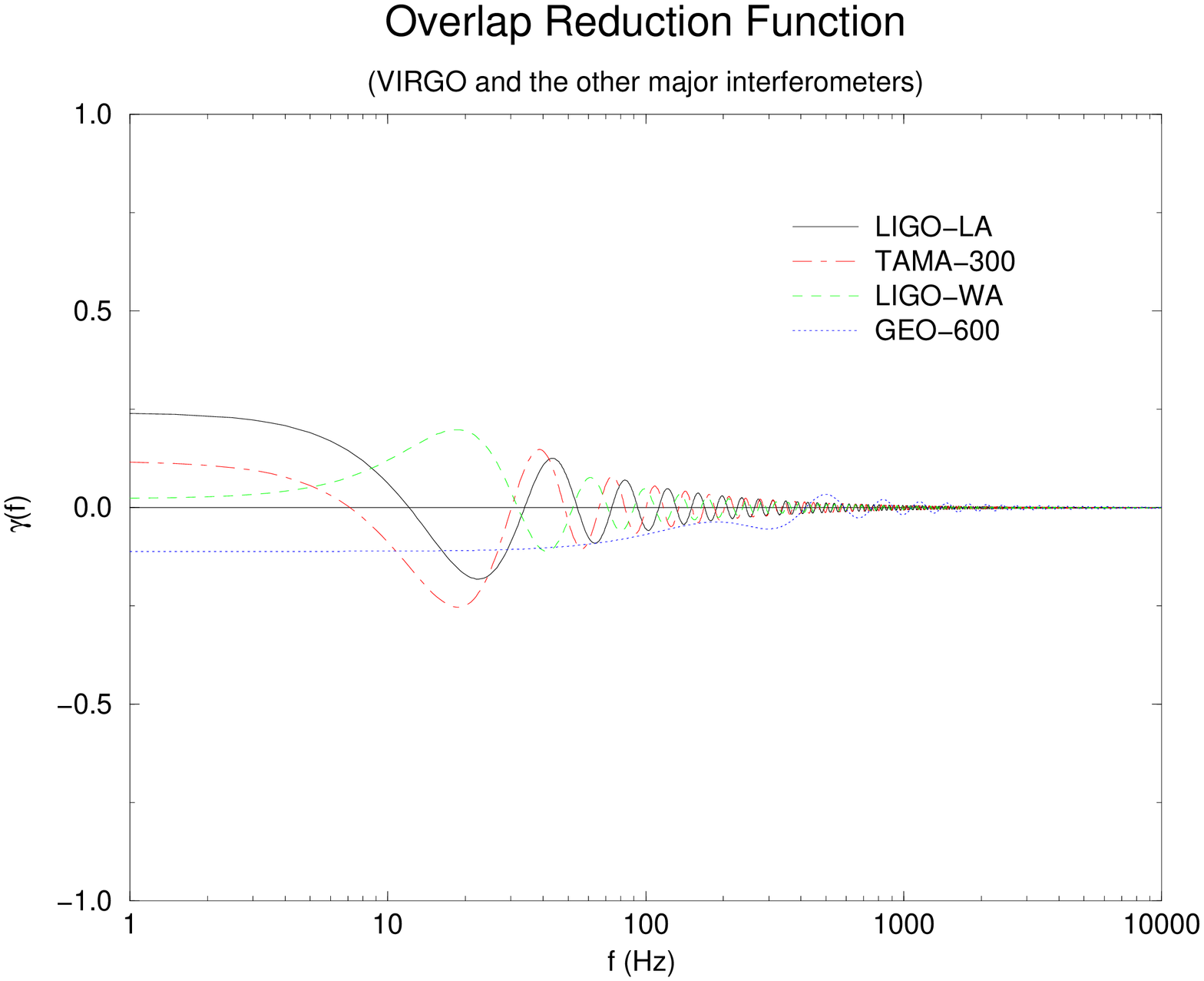,
angle=-90,width=3.4in,bbllx=25pt,bblly=50pt,bburx=590pt,bbury=740pt}}
\caption{\label{f:VIRGO_overlap}
The overlap reduction function $\gamma(f)$ for the VIRGO detector 
and the other major interferometers.
Note that the VIRGO and GEO-600 detectors are sensitive to almost
orthogonal polarizations.}
\end{center}
\end{figure}

\begin{figure}[htb!]
\begin{center}
{\epsfig{file=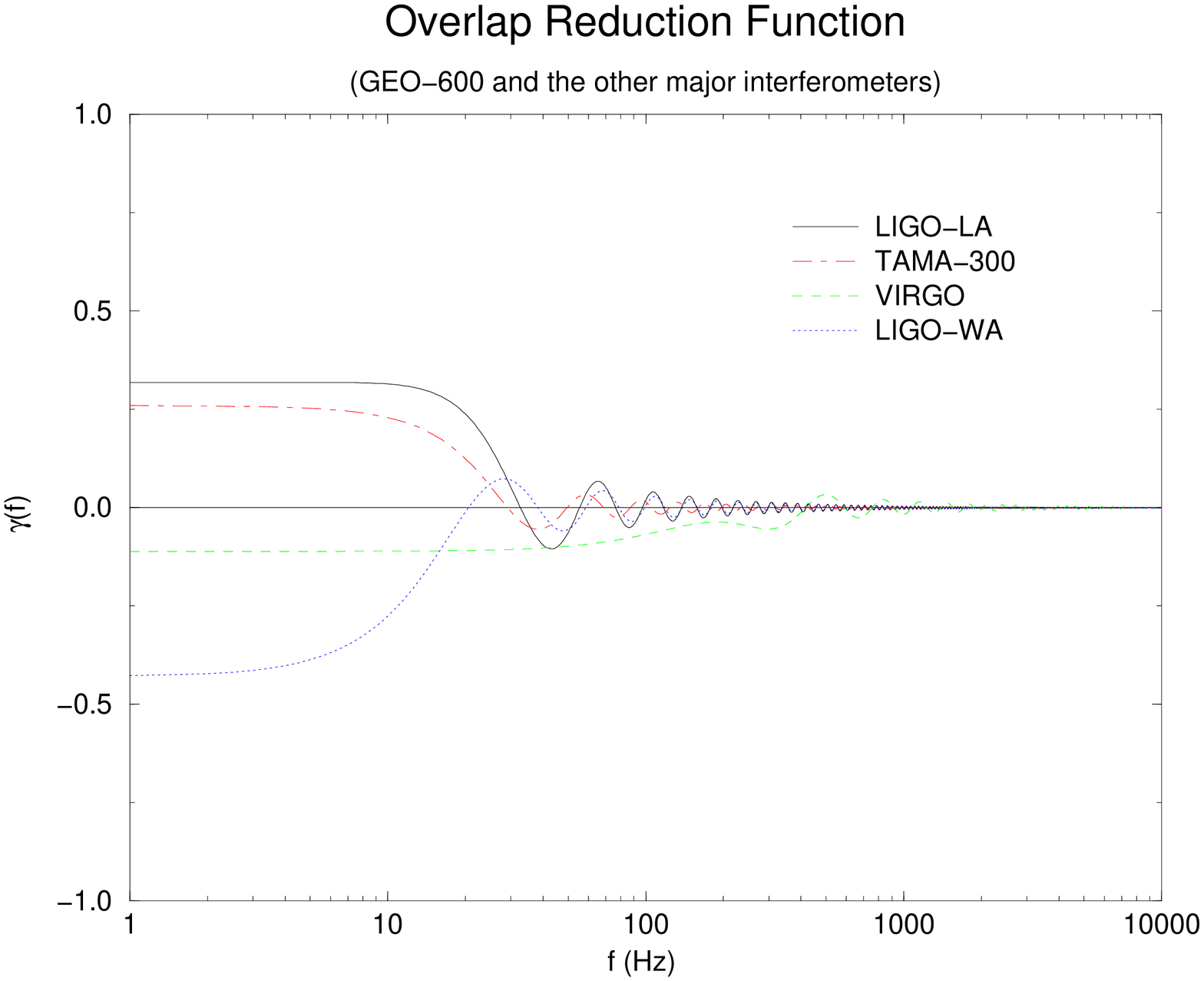,
angle=-90,width=3.4in,bbllx=25pt,bblly=50pt,bburx=590pt,bbury=740pt}}
\caption{\label{f:GEO-600_overlap}
The overlap reduction function $\gamma(f)$ for the GEO-600 detector 
and the other major interferometers.
Note that the VIRGO and GEO-600 detectors are sensitive to almost
orthogonal polarizations.}
\end{center}
\end{figure}

\begin{figure}[htb!]
\begin{center}
{\epsfig{file=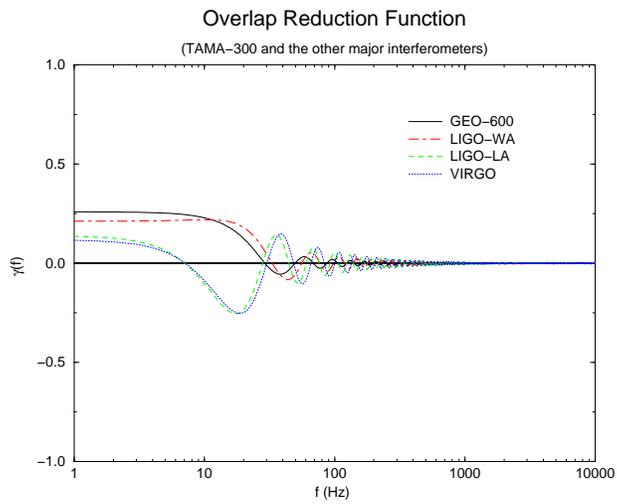,
angle=-90,width=3.4in,bbllx=25pt,bblly=50pt,bburx=590pt,bbury=740pt}}
\caption{\label{f:TAMA-300_overlap}
The overlap reduction function $\gamma(f)$ for the TAMA-300 detector 
and the other major interferometers.}
\end{center}
\end{figure}

\clearpage

\subsection{Optimal filter functions}
\label{subsec:off}

Figures~\ref{f:LIGO-WA_LIGO-LA_optimal}-\ref{f:GEO-600_TAMA-300_optimal} 
show the optimal filter functions $\tilde Q(f)$ for different detector 
pairs, for a stochastic background having a constant frequency spectrum
$\Omega_{\rm gw}(f)=\Omega_0$.
The optimal filter functions are normalized to have maximum magnitude 
of unity.

\begin{figure}[htb!]
\begin{center}
{\epsfig{file=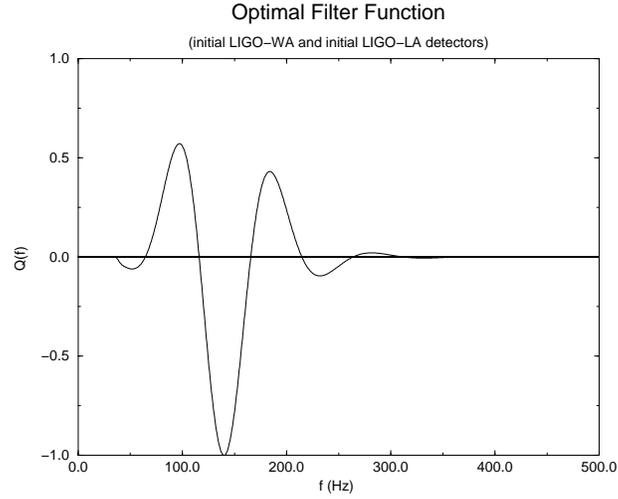,
angle=-90,width=3.4in,bbllx=25pt,bblly=50pt,bburx=590pt,bbury=740pt}}
\caption{\label{f:LIGO-WA_LIGO-LA_optimal}
The optimal filter function $\tilde Q(f)$ for the initial LIGO-WA and 
initial LIGO-LA detectors,
for a stochastic background having a constant frequency spectrum
$\Omega_{\rm gw}(f) =\Omega_0$.
The optimal filter function is normalized to have maximum magnitude of
unity.}
\end{center}
\end{figure}

\begin{figure}[htb!]
\begin{center}
{\epsfig{file=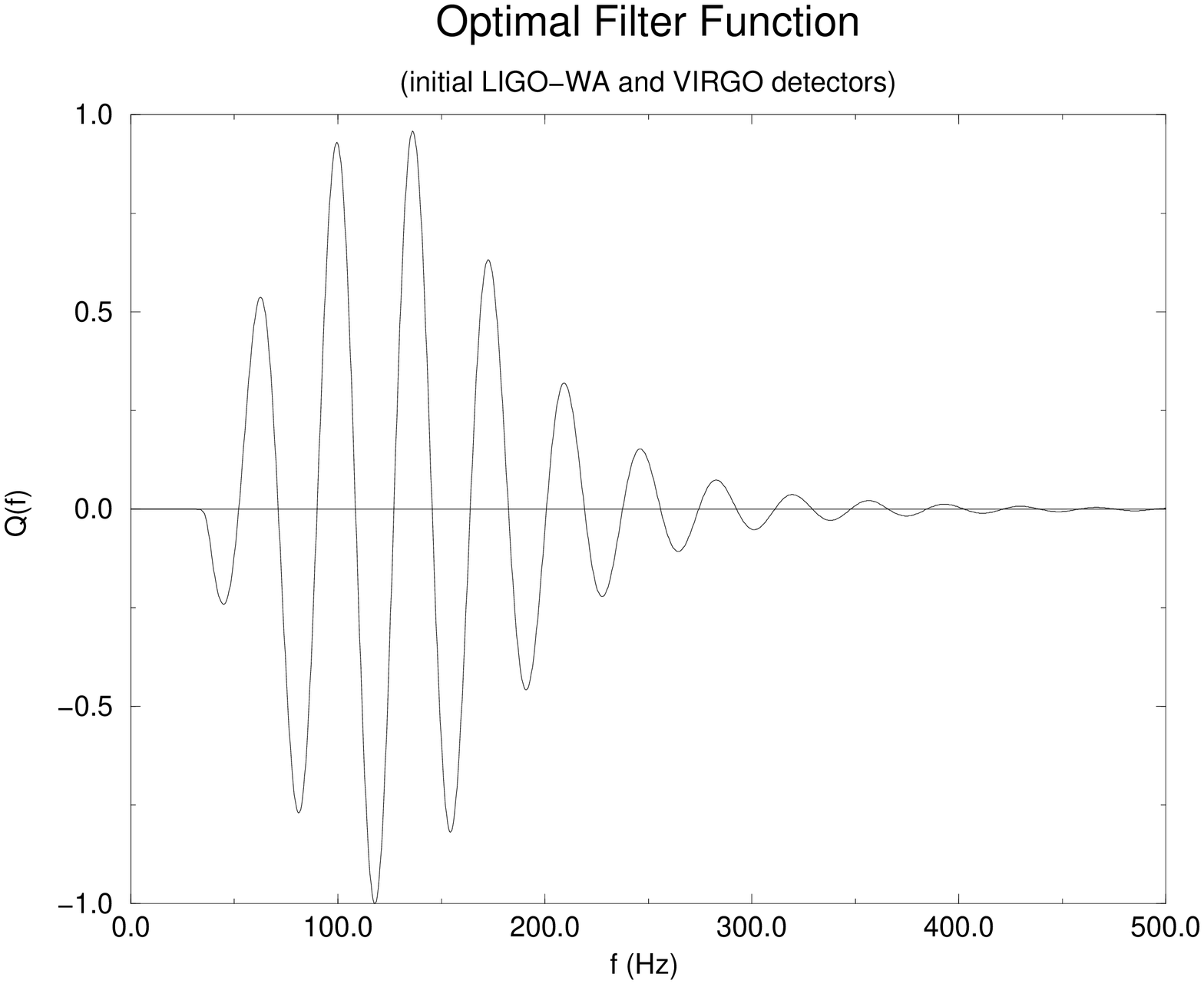,
angle=-90,width=3.4in,bbllx=25pt,bblly=50pt,bburx=590pt,bbury=740pt}}
\caption{\label{f:LIGO-WA_VIRGO_optimal}
The optimal filter function $\tilde Q(f)$ for the initial LIGO-WA and 
VIRGO detectors,
for a stochastic background having a constant frequency spectrum
$\Omega_{\rm gw}(f) =\Omega_0$.
The optimal filter function is normalized to have maximum magnitude of
unity.}
\end{center}
\end{figure}

\begin{figure}[htb!]
\begin{center}
{\epsfig{file=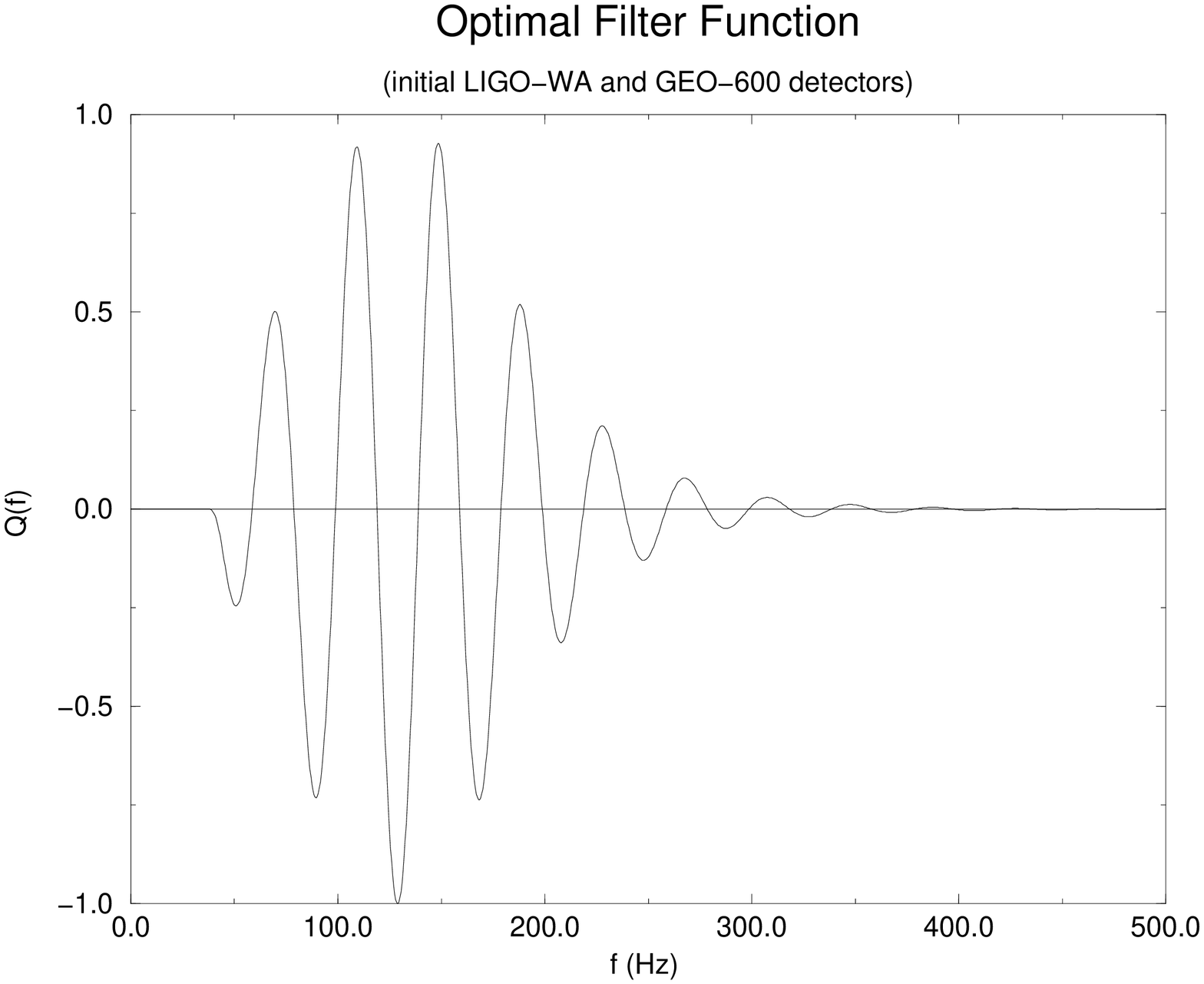,
angle=-90,width=3.4in,bbllx=25pt,bblly=50pt,bburx=590pt,bbury=740pt}}
\caption{\label{f:LIGO-WA_GEO-600_optimal}
The optimal filter function $\tilde Q(f)$ for the initial LIGO-WA and 
GEO-600 detectors,
for a stochastic background having a constant frequency spectrum
$\Omega_{\rm gw}(f) =\Omega_0$.
The optimal filter function is normalized to have maximum magnitude of
unity.}
\end{center}
\end{figure}

\begin{figure}[htb!]
\begin{center}
{\epsfig{file=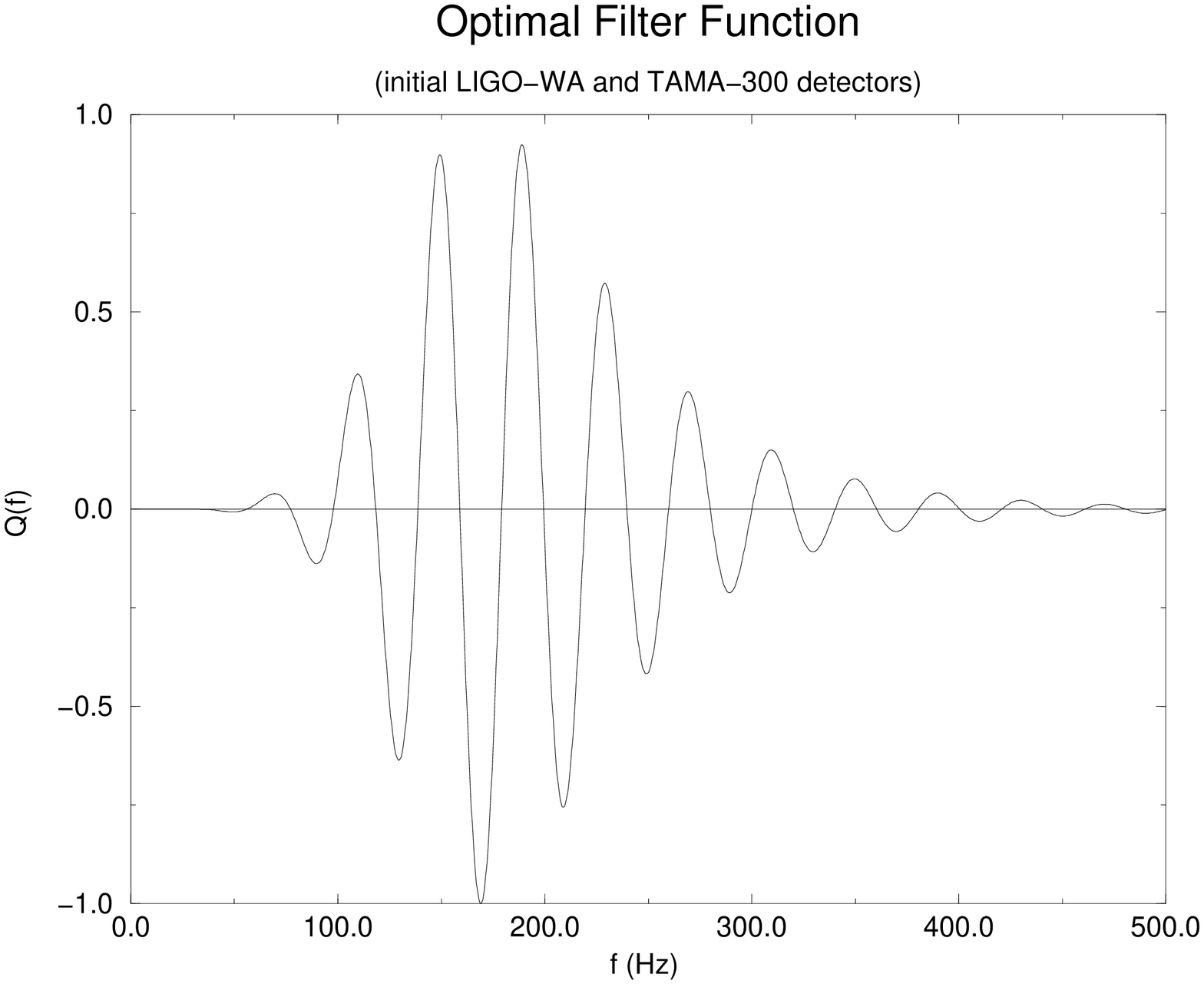,
angle=-90,width=3.4in,bbllx=25pt,bblly=50pt,bburx=590pt,bbury=740pt}}
\caption{\label{f:LIGO-WA_TAMA-300_optimal}
The optimal filter function $\tilde Q(f)$ for the initial LIGO-WA and 
TAMA-300 detectors,
for a stochastic background having a constant frequency spectrum
$\Omega_{\rm gw}(f) =\Omega_0$.
The optimal filter function is normalized to have maximum magnitude of
unity.}
\end{center}
\end{figure}

\begin{figure}[htb!]
\begin{center}
{\epsfig{file=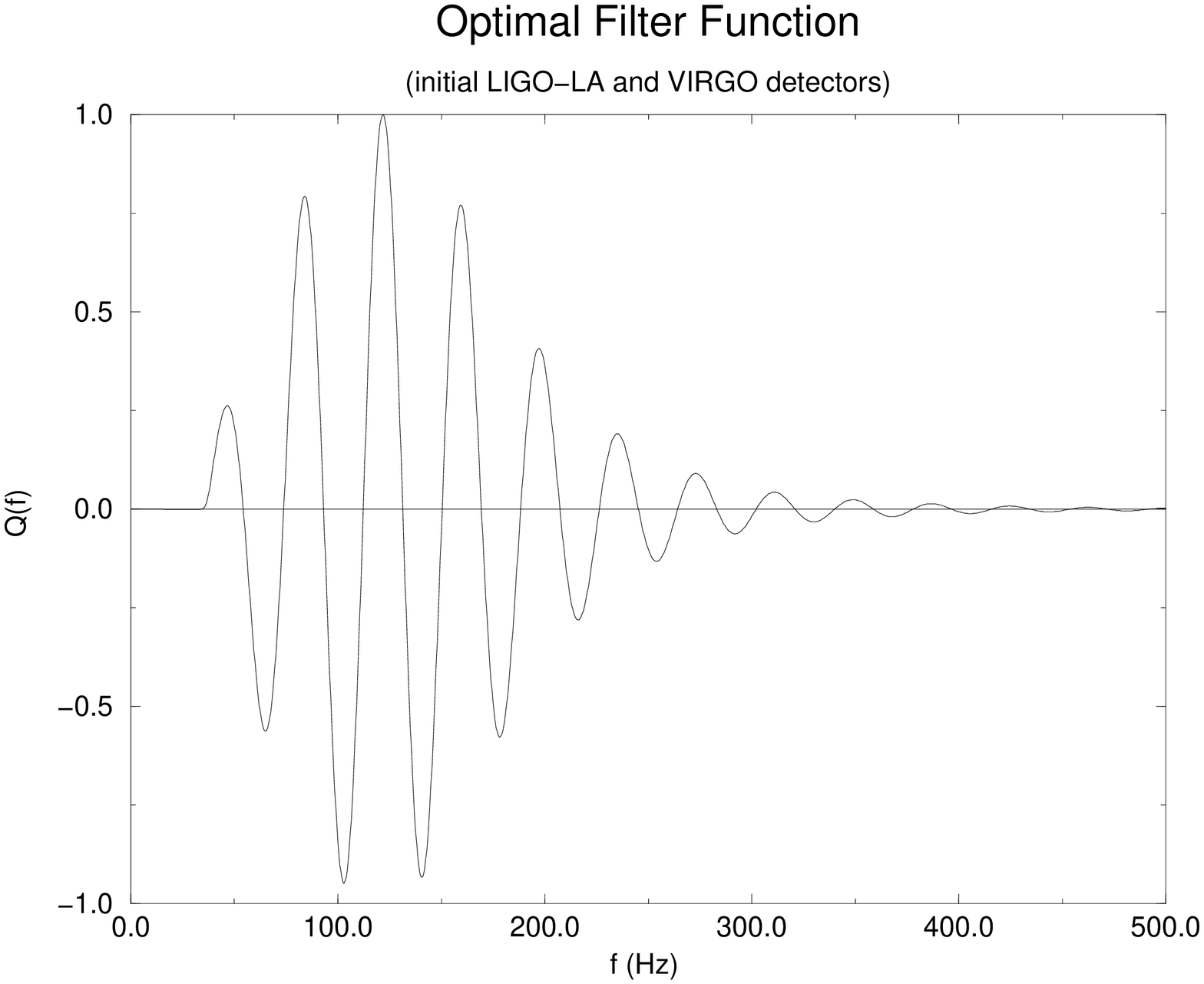,
angle=-90,width=3.4in,bbllx=25pt,bblly=50pt,bburx=590pt,bbury=740pt}}
\caption{\label{f:LIGO-LA_VIRGO_optimal}
The optimal filter function $\tilde Q(f)$ for the initial LIGO-LA and 
VIRGO detectors,
for a stochastic background having a constant frequency spectrum
$\Omega_{\rm gw}(f) =\Omega_0$.
The optimal filter function is normalized to have maximum magnitude of
unity.}
\end{center}
\end{figure}

\begin{figure}[htb!]
\begin{center}
{\epsfig{file=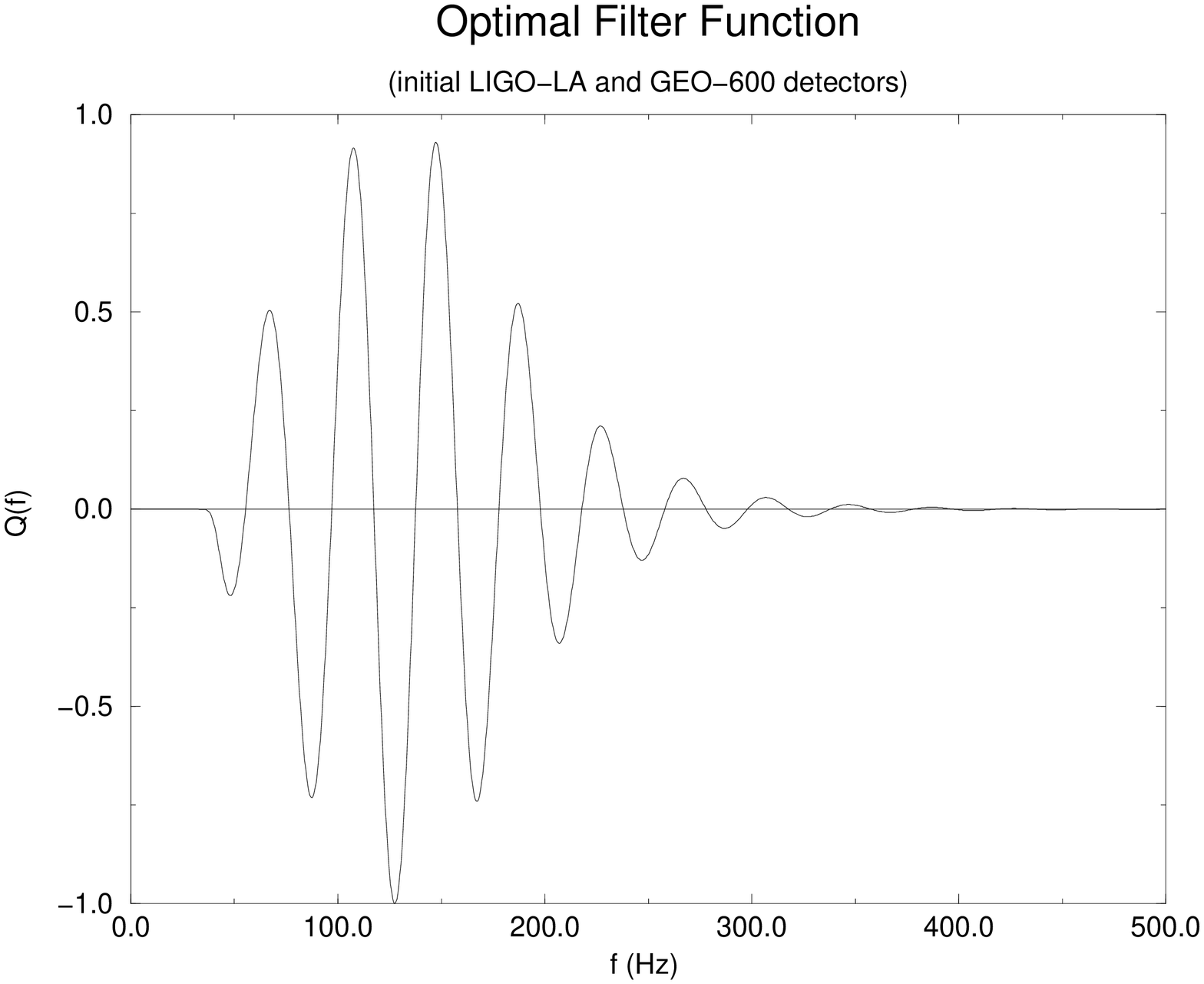,
angle=-90,width=3.4in,bbllx=25pt,bblly=50pt,bburx=590pt,bbury=740pt}}
\caption{\label{f:LIGO-LA_GEO-600_optimal}
The optimal filter function $\tilde Q(f)$ for the initial LIGO-LA and
GEO-600 detectors,
for a stochastic background having a constant frequency spectrum
$\Omega_{\rm gw}(f) =\Omega_0$.
The optimal filter function is normalized to have maximum magnitude of
unity.}
\end{center}
\end{figure}

\begin{figure}[htb!]
\begin{center}
{\epsfig{file=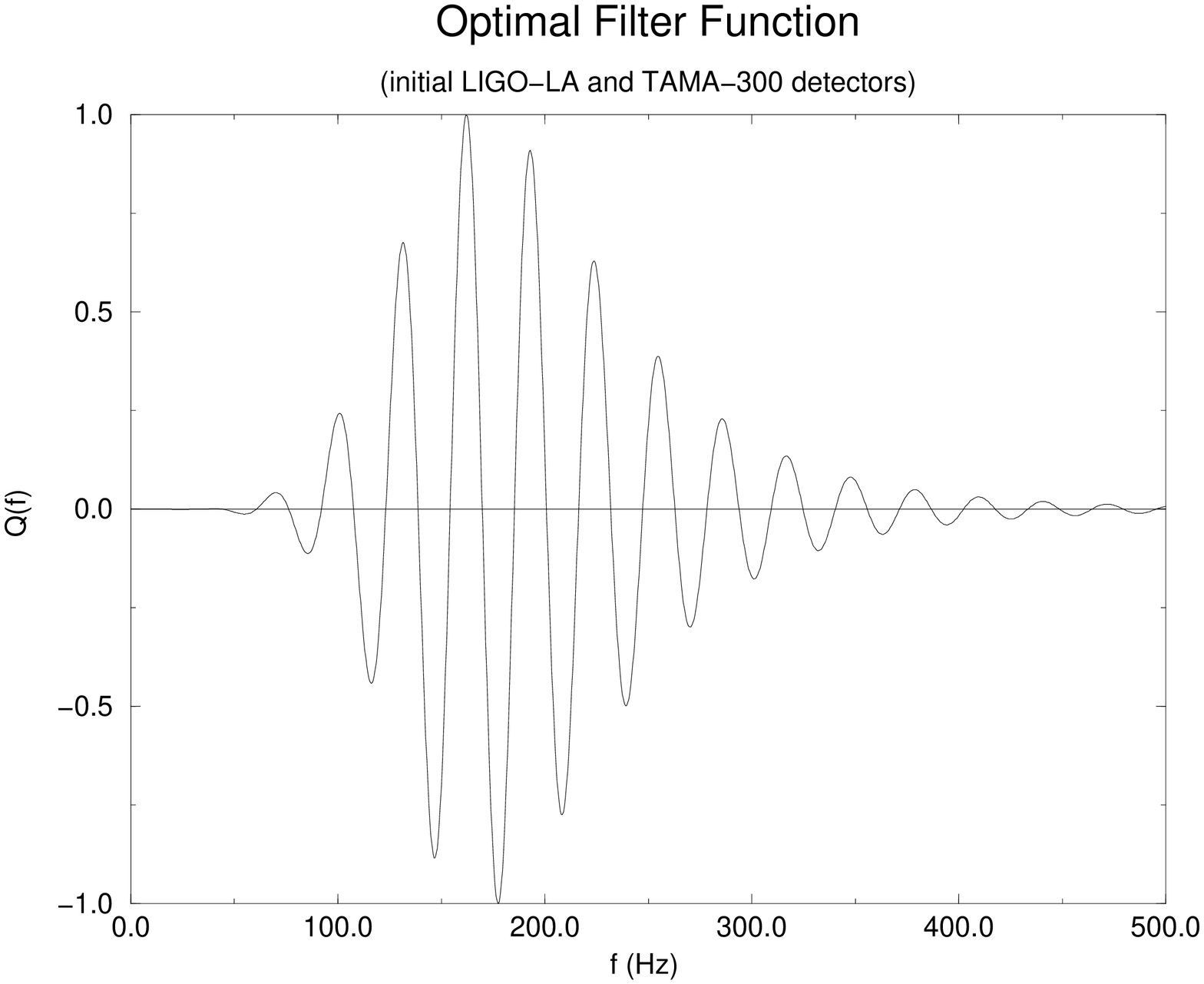,
angle=-90,width=3.4in,bbllx=25pt,bblly=50pt,bburx=590pt,bbury=740pt}}
\caption{\label{f:LIGO-LA_TAMA-300_optimal}
The optimal filter function $\tilde Q(f)$ for the initial LIGO-LA and 
TAMA-300 detectors,
for a stochastic background having a constant frequency spectrum
$\Omega_{\rm gw}(f) =\Omega_0$.
The optimal filter function is normalized to have maximum magnitude of
unity.}
\end{center}
\end{figure}

\begin{figure}[htb!]
\begin{center}
{\epsfig{file=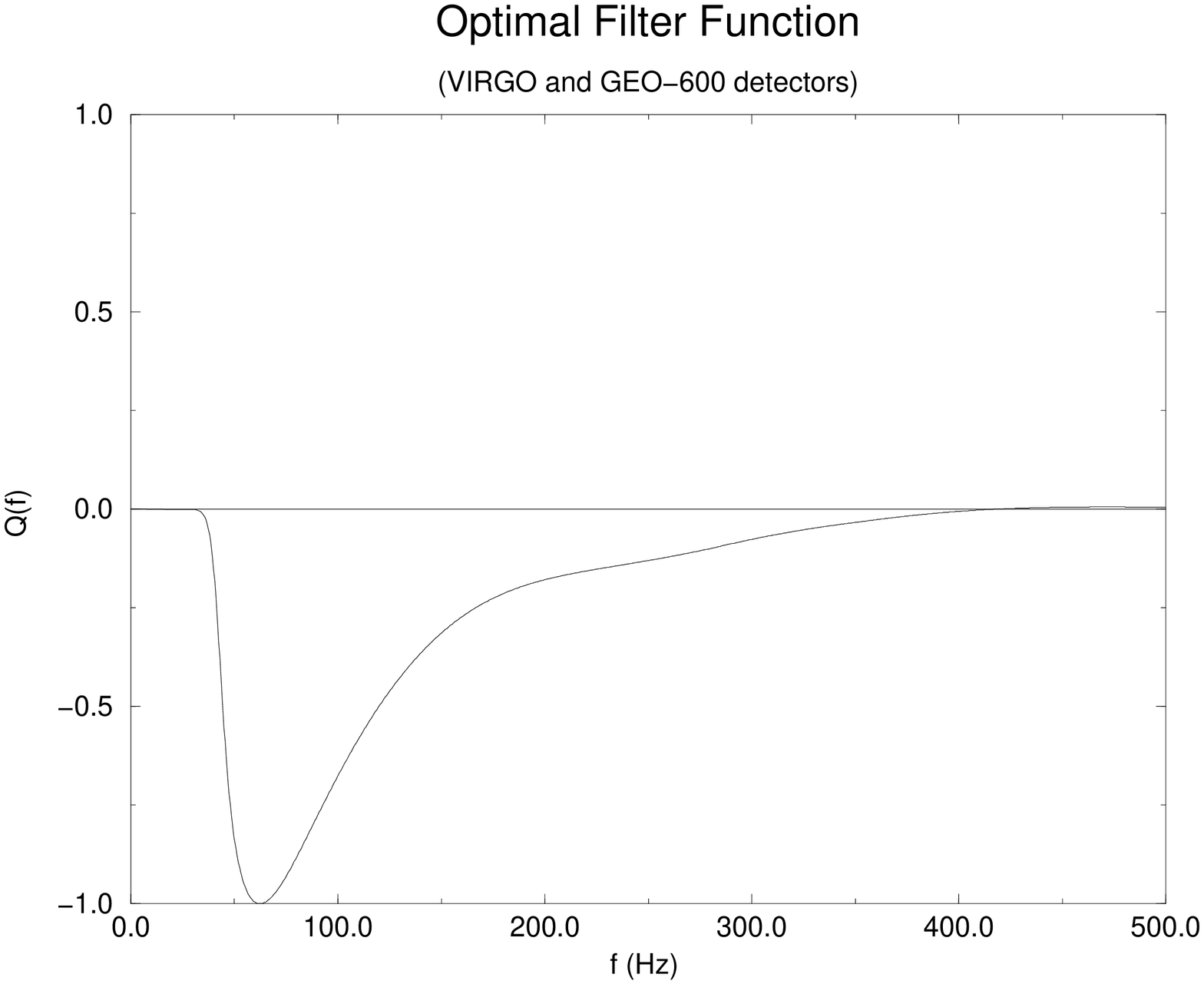,
angle=-90,width=3.4in,bbllx=25pt,bblly=50pt,bburx=590pt,bbury=740pt}}
\caption{\label{f:VIRGO_GEO-600_optimal}
The optimal filter function $\tilde Q(f)$ for the VIRGO and 
GEO-600 detectors,
for a stochastic background having a constant frequency spectrum
$\Omega_{\rm gw}(f) =\Omega_0$.
The optimal filter function is normalized to have maximum magnitude of
unity.}
\end{center}
\end{figure}

\begin{figure}[htb!]
\begin{center}
{\epsfig{file=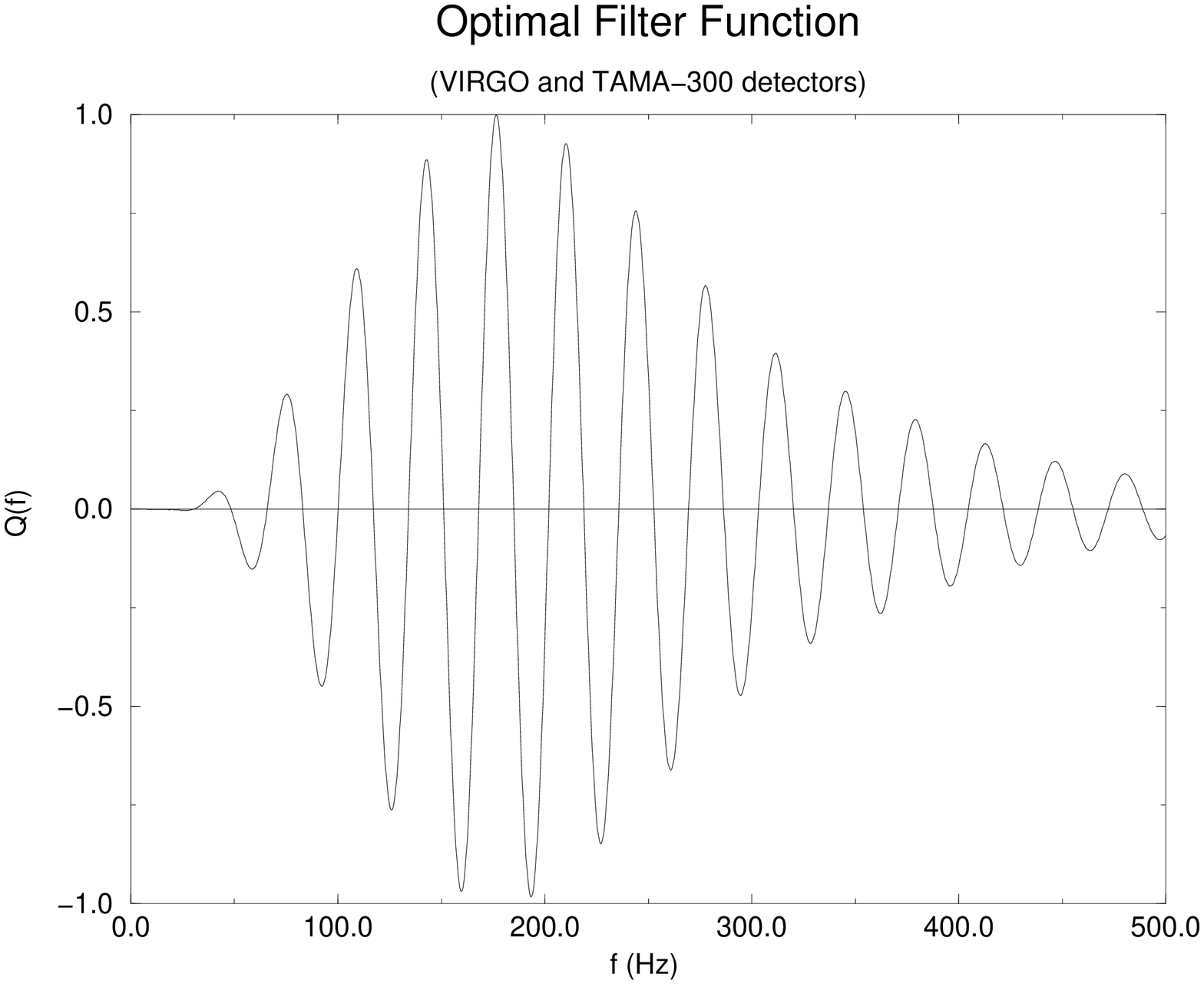,
angle=-90,width=3.4in,bbllx=25pt,bblly=50pt,bburx=590pt,bbury=740pt}}
\caption{\label{f:VIRGO_TAMA-300_optimal}
The optimal filter function $\tilde Q(f)$ for the VIRGO and 
TAMA-300 detectors,
for a stochastic background having a constant frequency spectrum
$\Omega_{\rm gw}(f) =\Omega_0$.
The optimal filter function is normalized to have maximum magnitude of
unity.}
\end{center}
\end{figure}

\begin{figure}[htb!]
\begin{center}
{\epsfig{file=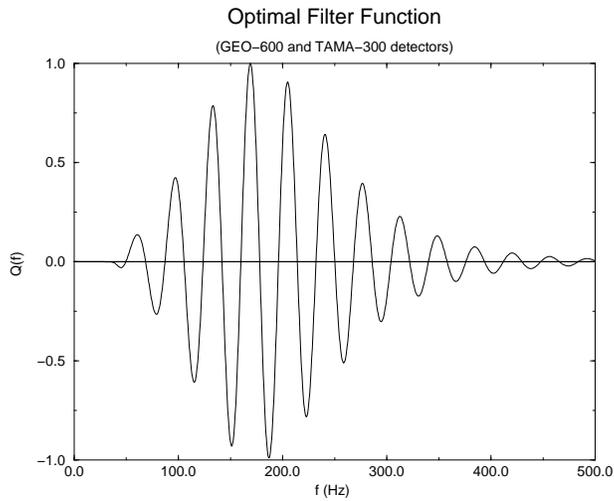,
angle=-90,width=3.4in,bbllx=25pt,bblly=50pt,bburx=590pt,bbury=740pt}}
\caption{\label{f:GEO-600_TAMA-300_optimal}
The optimal filter function $\tilde Q(f)$ for the GEO-600 and 
TAMA-300 detectors,
for a stochastic background having a constant frequency spectrum
$\Omega_{\rm gw}(f) =\Omega_0$.
The optimal filter function is normalized to have maximum magnitude of
unity.}
\end{center}
\end{figure}

\clearpage

\subsection{Signal-to-noise ratios and sensitivities}
\label{subsec:snrs}

Table~\ref{t:snr} contains the values of the theoretical signal-to-noise 
ratios after 4 months (i.e., $10^7$ sec) of observation, for the 
optimally-filtered cross-correlation signal between different detector pairs, 
for a stochastic background of gravitational radiation 
having a constant frequency spectrum
$\Omega_{\rm gw}(f)=\Omega_0=6\times 10^{-6}\ h_{100}^{-2}$.
Table~\ref{t:omega_min} contains the minimum values of
$\Omega_0\ h_{100}^{2}$ for 4 months of observation, for a false alarm 
rate equal to 5\%, and for a detection rate equal to 95\%,
for cross-correlation measurements between different detector pairs.
Tables~\ref{t:triples}-\ref{t:quadruples} contain the minimum values
$\Omega_0\ h_{100}^{2}$ for 4 months of observation, for a false alarm 
rate equal to 5\%, and for a detection rate equal to 95\%,
for the optimal combination of
cross-correlation measurements between multiple detector pairs, taken 
from all possible triples and quadruples of the five major interferometers.
Table~\ref{t:omega_min_4d} contains the minimum values of 
$\Omega_0\ h_{100}^{2}$ for 4 months of observation, for a false alarm 
rate equal to 5\%, and for a detection rate equal to 95\%,
for optimally-filtered 4-detector correlations.
(Note: The calculation of the signal-to-noise ratios and minimum values 
of $\Omega_0$ assume that the magnitude of the noise intrinsic to the
detectors is much larger than the stochastic gravity-wave background.
This corresponds to Eq.~(\ref{e:SNR_large_noise}) in the text.)

\mediumtext%
\begin{table}[htb!]
\begin{center}
\caption{\label{t:snr} Theoretical signal-to-noise 
ratios after 4 months of observation, for the optimally-filtered
cross-correlation signal between different detector pairs, 
for a stochastic background of gravitational radiation 
having a constant frequency spectrum
$\Omega_{\rm gw}(f)=\Omega_0=6\times 10^{-6}\ h_{100}^{-2}$.}
\begin{tabular}{l|ccccc}
          & LIGO-WA              & LIGO-LA              & VIRGO 
          & GEO-600              & TAMA-300             \\ \hline
LIGO-WA   & ---                  & 3.45                 & 1.74
          & $5.09\times 10^{-1}$ & $6.12\times 10^{-2}$ \\
LIGO-LA   & 3.45                 & ---                  & 2.10    
          & $7.66\times 10^{-1}$ & $9.16\times 10^{-2}$ \\
VIRGO     & 1.74                 & 2.10                 & ---
          & 1.56                 & $9.14\times 10^{-2}$ \\ 
GEO-600   & $5.09\times 10^{-1}$ & $7.66\times 10^{-1}$ & 1.56
          & ---                  & $1.32\times 10^{-2}$ \\
TAMA-300  & $6.12\times 10^{-2}$ & $9.16\times 10^{-2}$ & $9.14\times 10^{-2}$
          & $1.32\times 10^{-2}$ & ---                  \\
\end{tabular}
\end{center}
\end{table}

\begin{table}[htb!]
\begin{center}
\caption{\label{t:omega_min} Minimum values of 
$\Omega_0\ h_{100}^{2}$ for 4 months of observation, for a false alarm 
rate equal to 5\%, and for a detection rate equal to 95\%,
for cross-correlation measurements between different detector pairs.}
\begin{tabular}{l|ccccc}
          & LIGO-WA              & LIGO-LA              & VIRGO 
          & GEO-600              & TAMA-300             \\ \hline
LIGO-WA   & ---                  & $5.74\times 10^{-6}$ & $1.14\times 10^{-5}$
          & $3.89\times 10^{-5}$ & $3.24\times 10^{-4}$ \\
LIGO-LA   & $5.74\times 10^{-6}$ & ---                  & $9.45\times 10^{-6}$
          & $2.58\times 10^{-5}$ & $2.16\times 10^{-4}$ \\
VIRGO     & $1.14\times 10^{-5}$ & $9.45\times 10^{-6}$ & --- 
          & $1.27\times 10^{-5}$ & $2.17\times 10^{-4}$ \\
GEO-600   & $3.89\times 10^{-5}$ & $2.58\times 10^{-5}$ & $1.27\times 10^{-5}$
          & ---                  & $1.50\times 10^{-3}$ \\
TAMA-300  & $3.24\times 10^{-4}$ & $2.16\times 10^{-4}$ & $2.17\times 10^{-4}$
          & $1.50\times 10^{-3}$ & ---                  \\
\end{tabular}
\end{center}
\end{table}
\narrowtext\noindent%

\begin{table}[htb!]
\begin{center}
\caption{\label{t:triples} Minimum values of 
$\Omega_0\ h_{100}^{2}$ for 4 months of observation, for a false alarm 
rate equal to 5\%, and for a detection rate equal to 95\%,
for the optimal combination of cross-correlation measurements
between multiple detector pairs, taken from all possible triples of the 
five major interferometers.}
\begin{tabular}{lc}
Detectors & $\Omega_0^{95\%,5\%}\big|_{\rm optimal}\ h_{100}^2$  \\ \hline
LIGO-WA, LIGO-LA, VIRGO    & $4.50\times 10^{-6}$ \\
LIGO-WA, LIGO-LA, GEO-600  & $5.55\times 10^{-6}$ \\
LIGO-WA, LIGO-LA, TAMA-300 & $5.74\times 10^{-6}$ \\
LIGO-WA, VIRGO, GEO-600    & $8.28\times 10^{-6}$ \\
LIGO-WA, VIRGO, TAMA-300   & $1.14\times 10^{-5}$ \\
LIGO-WA, GEO-600, TAMA-300 & $3.86\times 10^{-5}$ \\
LIGO-LA, VIRGO, GEO-600    & $7.27\times 10^{-6}$ \\
LIGO-LA, VIRGO, TAMA-300   & $9.43\times 10^{-6}$ \\
LIGO-LA, GEO-600, TAMA-300 & $2.57\times 10^{-5}$ \\
VIRGO, GEO-600, TAMA-300   & $1.27\times 10^{-5}$ \\
\end{tabular}
\end{center}
\end{table}

\mediumtext%
\begin{table}[htb!]
\begin{center}
\caption{\label{t:quadruples} Minimum values of 
$\Omega_0\ h_{100}^{2}$ for 4 months of observation, for a false alarm 
rate equal to 5\%, and for a detection rate equal to 95\%,
for the optimal combination of cross-correlation measurements
between multiple detector pairs, taken from all possible quadruples of the 
five major interferometers.}
\begin{tabular} {lc}
Detectors & $\Omega_0^{95\%,5\%}\big|_{\rm optimal}\ h_{100}^2$ \\ \hline
LIGO-WA, LIGO-LA, VIRGO, GEO-600    & $4.17\times 10^{-6}$ \\
LIGO-WA, LIGO-LA, VIRGO, TAMA-300   & $4.50\times 10^{-6}$ \\
LIGO-WA, LIGO-LA, GEO-600, TAMA-300 & $5.54\times 10^{-6}$ \\
LIGO-WA, VIRGO, GEO-600, TAMA-300   & $8.28\times 10^{-6}$ \\
LIGO-LA, VIRGO, GEO-600, TAMA-300   & $7.27\times 10^{-6}$ \\
\end{tabular}
\end{center}
\end{table}

\mediumtext%
\begin{table}[htb!]
\begin{center}
\caption{\label{t:omega_min_4d} Minimum values of 
$\Omega_0\ h_{100}^{2}$ for 4 months of observation, for a false alarm 
rate equal to 5\%, and for a detection rate equal to 95\%,
for optimally-filtered 4-detector correlations.}
\begin{tabular} {lc}
Detectors & $\Omega_0^{95\%,5\%}\big|_{\rm optimal}\ h_{100}^2$ \\ \hline
LIGO-WA, LIGO-LA, VIRGO, GEO-600    & $6.49\times 10^{-6}$ \\
LIGO-WA, LIGO-LA, VIRGO, TAMA-300   & $2.51\times 10^{-5}$ \\
LIGO-WA, LIGO-LA, GEO-600, TAMA-300 & $5.44\times 10^{-5}$ \\
LIGO-WA, VIRGO, GEO-600, TAMA-300   & $4.68\times 10^{-5}$ \\
LIGO-LA, VIRGO, GEO-600, TAMA-300   & $3.84\times 10^{-5}$ \\
\end{tabular}
\end{center}
\end{table}

\narrowtext\noindent%

\clearpage

\section{Computer Simulation}
\label{sec:computer_simulation}

In Secs.~\ref{sec:detection}-\ref{sec:complications}, 
we described the data analysis and optimal signal processing required 
for the detection of a stochastic background of gravitational radiation.
This analysis was in terms of continuous time functions and their associated 
Fourier transforms.
But, in reality, when one performs the actual data analysis, continuous 
time functions will be replaced by {\em discrete\,} time-series and Fourier
transforms by their {\em discrete\,} frequency counterparts.
The discrete data can then be processed by computer code 
that takes the appropriate FFTs, constructs the optimal filters, whitens 
and windows the data, etc.
We have written a number of functions (in ANSI-C) to do precisely this.
These functions constitute part of a general-purpose data analysis package 
for gravitational-wave detection, called GRASP 
(Gravitational Radiation Analysis \& Simulation Package) \cite{GRASP}.
In this section, we describe a computer simulation (made up of these 
functions) that mimics the generation and detection of a simulated
stochastic gravity-wave signal in the presence of simulated detector noise. 
Documentation and further information about the code can be 
found in the GRASP user's manual.

\subsection{Purpose}
\label{subsec:purpose}

The main reason for writing the computer simulation was to verify many of
the theoretical calculations that were derived in the previous sections.
Specifically, we wanted to see if the theoretically predicted signal-to-noise 
ratio ${\rm SNR}$, for a stochastic background of gravitational radiation 
having a constant frequency spectrum $\Omega_{\rm gw}(f)=\Omega_0$,
would agree with an ``experimentally'' determined signal-to-noise ratio 
$\widehat{\rm SNR}$ produced by the simulation.
If the theoretical and experimental signal-to-noise ratios agreed, we
could be confident that the theoretical calculations were correct.
If they did not agree, we would know that something---either a theoretical 
calculation or a technical issue related to the simulation itself---needed 
further investigation.

The theoretical and experimental signal-to-noise ratios were said to 
be in agreement if the relative error defined by 
\begin{equation}
{\rm relative\ error}:=\left|{\widehat{\rm SNR}-{\rm SNR}\over{\rm SNR}}
\right|
\label{e:rel_error}
\end{equation}
was less than (or approximately equal to) the inverse of the theoretical 
signal-to-noise ratio after $n$ observation periods.%
\footnote{The total observation time is $T_{\rm tot}:=nT$, where $T$ is 
the duration of a single observation period.
For our simulations, $T=3.2768$ sec.}
This is the error that one would expect (approximately 68\% of the time) 
if we approximate the sample variance $\hat\sigma^2$ by the true variance 
$\sigma^2$, and use the fact that the sample mean
\begin{equation} 
\hat\mu:={1\over n}\sum_{i=1}^n S_i
\end{equation}
can itself be thought of as a random variable with mean $\mu$ and variance 
$\sigma^2/n$.
(Recall from Sec.~\ref{sec:detection_etc} 
that $S_1,S_2,\cdots,S_n$ are $n$ 
statistically independent measurements of the optimally-filtered
cross-correlation signal $S$, 
each associated with one of the $n$ observation periods.)
Thus, 
\widetext%
\begin{equation}
\widehat{\rm SNR}:={\hat\mu\over\hat\sigma}\approx
{\mu\pm \sigma/\sqrt{n}\over\sigma}=
{\mu\over\sigma}\pm{1\over\sqrt{n}}=
{\rm SNR}\pm{1\over\sqrt{n}}\ ,
\end{equation}
\narrowtext\noindent%
which implies
\begin{equation}
{\rm relative\ error}:=
\left|{\widehat{\rm SNR}-{\rm SNR}\over{\rm SNR}}\right|
\approx {1\over \sqrt{n}\ {\rm SNR}}\ .
\end{equation}
This criterion 
{\em was\,} satisfied by our simulation runs for both the initial and 
advanced LIGO detector pairs.
  
\subsection{Flow chart}
\label{subsec:flow_chart}

A ``flow chart'' for the simulation is as follows:

\begin{enumerate}
\item Input the parameters defining the simulation.
This can be done either interactively or via {\tt \#define} statements
at the beginning of the simulation program.
These parameters are:
\begin{itemize}
\item[(i)] the site identification numbers for the two detectors.
\item[(ii)] the number of time-series data points $N$ to be used when 
performing the data analysis (i.e., FFTs, cross-correlations, etc.).
$N$ should equal an integral power of 2.
\item[(iii)] the sampling period $\Delta t$ of the two detectors.
(Note: $T:=N\Delta t$ is the duration of a single observation period.)
\item[(iv)] the constant $\Omega_0$ that defines the stochastic 
background frequency spectrum: $\Omega_{\rm gw}(f)=\Omega_0$.
\item[(v)] the total number of runs $n$ that make up the simulation.
(Note: $T_{\rm tot}:=nT$ is the duration of the total observation 
period.)
\end{itemize}
\item Using the site identification numbers, obtain site location and 
orientation information, and information about the noise power spectrum 
and whitening filter of each detector.
(This information is contained in input data files.)
\item Using the site location and orientation information, construct 
the overlap reduction function $\gamma(f_i)$ for the two detectors.
(Here $f_i:=i/(N\Delta t)$ where $i=0,1,\cdots, N/2-1$.
By convention, we ignore the value of $\gamma(f)$, or any other function 
of frequency, at or above the Nyquist critical frequency 
$f_{\rm Nyquist}:=1/(2\Delta t)$.)
\item Simulate the generation of a stochastic background of gravitational
radiation having a constant frequency spectrum 
$\Omega_{\rm gw}(f)=\Omega_0$.
This can be done in the frequency domain by using a random number
generator to construct (complex-valued) Gaussian random variables 
$\tilde h_1(f_i)$ and $\tilde h_2(f_i)$ having zero mean and joint 
expectation values:
\begin{eqnarray}
\langle \tilde h_1^*(f_i)\tilde h_1(f_j)\rangle&=&
{3H_0^2\over 20\pi^2}\ T\ \delta_{ij}\ f_i^{-3}\ \Omega_0\ ,\\
\langle \tilde h_2^*(f_i)\tilde h_2(f_j)\rangle&=&
{3H_0^2\over 20\pi^2}\ T\ \delta_{ij}\ f_i^{-3}\ \Omega_0\ ,\\
\langle \tilde h_1^*(f_i)\tilde h_2(f_j)\rangle&=&
{3H_0^2\over 20\pi^2}\ T\ \delta_{ij}\ f_i^{-3}\ \Omega_0\ \gamma(f_i)\ .
\end{eqnarray}
These are just the discrete frequency versions of 
Eq.~(\ref{e:h_1(f)h_2(f')}) specialized to the case
$\Omega_{\rm gw}(f)=\Omega_0$.
Note that the Fourier amplitudes $\tilde h_1(f_i)$ and $\tilde h_2(f_i)$ 
fall-off like $f^{-3/2}$.
\item Simulate the generation of the noise intrinsic to the detectors,
using the information contained in the noise power spectrum data files.
This can be done in the frequency domain by using a random number
generator to construct (complex-valued) Gaussian random variables 
$\tilde n_1(f_i)$ and $\tilde n_2(f_i)$ having zero mean and joint 
expectation values:
\begin{eqnarray}
\langle\tilde n_1^*(f_i)\tilde n_1(f_j)\rangle&=&
{1\over 2}\ T\ \delta_{ij}\ P_1(f_i)\ ,\\
\langle\tilde n_2^*(f_i)\tilde n_2(f_j)\rangle&=&
{1\over 2}\ T\ \delta_{ij}\ P_2(f_i)\ ,\\
\langle\tilde n_1^*(f_i)\tilde n_2(f_j)\rangle&=&0\ .
\end{eqnarray}
These are just the discrete frequency versions of 
Eq.~(\ref{e:n_i(f)n_i(f')}).
\item Construct the Fourier amplitudes 
\begin{eqnarray}
\tilde s_1(f_i)&:=&\tilde h_1(f_i)+\tilde n_1(f_i)\ ,\\
\tilde s_2(f_i)&:=&\tilde h_2(f_i)+\tilde n_2(f_i)\ .
\end{eqnarray}
\item Whiten the data in the frequency domain by multiplying 
$\tilde s_1(f_i)$ and $\tilde s_2(f_i)$ by the frequency components
$\tilde W_1(f_i)$ and $\tilde W_2(f_i)$ of the whitening filters
of the two detectors:
\begin{eqnarray}
\tilde o_1(f_i)&:=&\tilde s_1(f_i)\ \tilde W_1(f_i)\ ,\\
\tilde o_2(f_i)&:=&\tilde s_2(f_i)\ \tilde W_2(f_i)\ .
\end{eqnarray}
This multiplication in the frequency domain corresponds to 
the convolution of $s_1(t_i)$ and $W_1(t_i)$ 
($s_2(t_i)$ and $W_2(t_i)$) in the time domain.
(Note: The purpose of whitening the data is to reduce the dynamic 
range of the corresponding power spectra.)
\item FFT $\tilde o_1(f_i)$ and $\tilde o_2(f_i)$ into the time domain 
to obtain the corresponding time-series $o_1(t_i)$ and $o_2(t_i)$.
(Here $t_i:=i\Delta t$ where $i=0,1,\cdots N-1$.)
\item Repeat steps 4-8 to obtain another set of time-series
data $o_1(t_i)$ and $o_2(t_i)$.
Distinguish these two different sets of data with superscripts:
${}^{(1)}\!o_1(t_i)$, ${}^{(1)}\!o_2(t_i)$, ${}^{(2)}\!o_1(t_i)$, 
${}^{(2)}\!o_2(t_i)$.
\item Offset ${}^{(1)}\!o_1(t_i)$ and ${}^{(2)}\!o_1(t_i)$  
(${}^{(1)}\!o_2(t_i)$ and ${}^{(2)}\!o_2(t_i)$) by $T/2$, and combine them 
with one another and with data left over from the previous observation period 
to produce a {\em continuous-in-time\,} data set $o_1(t_i)$ ($o_2(t_i)$).
$o_1(t_i)$ and $o_2(t_i)$ represent the ``raw'' (i.e., whitened) signal+noise 
data streams output by the two detectors.

(Note: Steps 4-10 make up the signal generation part of the simulation.)
\item Test the input data $o_1(t_i)$ and $o_2(t_i)$ to see if they have 
probability distributions consistent with that of a Gaussian random variable.
If either set fails this test, reject them both, and repeat steps 4-10 to 
obtain new input data $o_1(t_i)$ and $o_2(t_i)$.

\item Window the data streams $o_1(t_i)$ and $o_2(t_i)$ in the
time domain, using a Hann window function to reduce side-lobe
contamination of the corresponding power spectra.
\item FFT the windowed data into the frequency domain to obtain the 
corresponding Fourier amplitudes $\tilde o_1(f_i)$ and $\tilde o_2(f_i)$.
\item Unwhiten the data in the frequency domain by dividing 
$\tilde o_1(f_i)$ and $\tilde o_2(f_i)$ by the frequency components
$\tilde W_1(f_i)$ and $\tilde W_2(f_i)$ of the whitening filters
of the two detectors:
\begin{eqnarray}
\tilde s_1(f_i)&:=&{\tilde o_1(f_i)\over \tilde W_1(f_i)}\ ,\\
\tilde s_2(f_i)&:=&{\tilde o_2(f_i)\over \tilde W_2(f_i)}\ .
\end{eqnarray}
\item Shift the input data streams $o_1(t_i)$ and $o_2(t_i)$ forward
in time by $T/2$, and repeats steps (12-14), obtaining another set
of Fourier components $\tilde s_1(f_i)$ and $\tilde s_2(f_i)$.
Distinguish these two different sets of data with superscripts:
${}^{(1)}\!\tilde s_1(f_i)$, ${}^{(1)}\!\tilde s_2(f_i)$, 
${}^{(2)}\!\tilde s_1(f_i)$, ${}^{(2)}\!\tilde s_2(f_i)$.
\item Average 
${}^{(1)}\!\tilde s_1(f_i)$ and ${}^{(2)}\!\tilde s_1(f_i)$
(${}^{(1)}\!\tilde s_2(f_i)$ and ${}^{(2)}\!\tilde s_2(f_i)$) 
to produce $\tilde s_1(f_i)$ ($\tilde s_2(f_i)$).
$\tilde s_1(f_i)$ and $\tilde s_2(f_i)$ are the Fourier components of 
the unwhitened time-series data $s_1(t_i)$ and $s_2(t_i)$.
(Note: The purpose of this averaging is to reduce the variance in the
estimation of the spectra $\tilde s_1(f_i)$ and $\tilde s_2(f_i)$.)
\item Construct the optimal filter function $\tilde Q(f_i)$ with the 
overall normalization constant $\lambda$ chosen so that 
$\mu=\Omega_0\ T$, using the noise power spectra specified by the
input data files.
\item From $\tilde s_1(f_i)$, $\tilde s_2(f_i)$, and $\tilde Q(f_i)$ 
calculate the optimally-filtered cross-correlation signal $S$ 
corresponding to a single observation period $T$.

(Note: Steps 11-18 make up the signal analysis part of the simulation.)
\item Repeat steps 4-18 $n$ times, generating a set of optimally-filtered
cross-correlation signal values: $S_1,S_2,\cdots,S_n$. 
\item From $S_1,S_2,\cdots,S_n$ construct the sample mean
\begin{equation}
\hat\mu:={1\over n}\sum_{i=1}^n S_i
\end{equation}
and sample variance
\begin{equation}
\hat\sigma^2:={1\over n-1}\sum_{i=1}^n(S_i-\hat\mu)^2\ .
\end{equation}
The sample (or ``experimental'') signal-to-noise ratio produced by
the simulation is given by
\begin{equation}
\widehat{\rm SNR}:={\hat\mu\over\hat\sigma}\ .
\label{e:snr_expt}
\end{equation}
\item Calculate the theoretical signal-to-noise ratio, using a discrete
frequency approximation to the integral
\widetext%
\begin{equation}
{\rm SNR}=\Omega_0\ {3H_0^2\over 10\pi^2}\ \sqrt{T}\ 
\left[2\int_0^{f_{\rm Nyquist}} df\
{\gamma^2 (f)\over f^6 P_1(f) P_2(f)}
\right]^{1/2}\ .
\label{e:snr_theory}
\end{equation}
\narrowtext\noindent%
See Eq.~(\ref{e:SNR_large_noise}).
(Note: Since the data is discretely sampled, we should only integrate
up to the Nyquist critical frequency $f_{\rm Nyquist}:=1/(2\Delta t)$.)
\item From $\widehat{\rm SNR}$ and SNR, calculate the relative error
\begin{equation}
{\rm relative\ error}:=\left|{\widehat{\rm SNR}-{\rm SNR}\over{\rm SNR}}
\right|\ .
\end{equation}
As mentioned in Sec.~\ref{subsec:purpose}, this should be compared with 
the inverse of the theoretical signal-to-noise ratio after $n$ observation 
periods $1/(\sqrt{n}\ {\rm SNR})$.

(Note: Steps 20-22 make up the statistical analysis part of the
simulation.)
\end{enumerate}

Note: 
In order to obtain signal-to-noise ratios on the order of 10 
after $n=1600$ runs, we needed to use rather large values of $\Omega_0$
(e.g., $10^{-3}$ for the initial LIGO detectors, and $10^{-8}$ for 
the advanced LIGO detectors).
These large values meant that expression (\ref{e:snr_theory}) 
for the theoretical 
signal-to-noise ratio had to be modified to properly take into account
the contributions to the theoretical variance $\sigma^2$ that are due to 
a large stochastic gravity-wave signal.
(See Eqs.~(\ref{e:sigma^2_new}) and (\ref{e:R(f)}).)
Without these modifications, the theoretical and experimental 
signal-to-noise ratios would be more likely to disagree.
Thus, instead of (\ref{e:snr_theory}), we used a ``mixed'' expression
for the theoretical signal-to-noise ratio:
\widetext%
\begin{equation}
{\rm SNR}=\Omega_0\ {3H_0^2\over 10\pi^2}\ \sqrt{T}\ 
{\sqrt{2}\ 
\int_0^{f_{\rm Nyquist}} df\ {\gamma^2(f)\over f^6
P_1(f)P_2(f)}\over 
\left[\ \int_0^{f_{\rm Nyquist}} df\ {\gamma^2(f)\over f^6
P_1^2(f) P_2^2(f)} R(f)\ \right]^{1/2}}\ .
\end{equation}
\narrowtext\noindent%
This involves Eqs.~(\ref{e:sigma^2_new}) and (\ref{e:R(f)}) for the
variance $\sigma^2$, but the large noise expression (\ref{e:optimal}) 
for the optimal filter function $\tilde Q(f)$.

\section{Conclusion}
\label{sec:conclusion}

In this paper, we derived the optimal signal processing strategy
required for stochastic background searches.
We discussed signal detection, parameter estimation, and sensitivity 
levels from a frequentist point of view.
We also discussed the complications that arise when one 
considers:
(i) arbitrarily large stochastic backgrounds,
(ii) non-stationary detector noise,
(iii) multiple detector pairs, and 
(iv) correlated detector noise.
We explained how we verified some of the theoretical calculations 
by writing a computer simulation that mimics the generation and 
detection of a simulated stochastic gravity-wave signal in the 
presence of simulated detector noise. 
And we noted that the ``experimental'' results and theoretical
predictions agreed to within the expected error.
These results suggest that both the theoretical signal processing 
formulae and the implementation of these formulae in computer 
code are correct.
But we should not stop here.
For example, there are still a number of ways that we can 
improve the data analysis code before we use it to search for a 
{\em real\,} stochastic background in the outputs of {\em real\,} 
interferometers.
To conclude this paper, we list some of the desired improvements 
below:

(i) The first change that we would like to make is to calculate
{\em real-time\,} noise power spectra for the detectors, and to use 
this calculated data (rather than the information contained in the 
input noise power spectrum data files) to construct the optimal filter 
function $\tilde Q(f_i)$.
(See step 17 in the computer simulation described in 
Sec.~\ref{subsec:flow_chart}.)
Since the real-time noise power spectra will change slightly from
one measurement to the next, we could then apply the data analysis 
strategy discussed in Sec.~\ref{subsec:nonstationary_detector_noise}
for nonstationary detector noise.

(ii) In order to obtain accurate real-time noise power spectra
for the two detectors, it will probably be necessary to use more 
sophisticated spectral estimation techniques.
Currently, we use a Hann window to reduce side-lobe contamination, and
we average two overlapped data sets to reduce the variance, when forming 
our estimates of $\tilde s_1(f_i)$ and $\tilde s_2(f_i)$.
(See steps 12, 15, and 16 of the computer simulation.)
This procedure can be replaced by {\em multitaper} spectral estimation
methods, which use a special set of window functions---called Slepian
tapers---to form spectral estimates of time-series data.%
\footnote{See the original paper by Thomson \cite{Thomson} and the 
text by Percival and Walden \cite{PW} for more details.}
GRASP \cite{GRASP} contains a modified version of a public domain
package by Lees and Park \cite{LP} to perform the multitaper spectral
estimation.
In addition to providing better spectral estimates, multitaper methods
also provide nice techniques for ``spectral line'' parameter estimation 
and removal.
This feature will be extremely useful when analyzing data produced by
a real detector.
For example, one will be able to track contamination of a data set by
the line harmonic at 300 Hz, and remove a pendulum resonance at say,
590 Hz.
(See GRASP \cite{GRASP} for more information.)

(iii) In addition to being able to identify and to remove ``spectral 
lines'' from a real data set, one would also like to be able to test 
the data to see if the distribution of sampled values is consistent 
with normal detector operation.
For example, one might check the input data set to see if it has a 
probability distribution consistent with that of a Gaussian random
variable.
If the test reveals an exceptionally large number of ``outlier'' points,
then that particular data set can be rejected.
(See step 11 of the computer simulation.)
The GRASP data analysis package already contains a routine that performs
this Gaussian test.
But we would also like a more rigorously characterized test that
compares the distribution of the current data with that during ``normal'' 
detector operation, which most likely is {\em not} Gaussian.

(iv) Finally, even though it will still be a few years before we can 
analyze real data from any one of the major interferometers, real data 
from {\em prototypes}---like the Caltech 40-meter interferometer---can
be used in computer simulations.
For instance, rather than write a computer simulation 
(like the one we described in Sec.~\ref{sec:computer_simulation}) 
that mimics the generation and detection of a simulated stochastic 
gravity-wave signal in the presence of {\em simulated\,} detector noise, 
we can write a computer
simulation that mimics the generation and detection of a simulated 
stochastic gravity-wave signal in the presence of {\em real\,} detector 
noise.%
\footnote{The real detector noise would be provided by the prototype output.}
The fact that the noise level of a prototype interferometer is much 
larger than that of a major interferometer poses no problem;
we can simply ``dial-in'' a larger stochastic background signal to be
able to detect it in the same amount of observation time.
Another nice feature of this 
{\em fake\,} stochastic background/{\em real\,} detector noise 
simulation is that we can address all of the issues (i)-(iii) 
discussed above in a context where we can still compare ``experimental''
(i.e., simulation) performance against theoretical expectations.
We must be totally convinced that the data analysis code is working as 
expected, before we can trust it when searching for a real
stochastic background in the outputs of real interferometers.

We are currently working on all of the above topics.
We hope to complete and report on these projects sometime in the near 
future.

\acknowledgments
This work has been partially supported by NSF grant PHY95-07740.
We would like to thank Luca Gammaitoni, Albrecht Rudiger, 
Kenneth Strain, Masa-Katsu Fujimoto, and Rainer Weiss for kindly supplying 
the numerical data for the noise power spectra for the VIRGO, GEO-600, 
TAMA-300, and ``enhanced'' LIGO detectors.
We would also like to thank Sam Finn for carefully proofreading
Sec.~\ref{sec:detection_etc}, and for explaining the differences 
between the Bayesian and frequentist approaches to data analysis,
and Eric Key for bringing the Law of the Iterated Logarithm to our 
attention.
BA gratefully acknowledges the LIGO visitors program for support 
under NSF grant PHY96-03177, and useful conversations with 
Kent Blackburn, Ron Dreever, Eanna Flanagan, B.S.~Sathyaprakash, 
David Shoemaker, Kip Thorne, Robbie Vogt, Rainer Weiss, and Stan Whitcomb.
\clearpage



\begin{references}

\bibitem{science92}
A.~Abramovici et al., 
{\em Science} {\bf 256}, 325 (1992).

\bibitem{virgo}
B.~Caron et al., 
in {\em Gravitational Wave Experiments}, proceedings of the Edoardo Amaldi 
Conference, World Scientific, 86 (1995).

\bibitem{geo600}
K.~Danzmann et al.,
in {\em Gravitational Wave Experiments}, proceedings of the Edoardo Amaldi 
Conference, World Scientific, 100 (1995).

\bibitem{tama300}
K.~Tsubono,
in {\em Gravitational Wave Experiments}, proceedings of the Edoardo Amaldi 
Conference, World Scientific, 112 (1995).

\bibitem{mich}
P.~F.~Michelson, 
{\em Mon. Not. Roy. Astron. Soc.} {\bf 227}, 933 (1987).

\bibitem{chris}
N.~Christensen, 
{\em Phys. Rev.} {\bf D46}, 5250 (1992).

\bibitem{flan}
E.~Flanagan, 
{\em Phys. Rev.} {\bf D48}, 2389 (1993).

\bibitem{leshouches}
B.~Allen,
in {\em Proceedings of the Les Houches School on Astrophysical Sources of 
Gravitational Waves}, 
eds. J.A.~Marck and J.P.~Lasota, Cambridge, 373 (1997).

\bibitem{kolbturner}
E.W.~Kolb and M.~Turner,
{\em The Early Universe},
(Frontiers in Physics, Addison Wesley, 1990).

\bibitem{allenottewill}
B.~Allen and A.C.~Ottewill,
{\em Phys. Rev.} {\bf D55}, 15 (1997).

\bibitem{Blair}
D.G.~Blair, R.~Burman, L.~Ju, S.~Woodings, M.~Mulder, M.G.~Zadnik,
``The supernova cosmological background of gravitational waves,''    
Preprint, University of Western Australia, 1997;
%
V. Ferraria, 
in {\em Proceedings of the XII Italian conference on GR and Gravitational
Physics}, 
(World Scientific, 1997);
%
D.G.~Blair and L.~Ju, 
``A cosmological background of gravitational waves produced by supernovae 
in the early universe,'' 
{\em Mon. Not. R. Astron. Soc.}, to appear, (1996).

\bibitem{MTW}
C.W.~Misner, K.S.~Thorne, and J.A.~Wheeeler,
{\em Gravitation},
(W.H.~Freeman and Company, San Francisco, 1973).

\bibitem{cobea}
G.F.~Smoot et al.,
{\em Astrophys. J.} {\bf 396}, L1 (1992).

\bibitem{cobeb}
C.L.~Bennett et al.,
{\em Astrophys. J.} {\bf 396}, L7 (1992).

\bibitem{cobe2}
C.L.~Bennett et al., 
{\em Astrophys. J.} {\bf 436}, 423 (1994).

\bibitem{cobe4}
C.L.~Bennett et al.,
{\em Astrophys. J.} {\bf 464}, L1 (1996).

\bibitem{Taylor}
V.~Kaspi, J.~Taylor, and M.~Ryba,
{\em Astrophys. J.} {\bf 428}, 713 (1994).

\bibitem{lsf1}
L.S.~Finn,
``Gravitational-wave data analysis with multiple detectors:
The gravitational-wave receiver. I. Deterministic sources,''
in preparation, (1997).

\bibitem{lsf2}
L.S.~Finn,
``Gravitational-wave data analysis with multiple detectors:
The gravitational-wave receiver. II. Stochastic signals,''
in preparation, (1997).

\bibitem{mf} 
I.~Miller and J.E.~Freund, 
{\em Probability and Statistics for Engineers}, 
(Prentice-Hall, Inc., Englewood Cliffs, NJ, 1985).

\bibitem{handbook}
W.H.~Beyer,
{\em CRC Standard Probability and Statistics Tables and Formulae},
(CRC Press, Boca Raton, 1991).
 
\bibitem{helstrom}
C.W.~Helstrom,
{\em Statistical theory of signal detection}, 2nd edition,
(Pergamon Press, Oxford, 1968).

\bibitem{feller}
W.~Feller,
{\em An Introduction to Probability Theory and Its Applications},
Volume One, 
(John Wiley, New York, 1950).

\bibitem{christensen_thesis}
N.~Christensen, 
``On measuring the stochastic gravitational radiation background
with laser interferometric antennas,''
Ph.D. Thesis, Massachusetts Institute of Technology, 1990.

\bibitem{coyne}
D.~Coyne, 
LIGO project, private communication, 1997.

\bibitem{noise_curves}
The data for the predicted noise power spectra for the initial and 
advanced LIGO detectors were taken from \protect\cite{science92}.
Those for the VIRGO detector were supplied by Luca Gammaitoni 
(e-mail: {\tt gammaitoni@perugia.infn.it});
the GEO-600 detector by Albrecht Rudiger and Kenneth Strain
(e-mail: {\tt atr@mpq.mpg.de} and {\tt kstrain@physics.gla.ac.uk});
and the TAMA-300 detector by Masa-Katsu Fujimoto
(e-mail: {\tt fujimoto@gravity.mtk.nao.ac.jp}).
The predicted noise power spectra for the ``enhanced'' LIGO detectors 
were supplied by Rainer Weiss
(e-mail: {\tt weiss@tristan.mit.edu}).

\bibitem{GRASP}
B.~Allen,
``GRASP: a data analysis package for gravitational wave detection,''
(1997).
An up-to-date version of the users manual may be obtained at:
{\tt http://www.ligo.caltech.edu/LIGO\_web/Collaboration/manual.pdf} or
{\tt http://www.ligo.caltech.edu/LIGO\_web/Collaboration/lsc\_interm.html}.
The software package is available upon request.

\bibitem{Thomson}
D.J.~Thomson, 
{\em Proceedings of the IEEE} {\bf 70}, 1055 (1982).

\bibitem{PW}
D.B~Percival and A.T.~Walden,
{\em Spectral analysis for physical applications},
(Cambridge University Press, Cambridge, 1993).
 
\bibitem{LP}
J.M.~Lees and J.~Park,
{\em Computers \& Geology} {\bf 21}, 199 (1995).
The multitaper spectral estimation public domain package can be found at 
the website {\tt http://love.geology.yale.edu/mtm/}.

\end{references}
\end{document}